\documentclass[reprint,aps,prb,nofootinbib,superscriptaddress,longbibliography,preprintnumbers]{revtex4-1}
\usepackage{amssymb}
\usepackage{setspace}
\usepackage{amsmath}
\usepackage{amsfonts}
\usepackage{hyperref}
\usepackage{graphicx}
\usepackage{float}
\usepackage{subfigure}
\usepackage{color}

\usepackage{feynmp}
\DeclareMathOperator{\Tr}{Tr}
\DeclareMathOperator{\im}{Im}

\newcommand{\vect}[1]{\boldsymbol{#1}}

\newcommand{\id}{\mathbb{I}}

\newcommand{\Mi}{{\bf{M}}_{i}}
\newcommand{\cMi}{{\bf{M}}_{i}^{*}}
\newcommand{\cur}{{\bf{\Phi}}_{i}}
\newcommand{\ccur}{{\bf{\Phi}}_{i}^{*}}
\newcommand{\vs}{\vec{\sigma}}
\newcommand{\dg}{\dagger}

\newcommand{\mc}[1]{\mathcal{#1}}

\newcommand{\cd}{c^{\dagger}}

\newcommand{\ed}{e^{\dagger}}
\newcommand{\fd}{f^{\dagger}}
\newcommand{\pd}{p^{\dagger}}

\newcommand{\bK}{{\bf{K}}}

\newcommand{\kv}{{\boldsymbol{k}}}

\newcommand{\pv}{{\boldsymbol{p}}}
\newcommand{\qv}{{\boldsymbol{q}}}

\newcommand{\be}{\begin{equation}}
\newcommand{\ee}{\end{equation}}
\newcommand{\non}{\nonumber}

\newcommand*\diff{\mathop{}\!\mathrm{d}}
\newcommand*\Diff[1]{\mathop{}\!\mathrm{d^#1}}
\newcommand{\sdw}[1]{{\bf M}_{#1}}
\newcommand{\bsdw}[1]{{\bar {\bf M}}_{#1}}
\newcommand{\br}{{\bf {r}}}
\allowdisplaybreaks

\begin{document}
\title{Itinerant fermions on a triangular lattice:  \\ unconventional magnetism and other ordered states}\author{Mengxing Ye}
\affiliation{School of Physics and Astronomy, University of Minnesota, Minneapolis, MN 55455}
\affiliation{Kavli Institute for Theoretical Physics, University of California Santa Barbara, Santa Barbara, CA 93106}
\author{Andrey V. Chubukov}
\affiliation{School of Physics and Astronomy, University of Minnesota, Minneapolis, MN 55455}
\date{\today}
\begin{abstract}
We consider a system of 2D fermions on a triangular lattice with well separated electron and hole pockets of similar sizes, centered at certain high-symmetry-points in the Brillouin zone. We first analyze Stoner-type spin-density-wave (SDW) magnetism.  We show that SDW order is degenerate at the mean-field level. Beyond mean-field, the degeneracy is lifted  and is either $120^{\circ}$ ``triangular" order (same as for localized spins), or a collinear order with antiferromagnetic spin arrangement on two-thirds of sites, and non-magnetic on the rest of sites.
We also study a time-reversal symmetric directional spin bond order, which emerges when some interactions are repulsive and some are attractive.  We show that this order is also degenerate at a mean-field level, but beyond mean-field the degeneracy is again lifted. We next consider the evolution of a magnetic order in a magnetic  field starting  from an SDW state in zero field. We show that a field gives rise to a canting of an SDW spin configuration. In addition, it necessarily triggers the directional bond order, which, we argue,  is linearly coupled to the SDW order in a finite field. We derive the corresponding term in the Free energy. Finally, we consider the interplay between an SDW order and superconductivity and charge order. For this, we analyze the flow of the couplings within parquet renormalization group (pRG) scheme. We show that magnetism wins if all interactions are repulsive and there is little energy space for pRG to develop. However,  if system parameters are such that pRG runs over a wide range of energies, the system may develop either superconductivity or an unconventional charge order, which breaks time-reversal symmetry.
  \end{abstract}
\maketitle

\section{Introduction}

The nature of a magnetic order in itinerant electron systems and the interplay between magnetism, superconductivity, and charge order has attracted a substantial interest in the last decade~\cite{Ishida_review,Graser09,Johnston10,Paglione10,Stewart11,Canfield_review,Vorontsov10,RMF14,Chubukov_review, Chubukov_book,HHWen_review,DHLee_review,Thomale_review,Scalapino_review,Coldea2018,Keimer2015,Fradkin2010,Lacroix2010}, chiefly in the context of the analysis of cuprate and iron-based superconductors (FeSCs). Recently, studies of itinerant magnetism and its interplay with other orders have been extended to include itinerant systems on hexagonal lattices, like doped graphene~\cite{Batista2008,Chubukov2012,Batista2012,Starykh2013,Thomale_12} and transition metal dichalcogenides (TMDs)~\cite{Ganesh2014}.  In localized spin system, a magnetic order on a hexagonal lattice (a triangular, honeycomb, or a Kagome lattice) is strongly influenced by geometrical frustration~\cite{Sachdev_92b,Chubukov1992K,IntroFM,Moessner2001,Starykh2015}, and in certain cases a classical ground state magnetic configuration can be infinite degenerate, like in, e.g., an  antiferromagnet on a Kagome lattice with nearest-neighbor Heisenberg interaction. However, such degeneracy is almost certainly  lifted by interactions involving further neighbors~\cite{Chubukov1992K,Domenge2005}.

In itinerant systems, relevant interactions are in general long-ranged in real space as they involve fermions near particular $k-$points in the Brillouin zone, where Fermi surfaces (FSs) are located. Yet, magnetism in itinerant systems also shows a strong frustration, this time because of competition between several symmetry-equivalent magnetic orderings between different FSs.  This holds already in systems on non-frustrated lattices, e.g., in square lattice systems with a circular hole FS at $(0,0)$ and electron FSs at $(0,\pi)$ and $(\pi,0)$ (similar to parent compounds of Fe-pnictides).  A dipole spin-density-wave (SDW) order parameter in such a system can be ${\bf M}_1$ with momenta $(\pi,0)$ or ${\bf M}_2$ with momentum $(0,\pi)$.  At a mean-field level, the Free energy depends on ${\bf M}^2_1 + {\bf M}^2_2$, i.e., the ground state is infinitely degenerate.   The degeneracy is lifted  either by changing the FS geometry, e.g.,  making the electron pockets non-circular,  or by adding other interactions between fermions near hole and electron pockets~\cite{Lorenzana08,Chubukov2010M,Fernandes12,Fernandes2016}, which do not contribute to SDW instability at the mean-field level, but distinguish between different ordered states from a degenerate manifold.

\begin{figure}
  \centering
\includegraphics[width =0.9\linewidth]{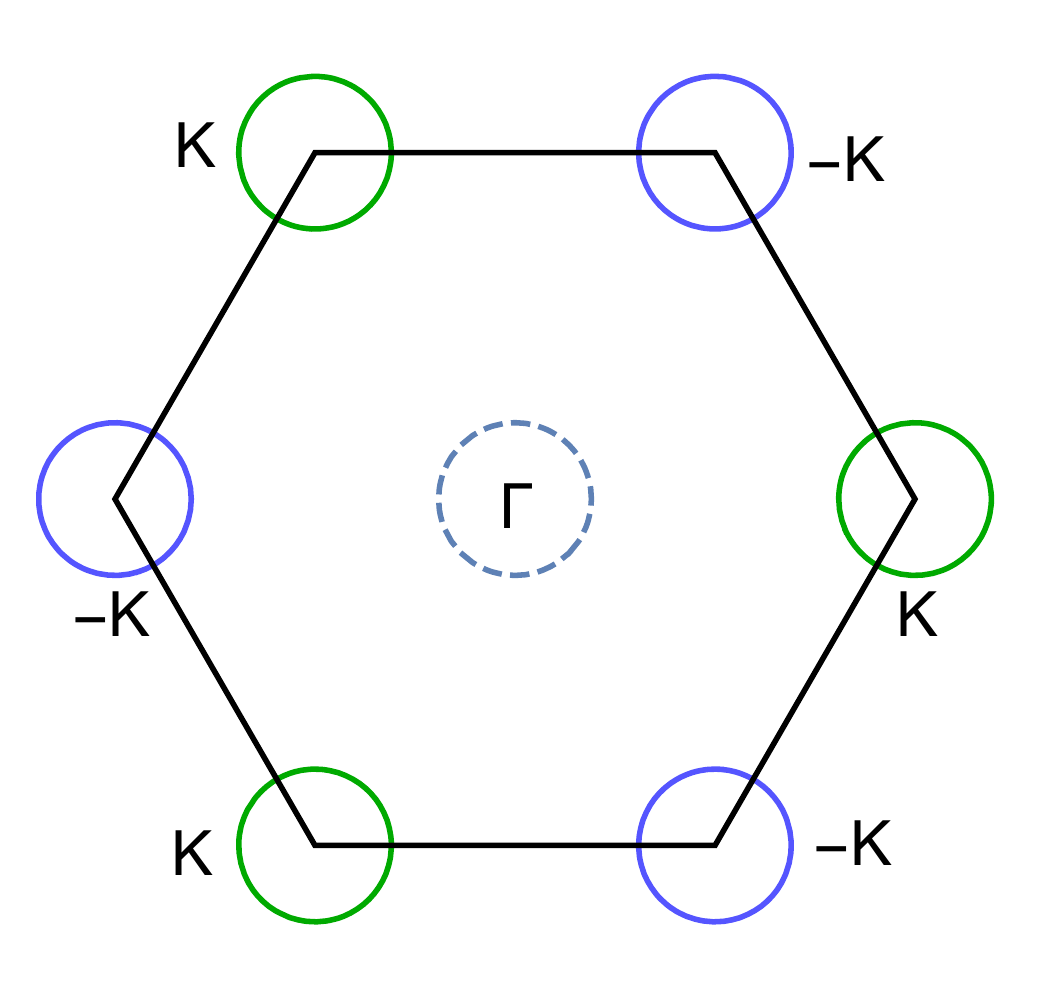}
\caption{ The Brillouin zone and the locations of the Fermi surfaces.  There is one hole pocket, centered at $\Gamma$, (shown by the dashed line) and two electron pockets, centered at ${\bf K}$ (green solid line) and $-{\bf K}$ (blue solid line).
\label{fig:BZ3p}
 }
\end{figure}

\begin{figure}
  \centering
\subfigure[]{\includegraphics[width=0.45\linewidth]{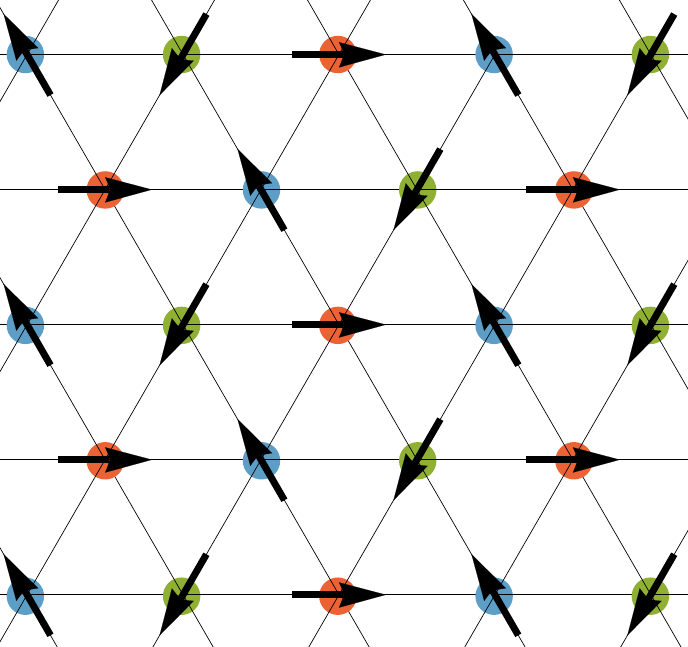}\label{fig:SDWconfig1}}\quad
  \subfigure[]{\includegraphics[width=0.45\linewidth]{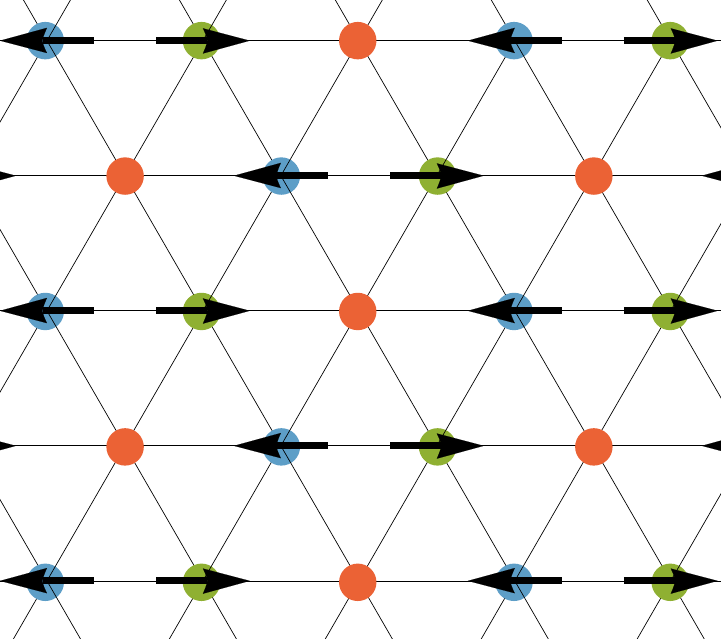}\label{fig:SDWconfig2}}
\caption{Real space structure of on-site SDW order ${\bf M}_{\pm K} = {\bf M}_r \pm i {\bf M}_i$.
At the mean-field level the ground state is infinitely degenerate for circular pockets (the ground state energy depends only on ${\bf M}^2_r +{\bf M}^2_i$),  but beyond mean-field and/or for non-circular (but $C_3$-symmetric) pockets, the degeneracy is lifted.
Panels {\bf(a)} and {\bf(b)} -- the two SDW configurations selected in the model --  the $120^\circ$ spiral order (the same as for localized spins)  (panel {\bf(a)}) and the  collinear magnetic order with antiferromagnetic spin arrangement on two-thirds of sites, and no magnetization on the remaining one-third of the sites  (panel {\bf(b)}).
 The three colors  indicate the three-sublattice structure of the SDW order.}
\end{figure}

   In this communication we analyze the structure of an SDW order in a system of 2D itinerant fermions on a triangular lattice. We consider a band metal with a hole pocket at $\Gamma = (0,0)$ ($c$-band) and two electron pockets at $\pm {\bf K}$ ($f$-band), where ${\bf K} = (4\pi/3, 0)$ (see Fig.~\ref{fig:BZ3p}).  We discuss the electronic structure and interactions in Sec.~\ref{structure}. In such a system an SDW order parameter can be either with momentum ${\bf K}$ or with $-{\bf K}$.  The SDW order parameters with ${\bf K}$  and  $-{\bf K}$  are $\sdw{\pm K}=\frac{1}{2}({\bf{\Delta}}_{\pm K}+{\bf{\Delta}}_{\mp K}^*)$, where ${\bf{\Delta}}_K= \sum_{\pv}\langle\fd_{{\bf{K}}+\pv}\vec{\sigma} c_{\pv}\rangle$ and  ${\bf{\Delta}}_{-K}= \sum_{\pv}\langle\fd_{-{\bf{K}}+\pv}\vec{\sigma} c_{\pv}\rangle$.  The two underlying order parameters ${\bf{\Delta}}_K$ and ${\bf{\Delta}}_{-K}$ are coupled within a set self-consistent equations for a magnetic order, and in zero magnetic field turn out to be complex conjugate to each other (Sec.~\ref{sec:SDW}). Then $\sdw{K}={\bf{\Delta}}_{K}, \, \sdw{-{K}}={\bf{\Delta}}^*_{K}$. However, because ${\bf K}$ and $-{\bf K}$ are in-equivalent points (a reciprocal lattice vector is $3{\bf K}$, not $2{\bf K}$ like in  systems on a square lattice), ${\bf M}_{K} = {\bf{\Delta}}_{K}$ is a complex variable:  $\sdw{K}=\sdw{-K}^*=\sdw{r}+i \sdw{i}$ (the magnetization at site $\br$ is ${\bf M}(\br)=\sdw{r}\cos \bK \br +\sdw{i}\sin \bK \br$).  Keeping only the interactions in the SDW channel, we find in Sec. III that the ground state manifold is degenerate and the Free energy depends only on $\sdw{r}^2+\sdw{i}^2$. A unique SDW order is selected by either interactions outside of SDW channel, or by the anisotropy of the pockets, or, potentially, by other perturbations.  We show in  Sec.~\ref{sec:SDW}  that these additional terms stabilize either a $120^\circ$ spiral order with three-fold rotation symmetry  (${\bf M}_r \perp {\bf M}_i$, $|{\bf M}_r| = |{\bf M}_i|$), or a collinear SDW with non-equal magnitude of magnetization on different lattice sites (${\bf M}_r \parallel {\bf M}_i$, or ${\bf M}_r = 0$, or ${\bf M}_i =0$). In particular, when  $|{\bf M}_r| =0$, the SDW order is antiferromagnetic on two-third of sites and there is no magnetization  on the remaining one-third of sites. We show SDW configurations in real space for these two types of order in Figs.~\ref{fig:SDWconfig1} and  Fig.~\ref{fig:SDWconfig2}.

       We also consider in Section~\ref{sec:SDW} another type of magnetic order, with the order parameter ${\bf \Phi}_{\pm K}=\frac{1}{2}({\bf{\Delta}}_{\pm K}-{\bf{\Delta}}_{\mp K}^*)$.  At zero magnetic field, self-consistent equations for ${\bf \Phi}_{\pm K}$ and ${\bf M}_{\pm K}$ decouple.  The one for ${\bf \Phi}_{\pm K}$ yields ${\bf\Delta}_{\pm K} = - {\bf\Delta}^*_{\mp K}$, i.e., ${\bf \Phi}_{K} = {\bf\Delta}_K$, ${\bf \Phi}_{-K} = -{\bf\Delta}^*_K$.
       For  repulsive interactions between low-energy fermions, the Free energy for ${\bf \Phi}$ order is higher than for ${\bf M}$ (SDW) order, i.e., the leading  instability is SDW. However,  ${\bf\Phi}$ order  wins when some interactions are repulsive and some are attractive.  Like for SDW, the ${\bf \Phi}$ order parameter is a complex vector,  ${\bf \Phi}_K =  {\bf \Phi}_r + i{\bf \Phi}_i$.  At a mean-field level, the Free energy for the ${\bf\Phi}$ depends on $|{\bf \Phi}|^2$, i.e., the ground state manifold is degenerate. The degeneracy is lifted by other interactions,  like for an SDW order,  and the selected states are the analogs of $120^\circ$ and collinear SDW states.

\begin{figure}
\centering
{\includegraphics[width=0.9\linewidth]{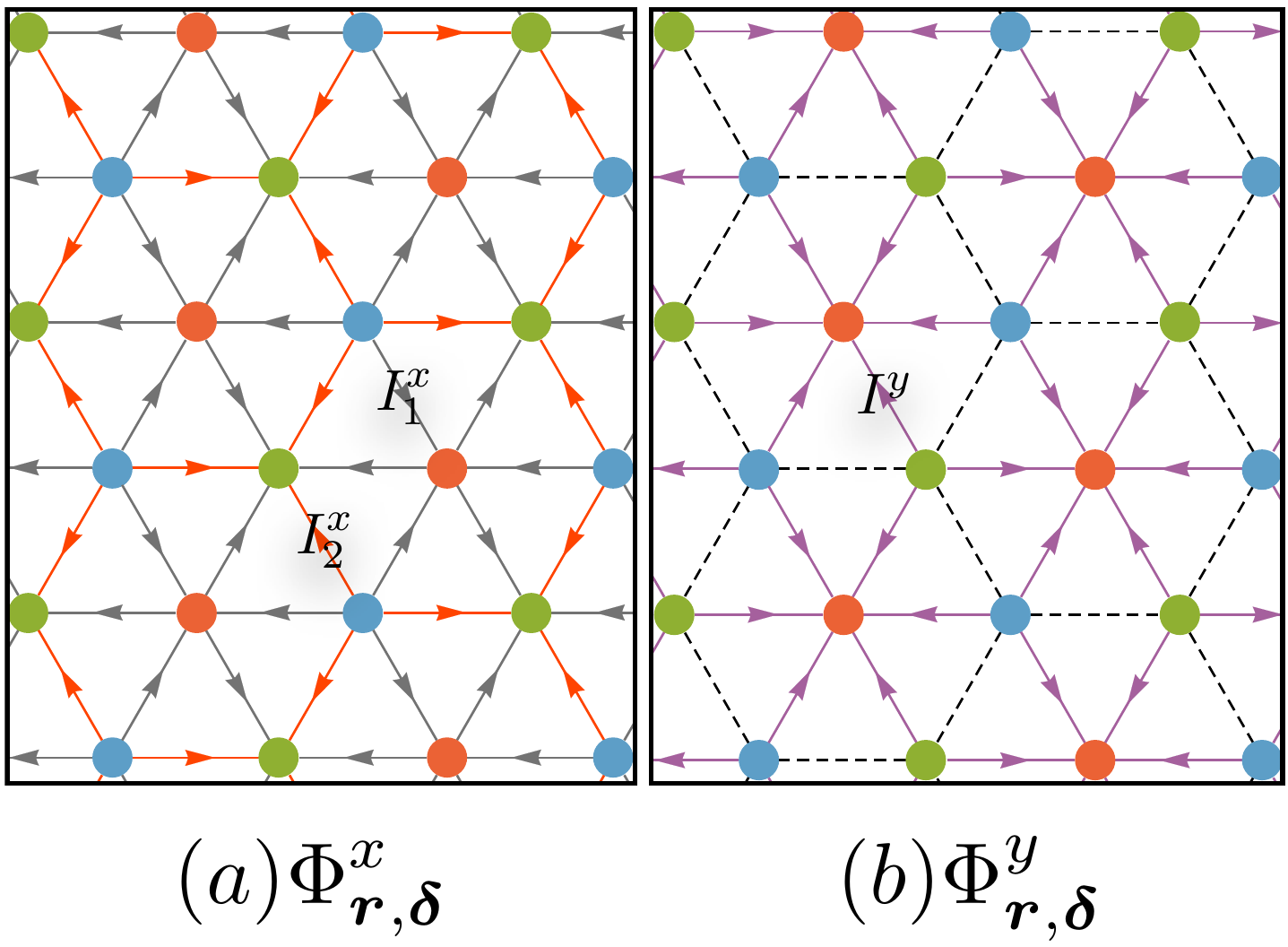}\\
\includegraphics[width=0.9\linewidth]{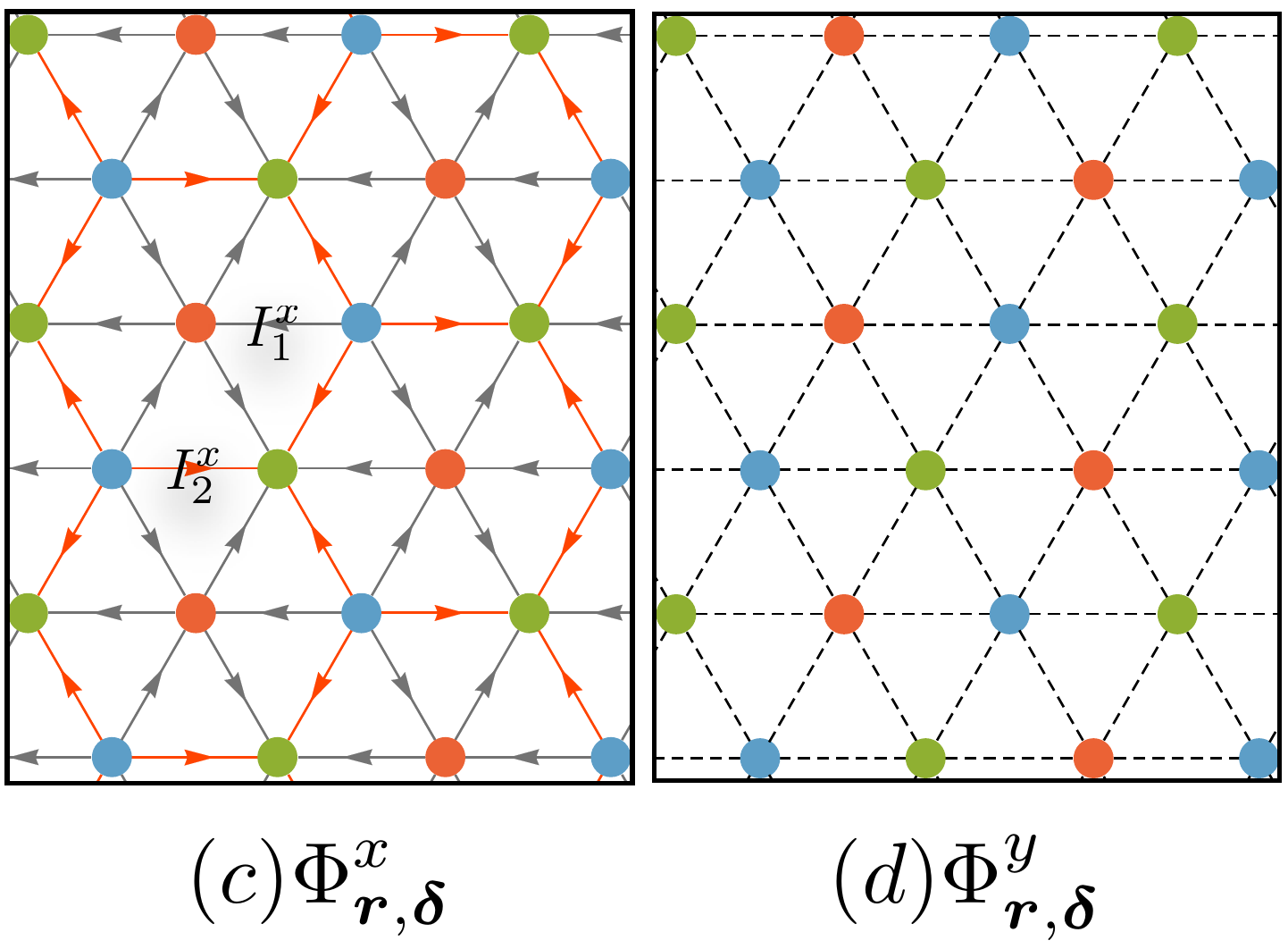}}
\caption{ Real space structure of imaginary spin bond  order ${\bf \Phi}_{\pm K} = {\bf \Phi}_r \pm i {\bf \Phi}_i$ (labeled as ISB order in the text).  The order on the bonds between  nearest neighbors is shown.  At the mean-field level the ground state is infinitely degenerate for circular pockets (the ground state energy depends only on ${\bf \Phi}^2_r +{\bf \Phi}^2_i$,  but beyond mean-field and/or for non-circular (but $C_3$-symmetric) pockets, the degeneracy is lifted. In panels $({\bf a})$ - $({\bf d})$  we show two selected ISB configurations. Panels $({\bf a})$ and $({\bf b})$ show ISB order, analogous to the $120^{\circ}$ spiral SDW order  from Fig.~\ref{fig:SDWconfig1}. This order corresponds to ${\bf \Phi}_r \perp {\bf \Phi}_i$, $|{\bf \Phi}_r| = |{\bf \Phi}_i|$ ($\varphi_x=0,\varphi_y=\pi/2$ in Eq.~\ref{eq:ISB1}.  In units of $I_0\sim \frac{h\Delta}{\hbar\mu}$, the magnitude of the ISB order is $I^x_1=\frac{\sqrt{3}}{4}I_0$ on a grey arrow and $I^x_2=\frac{\sqrt{3}}{2}I_0$ on an orange arrow in panel $({\bf a})$, and  $I^y=\frac{3}{4}I_0$ on a purple arrow in $({\bf a})$. Panels $({\bf c})$ and $({\bf d})$ show ISB order analogous to the partial collinear SDW order from Fig.~\ref{fig:SDWconfig2}.  This ISB order configuration corresponds to  $ {\bf \Phi}_i=0$. A dashed lines denote bonds with zero magnitude of ISB order. Notice that $\Phi ^{x}_{{\bf r},{\vect \delta}}$ in $({\bf c})$ has the same pattern as in $({\bf a})$, but $\Phi ^{y}_{{\bf r},{\vect \delta}}$ in $({\bf b})$ and  $({\bf d})$ are very different.\label{fig:ISBconfig}
 }
\end{figure}

\begin{figure}
\centering
\includegraphics[width=0.6\linewidth]{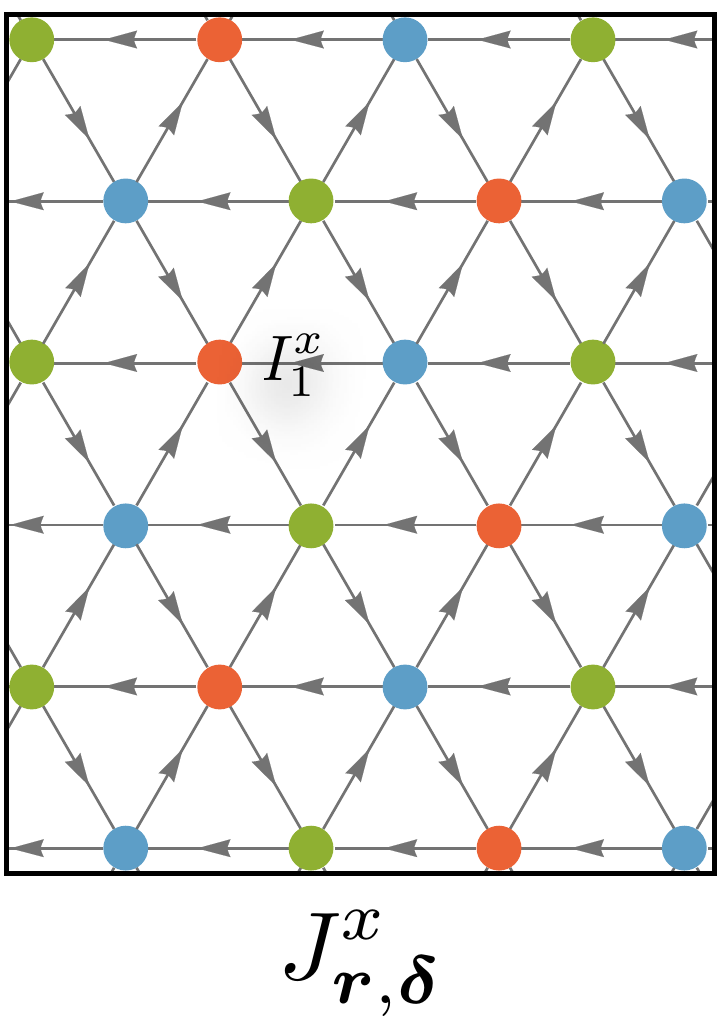}
\caption{ A potential circular spin current configuration generated from the ISB order for a proper  symmetry of hopping integrals. Such behavior may hold in a multi-orbital 3 pocket model. The figure is obtained by changing the direction of all red bonds directed towards green sites of panel Fig.~\ref{fig:ISBconfig} $({\bf a})$ and by changing by half the magnitude of ISB order on these bonds.
  \label{fig:ISBconfig_1}
 }
\end{figure}

The order parameter  ${\bf \Phi}_{\pm K}$ preserves the sign under time reversal and is similar to $iSDW$ order on a square lattice, discussed in the context of FeSCs ~\cite{Vafek13,Chubukov_prx,*Khodas2016,Klug2017} (the direct analogy holds when ${\bf{\Delta}}_{\pm K}$ is purely imaginary and ${\bf \Phi}_{-K} =- {\bf \Phi}^*_{K}$). In real space, a non-zero ${\bf \Phi}_{\pm K}$ does not give rise to either site or bond real magnetic order, but it gives rise to a non-zero order parameter $\Phi$, which is expressed via the imaginary part of the  expectation value of a  spin operator on a bond between  ${\bf r}+{\vect \delta}/2$ and ${\bf r}-{\vect \delta}/2$: \begin{align}\label{eq:ISBdef}
\Phi ^{\alpha}_{{\bf r},{\vect \delta}}=\frac{i}{\hbar}\hat{\vect{\delta}}\langle \fd_{{\bf r}+{\vect \delta}/2}\sigma^{\alpha}c_{{\bf r}-{\vect \delta}/2}+\cd_{{\bf r}+{\vect \delta}/2}\sigma^{\alpha}f_{{\bf r}-{\vect \delta}/2}-h.c.\rangle
\end{align}
 We label the order with a non-zero $\Phi$  as ``imaginary" spin bond (ISB) order. We show that one can associate $\Phi^\alpha_{\br,\vect\delta}$ with a vector directed either along or opposite to $\vect\delta$, depending on the sign of $\Phi^{\alpha}_{{\bf r},{\vect \delta}}$.   In Fig.~\ref{fig:ISBconfig} we display graphically ISB order parameter in real space for two $\Phi$ states -- one is the analog of the $120^\circ$ SDW order [${\bf \Phi}_r \perp {\bf \Phi}_i$, $|{\bf \Phi}_r| = |{\bf \Phi}_i|$, panels (a) and (b) in Fig.~\ref{fig:ISBconfig}];
  the other is the analog of a partial collinear SDW order [the case  ${\bf \Phi}_i =0$, panels (c) and (d) in Fig.~\ref{fig:ISBconfig}].
 In a multi-band system an ISB  order may give rise to circulating spin current~\cite{Klug2017}
 $J^{\alpha}_{{\bf r},{\vect \delta}} \sim \sum_{(a,b)}t^{(a,b)}_{{\bf r},{\vect \delta}} \Phi ^{\alpha(a,b)}_{{\bf r},{\vect \delta}}$, if the hopping $t^{(a,b)}_{{\bf r},{\vect \delta}}$ has a proper form ( $a$, $b$ label orbitals of $f$- and $c$-fermions in Eq.~\ref{eq:ISBdef}).
This does not hold in our model, where a potential  multi-orbital composition of low-energy states are neglected. We show a potential circulating spin-current order in Fig.~\ref{fig:ISBconfig_1}.

   We next return to SDW order and analyze in Sec.~\ref{sec:MagneticField} its evolution in a small magnetic field. We show that the $120^\circ$ spiral order becomes  cone-like, i.e. the order in the plane transverse to the field remains $120^\circ$ spiral, and the
    order in the direction of the field is ferromagnetic,  due to an imbalance of spin up and down electrons. In this respect, the field evolution of the $120^\circ$  order in an itinerant system is different from the one in the  Heisenberg model with nearest neighbor exchange, where spins remain in the same plane during the field evolution and pass through an intermediate up-up-down phase~\cite{Chubukov1991,Starykh2009,Ye2017a,*Ye2017b}.  We next argue that in a field, spin-polarization operators for spin components along and transverse to the field become different, and the bubbles made out of spin-up $c-$fermion and spin-down $f-$fermion and out of spin-down $c-$fermion and spin-up $f-$fermion also become different.
     The first discrepancy keeps ${\bf \Delta}_{\pm { K}}$ in the plane perpendicular to a field, the second breaks the equivalence between ${\bf{\Delta}}_{K}$ and ${\bf{\Delta}}_{-K}^*$.  As the consequence, SDW and ISB orders get linearly coupled. We explicitly derive the bilinear coupling term $F_{cross}({\bf M}, {\bf \Phi})$ in the Free energy. Because of the linear coupling of ${\bf M}$ and ${\bf \Phi}$, {\it an itinerant system in a field necessarily possesses both SDW and ISB orders}, even if only SDW order was present in zero field (and vice versa).

Finally, in Sec.~\ref{sec:RG} we return to zero field and consider a model with purely repulsive interactions, when the magnetic order is SDW.
We use parquet renormalization group (pRG) approach and analyze the competition between SDW magnetism and other orders bilinear in fermions, such as superconductivity and conventional and unconventional charge density-wave orders. Magnetism is an expected winner in an itinerant system, if the corresponding instability temperature is high enough, because at relatively high energies the only attractive 4-fermion interaction is in the SDW channel. However, if an instability develops  at a smaller energy/temperature, other channels compete with SDW because in the process of the flow from higher to lower energies, partial components of the interaction in some superconducting and charge-density-wave channels change sign and become attractive. As the consequence, the system may develop superconductivity or charge order instead of SDW magnetism.  We show that this actually happens, at least in some range of input parameters,  and   the system develops either $s^{\pm}$-wave superconductivity, or an  unconventional charge-order, which breaks time-reversal symmetry.

We present the  summary of our results in Sec.~\ref{sec:Summary}.
\section{Electronic structure and interactions}
\label{structure}

We consider a system of 2D itinerant fermions on a triangular lattice, with hole and electron FSs.  The hole FS is centered at $\Gamma = (0,0)$,  and the two in-equivalent electron pockets are centered at $\pm {\bf K}$ ($f$-band), where ${\bf K} = (4\pi/3, 0)$.  We show the Brillouin zone and the FSs  in Fig.~\ref{fig:BZ3p}.   We label fermionic operators  with momenta near $\Gamma$ as $c_\pv$ and the ones near $\pm {\bf K}$ as $f_{\pm \bK +\pv}$. The electronic dispersion is this three-pocket ($3p$) model can be
 approximated as $\epsilon_{\Gamma,\kv}=-\frac{k^2}{2m_h}+\mu_h$ and $\epsilon_{\pm \bK+\kv}=\frac{k^2}{2m_e}-\mu_e$.  The quadratic Hamiltonian  in zero field can be expressed via a $6$-component electronic spinor $\Psi_{\kv}=\{c_{\kv,\sigma},f_{\bK+\kv,\sigma}, f_{-\bK+\kv,\sigma}\}^T$ as
\begin{align}  \label{eq:noninteracting3p}
\mc{H}_0&=\Psi_{\kv}^{\dg}H_0\Psi_{\kv},\non\\
H_0&=
\begin{pmatrix}
\epsilon_{\Gamma,\kv}\id & 0 & 0\\
0 & \epsilon_{\bK+\kv}\id & 0\\
0 & 0 & \epsilon_{-\bK+\kv}\id
\end{pmatrix}.
\end{align}
where each time $\kv$ is shifted from the center of a FS and   $\id$ is the $2\times2$ identity matrix in spin space.

There are 8 different four-fermion interactions between low-energy fermionic states near hole and electron pockets. We show the fermion propagators and four-fermion interactions graphically in Eqs.~\ref{eq:diag-propagator} and ~\ref{eq:diag-interaction}.  These 8  terms include inter-pocket and  exchange interactions between  fermions near a hole pocket and an electron pocket ($g_1$ and $g_2$ terms, respectively), a pair hopping  from a hole pocket into electron pockets at ${\bf K}$ and $-{\bf K}$ ($g_3$ term),  intra-pocket interactions between fermions near a hole pocket and one of electron pockets ($g_4$ and $g_5$ terms, respectively), inter-pocket density-density and exchange interactions between fermions near the two electron pockets ($g_6$ and $g_7$ terms, respectively), and umklapp interaction in which incoming  fermions are near a hole pocket and  one of electron pockets and outgoing fermions are near the other electron pocket. This last interaction is allowed because $3 {\bf K}$ is a reciprocal lattice vector.   We do not consider in this work potential multi-orbital composition of the excitations around hole and electron pockets, like in Fe-based superconductors. Accordingly, we  treat $g_i$  as some constants, independent on the angles along the FSs.   For most of the paper we assume that all $g_i >0$, i.e,  all interactions are repulsive.

\begin{align}\label{eq:diag-propagator}
&
\begin{fmffile}{diag-prop}
\fmfset{arrow_len}{2mm}
\fmfset{dash_len}{2mm}
\begin{fmfgraph*}(60,20)
\fmfleft{l1}
\fmfright{r1}
\fmf{scalar,label=$\Gamma$}{l1,r1}
\end{fmfgraph*}\quad
\begin{fmfgraph*}(60,20)
\fmfleft{l1}
\fmfright{r1}
\fmf{fermion,label=$K$}{l1,r1}
\end{fmfgraph*}\quad
\begin{fmfgraph*}(60,20)
\fmfleft{l1}
\fmfright{r1}
\fmf{heavy,label=$-K$}{l1,r1}
\end{fmfgraph*}
\end{fmffile}
\end{align}
\begin{align}\label{eq:diag-interaction}
&
\begin{fmffile}{diag-int1}
\fmfset{arrow_len}{2mm}
\fmfset{dash_len}{2mm}
\begin{gathered}
\begin{fmfgraph*}(60,30)
\fmfleftn{l}{2}\fmfrightn{r}{2}
\fmf{fermion}{l1,v1}
\fmf{scalar}{l2,v2}
\fmf{fermion}{v1,r1}
\fmf{scalar}{v2,r2}
\fmf{photon,label=${{g}_{1}}$,tension=0}{v1,v2}
\end{fmfgraph*}
\end{gathered}\quad
\begin{gathered}
\begin{fmfgraph*}(60,30)
\fmfleftn{l}{2}\fmfrightn{r}{2}
\fmf{fermion}{l1,v1}
\fmf{scalar}{l2,v2}
\fmf{scalar}{v1,r1}
\fmf{fermion}{v2,r2}
\fmf{photon,label=${{g}_{2}}$,tension=0}{v1,v2}
\end{fmfgraph*}
\end{gathered}\quad
\begin{gathered}
\begin{fmfgraph*}(60,30)
\fmfleftn{l}{2}\fmfrightn{r}{2}
\fmf{scalar}{l1,v1}
\fmf{scalar}{l2,v2}
\fmf{fermion}{v1,r1}
\fmf{heavy}{v2,r2}
\fmf{photon,label=${{g}_{3}}$,tension=0}{v1,v2}
\end{fmfgraph*}
\end{gathered}
\end{fmffile}\non\\
&\begin{fmffile}{diag-int2}
\fmfset{arrow_len}{2mm}
\fmfset{dash_len}{2mm}
\begin{gathered}
\begin{fmfgraph*}(60,30)
\fmfleftn{l}{2}\fmfrightn{r}{2}
\fmf{fermion}{l1,v1}
\fmf{fermion}{l2,v2}
\fmf{fermion}{v1,r1}
\fmf{fermion}{v2,r2}
\fmf{photon,label=${{g}_{4}}$,tension=0}{v1,v2}
\end{fmfgraph*}
\end{gathered}\quad
\begin{gathered}
\begin{fmfgraph*}(60,30)
\fmfleftn{l}{2}\fmfrightn{r}{2}
\fmf{scalar}{l1,v1}
\fmf{scalar}{l2,v2}
\fmf{scalar}{v1,r1}
\fmf{scalar}{v2,r2}
\fmf{photon,label=${{g}_{5}}$,tension=0}{v1,v2}
\end{fmfgraph*}
\end{gathered}\quad
\begin{gathered}
\begin{fmfgraph*}(60,30)
\fmfleftn{l}{2}\fmfrightn{r}{2}
\fmf{heavy}{l1,v1}
\fmf{fermion}{l2,v2}
\fmf{heavy}{v1,r1}
\fmf{fermion}{v2,r2}
\fmf{photon,label=${{g}_{6}}$,tension=0}{v1,v2}
\end{fmfgraph*}
\end{gathered}
\end{fmffile}\non\\
&\begin{fmffile}{diag-int3}
\fmfset{arrow_len}{2mm}
\fmfset{dash_len}{2mm}
\begin{gathered}
\begin{fmfgraph*}(60,30)
\fmfleftn{l}{2}\fmfrightn{r}{2}
\fmf{heavy}{l1,v1}
\fmf{fermion}{l2,v2}
\fmf{fermion}{v1,r1}
\fmf{heavy}{v2,r2}
\fmf{photon,label=${{g}_{7}}$,tension=0}{v1,v2}
\end{fmfgraph*}
\end{gathered}\quad
\begin{gathered}
\begin{fmfgraph*}(60,30)
\fmfleftn{l}{2}\fmfrightn{r}{2}
\fmf{fermion}{l1,v1}
\fmf{scalar}{l2,v2}
\fmf{heavy}{v1,r1}
\fmf{heavy}{v2,r2}
\fmf{photon,label=$g_{8}$,tension=0}{v1,v2}
\end{fmfgraph*}
\end{gathered}
\end{fmffile}
\end{align}

\section{Magnetic order and its selection  by electronic correlations}\label{sec:SDW}

At low enough temperature  interactions may give rise to an instability of the normal state towards some form of electronic order.  Like we said, the most natural candidate  for the ordered state is SDW magnetism, because a magnetic order develops when electron-electron interaction is repulsive, while other instabilities, like superconductivity and charge order, require an attraction in some partial channel. This is particularly true if  the instability develops at a relatively high energy, before interactions get modified in the RG flow. In this section we assume that itinerant SDW magnetism is the leading instability and  study the structure of SDW order in zero magnetic field. We also consider the case when $g_3$ interaction is attractive, in which case the leading magnetic instability is towards ISB order.

\subsection{The development of a magnetic order}
 We introduce two complex spin operators, bilinear in fermions, with transferred momentum near ${\bf K}$ and $-{\bf K}$:
\begin{align}
\hat{\bf \Delta}_{\bK +\qv}=\sum_{\pv}\fd_{\bK+\pv+\qv}\vec{\sigma} c_{\pv}, ~~ \hat{\bf \Delta}_{-\bK+\qv}=\sum_{\pv}\fd_{-\bK+\pv+\qv}\vec{\sigma} c_{\pv}.
\label{me_1}
\end{align}
Each order parameter is constructed out of a fermion near a hole pocket and near an electron pocket.
The  SDW order parameters with momenta $\pm {\bf K}$ are
\begin{align}
{\bf M}_{\pm  K} &= \Big\langle\frac{1}{2} \sum_{{\bf p}, \alpha, \beta} \left( \fd_{\pm{\bf K}+{\bf p}, \alpha}\vec{\sigma}_{\alpha \beta} c_{{\bf p},\beta} + \cd_{{\bf p}\alpha} \vec{\sigma}_{\alpha \beta}f_{\mp {\bf K}+{\bf p}, \beta}\right)\Big\rangle\non\\
 & = \frac{1}{2} \left({\bf \Delta}_{\pm { K}} + {\bf \Delta}^*_{ \mp { K}} \right)
 \end{align}
where ${\bf \Delta}_{\pm { K}} = \langle \hat{\bf \Delta}_{\pm K}\rangle$.
In real space, ${\bf M} ({\bf r}) = {\bf M}_{ K} e^{i {\bf K} {\bf r}} + {\bf M}_{- K} e^{-i {\bf K} {\bf r}}$.
The ISB order parameters are
\begin{align}
{\bf \Phi}_{\pm  K} &= \Big\langle\frac{1}{2} \sum_{{\bf p}, \alpha, \beta} \left( \fd_{\pm{\bf K}+{\bf p}, \alpha}\vec{\sigma}_{\alpha \beta} c_{{\bf p},\beta} - \cd_{{\bf p}\alpha} \vec{\sigma}_{\alpha \beta}f_{\mp {\bf K}+{\bf p}, \beta}\right)\Big\rangle\non\\
 & = \frac{1}{2} \left({\bf \Delta}_{\pm { K}} - {\bf \Delta}^*_{ \mp { K}} \right)
 \end{align}

Out of eight interactions, the two, $g_1$ and $g_3$, can be re-expressed as the interactions between $\hat\Delta$s as
\begin{align}\label{eq:H4Spin}
\mc{H}_4=&\sum_{p,p',q,\sigma,\sigma'}g_3\big(\cd_{p+q,\sigma}\cd_{p'-q,\sigma'}f_{K+p',\sigma'}f_{-K+p,\sigma}+h.c.\big)\non\\
&g_1\big(\cd_{p+q,\sigma}\fd_{K+p'-q,\sigma'}f_{K+p',\sigma'}c_{p,\sigma}+(K\rightarrow -K)\big)\non\\
=&-\frac{g_3}{2}\big(\hat{\bf \Delta}_{K-q}\hat{\bf \Delta}_{-K+q}+h.c.)\big)\non\\
&-\frac{g_1}{2}\big(\hat{\bf \Delta}^{\dg}_{-K-q}\hat{\bf \Delta}_{K+q}+(K\rightarrow -K)\big)+...,
\end{align}
The self-consistent equations on infinitesimal  ${\bf \Delta}_{K}$  and ${\bf \Delta}_{-K}$ are obtained by summing up series of ladder diagrams:
\begin{align}
&\begin{fmffile}{diag-tri1s}
\fmfset{arrow_len}{2mm}
\fmfset{dash_len}{2mm}
\begin{gathered}
\begin{fmfgraph*}(60,40)
\fmfleftn{l}{1}\fmfrightn{r}{2}
\fmfpolyn{shaded}{G}{3}
\fmf{photon,label=${{\bf{\Delta}}_{K}^{*}}$,tension=2}{l1,G1}
\fmf{phantom,label=${\vec{\sigma}}$,tension=0}{G2,G3}
\fmf{scalar,tension=0.65}{r1,G2}\fmf{fermion,tension=0.65}{G3,r2}
\end{fmfgraph*}
\end{gathered}=
\begin{gathered}
\begin{fmfgraph*}(60,40)
\fmfleftn{l}{1}\fmfrightn{r}{2}
\fmfpolyn{shaded}{G}{3}
\fmf{photon,label=${{\bf{\Delta}}_{K}^{*}}$,tension=2}{l1,G1}
\fmf{scalar,tension=0.65}{v1,G2}\fmf{fermion,tension=0.65}{G3,v2}
\fmf{photon,label=$g_1$,tension=0}{v2,v1}
\fmf{scalar,tension=0.65}{r1,v1}\fmf{fermion,tension=0.65}{v2,r2}
\end{fmfgraph*}
\end{gathered}+
\begin{gathered}
\begin{fmfgraph*}(60,40)
\fmfleftn{l}{1}\fmfrightn{r}{2}
\fmfpolyn{shaded}{G}{3}
\fmf{photon,label=${{\bf{\Delta}}_{-K}}$,tension=2}{l1,G1}
\fmf{heavy,tension=0.65}{v1,G2}\fmf{scalar,tension=0.65}{G3,v2}
\fmf{photon,label=$g_3$,tension=0}{v2,v1}
\fmf{scalar,tension=0.65}{r1,v1}\fmf{fermion,tension=0.65}{v2,r2}
\end{fmfgraph*}
\end{gathered}
\end{fmffile},\non\\
&\begin{fmffile}{diag-tri2s}
\fmfset{arrow_len}{2mm}
\fmfset{dash_len}{2mm}
\begin{gathered}
\begin{fmfgraph*}(60,40)
\fmfleftn{l}{1}\fmfrightn{r}{2}
\fmfpolyn{shaded}{G}{3}
\fmf{photon,label=${{\bf{\Delta}}_{-K}}$,tension=2}{l1,G1}
\fmf{heavy,tension=0.65}{r1,G2}\fmf{scalar,tension=0.65}{G3,r2}
\fmf{phantom,label=${\vec{\sigma}}$,tension=0}{G2,G3}
\end{fmfgraph*}
\end{gathered}=
\begin{gathered}
\begin{fmfgraph*}(60,40)
\fmfleftn{l}{1}\fmfrightn{r}{2}
\fmfpolyn{shaded}{G}{3}
\fmf{photon,label=${{\bf{\Delta}}_{-K}}$,tension=2}{l1,G1}
\fmf{heavy,tension=0.65}{v1,G2}\fmf{scalar,tension=0.65}{G3,v2}
\fmf{photon,label=$g_1$,tension=0}{v2,v1}
\fmf{heavy,tension=0.65}{r1,v1}\fmf{scalar,tension=0.65}{v2,r2}
\end{fmfgraph*}
\end{gathered}+
\begin{gathered}
\begin{fmfgraph*}(60,40)
\fmfleftn{l}{1}\fmfrightn{r}{2}
\fmfpolyn{shaded}{G}{3}
\fmf{photon,label=${{\bf{\Delta}}_{K}^{*}}$,tension=2}{l1,G1}
\fmf{scalar,tension=0.65}{v1,G2}\fmf{fermion,tension=0.65}{G3,v2}
\fmf{photon,label=$g_3$,tension=0}{v2,v1}
\fmf{heavy,tension=0.65}{r1,v1}\fmf{scalar,tension=0.65}{v2,r2}
\end{fmfgraph*}
\end{gathered}
\end{fmffile}\label{eq:LinearSDW}
\end{align}
 At zero  magnetic field the equations for all three spin components of ${\bf \Delta}_{\pm K}$ are the same, and we have
\begin{align}\label{appeq:LinearTr}
{\bf \Delta}^*_{K}&=-(g_1\Pi_{(+K)}{\bf \Delta}^*_{K} + g_3 \Pi_{(-K)}{\bf \Delta}_{-K}),\non\\
{\bf \Delta}_{-K}&=-(g_3\Pi_{(+K)}{\bf \Delta}^*_{K} + g_1 \Pi_{(-K)}{\bf \Delta}_{-K}),
\end{align}
where $\Pi_{(\pm K)}=T\sum_{\omega_n}\int \frac{\diff^2 k}{\mc{A}_{B.Z.}}\mc{G}^{f}(\kv\pm \bK)\mc{G}^{c}(\kv)$, and $\mc{A}_{B.Z.}$ is the area of the Brillouin zone.
  Because the dispersions near ${\bf K}$ and $-{\bf K}$ are identical, $\Pi_{(+K)} = \Pi_{(-K)} = \Pi$.  Eq.~\ref{appeq:LinearTr} then decouples into
 \begin{align}\label{appeq:LinearTr_1}
{\bf \Delta}^*_{{ K}} + {\bf \Delta}_{-{ K}} &= -(g_1 + g_3)  \Pi \left({\bf \Delta}^*_{{ K}} +{\bf \Delta}_{-{ K}}\right),\non\\
{\bf \Delta}^*_{{ K}} - {\bf \Delta}_{-{ K}} &= -(g_1 - g_3)  \Pi \left({\bf \Delta}^*_{{ K}} -{\bf \Delta}_{-{ K}}\right)
\end{align}
 or
 \begin{align}\label{appeq:LinearTr_2}
{\bf M}_{\pm K} &= -(g_1 + g_3)  \Pi \,{\bf M}_{\pm K}, \non\\
{\bf \Phi}_{\pm K} &= -(g_1 - g_3)  \Pi \,{\bf \Phi}_{\pm K},
\end{align}
 We see that ${\bf M}$ and ${\bf \Phi}$ channels are decoupled.

 One can easily verify that (i) $\Pi <0$ and (ii) its magnitude  grows logarithmically with decreasing $T$ due to opposite signs of dispersions near $\Gamma$ and near $\pm {\bf K}$, even if the masses of the two dispersions are different (i.e., even if there is no true nesting).  We found numerically that the logarithmic enhancement holds  down to $T \sim |\mu_1-\mu_2|/5$,  below which $\Pi$ saturates.  The combination of (i) and (ii)  implies that the magnetic instability develops already for small values of $g_1, g_3$, but still the interaction should be above the threshold.

\subsection{The SDW order}

    When both $g_1$ and $g_3$ are positive, the leading instability occurs when $(g_1+g_3) |\Pi| =1$, and the emerging order is SDW with ${\bf \Delta}_{K} = {\bf \Delta}^*_{-K}$, i.e., ${\bf M}_{ K} = {\bf \Delta}_{ K} = {\bf M}^*_{- K}$.     We verified that the condition  ${\bf \Delta}_{K} = {\bf \Delta}^*_{-K}$ holds also for the solution of the full non-linear self-consistent equation at a finite SDW order parameter.

    Keeping  ${\bf M}_{ K} = {\bf M}^*_{- K}$ and adding to the quadratic Hamiltonian the SDW terms ${\bar {\bf M}}_{\pm K} = \frac{g_{sdw}}{2} {\bf M}_{\pm K}$, where $g_{sdw} = g_1 + g_3$, we found that $H_0$  modifies to
\begin{align}  \label{eq:HSDW3p}
\mc{H}_{\bf{M}}&=\Psi_{\kv}^{\dg}H_{\bf{M}}\Psi_{\kv}\non\\
H_{\bf{M}}&=
\begin{pmatrix}
\epsilon_{\Gamma\kv}\id & -\bsdw{K}\cdot\vec{\sigma} & -\bsdw{-K}\cdot\vec{\sigma}\\
-\bsdw{K}^*\cdot\vec{\sigma} & \epsilon_{\bK+\kv}\id & 0\\
-\bsdw{-K}^*\cdot\vec{\sigma} & 0 & \epsilon_{-\bK+\kv}\id
\end{pmatrix},
\end{align}
Eq.~\ref{eq:HSDW3p} can be also obtained via Hubbard-Stratonovich transformation using the interaction terms projected to the SDW channel. We show the derivation in Appendix~\ref{app:HubbardStratonovich}. We emphasize that each component of  $\bsdw{K}$ is a complex variable because in our case
 $\bK$ and $-\bK$ are not separated by a reciprocal lattice vector.  In this respect, SDW on a hexagonal lattice differs from commensurate SDW with $\bsdw{Q}$ on a square lattice as for the latter $\bsdw{Q}$ is real because $Q$ and -$Q$ differ by a reciprocal lattice vector.  For convenience, we separate $\bsdw{ K}$ and $\bsdw{-{ K}}$ into $\bsdw{K}=\bsdw{r}+i\bsdw{i}$ and $\bsdw{-K}=\bsdw{r}-i\bsdw{i}$.

 The quadratic Hamiltonian $\mc{H}_{\bf M}$ can be diagonalized by two subsequent Bogolyubov transformations (see Appendix~\ref{app:Diagonalization} for details).
The result is
\begin{align}\label{eq:quadraticH}
\mc{H}_{\bf M}=&\sum_{\kv, \alpha} E^{+}_{\kv} \ed_{\kv, \alpha}e_{\kv, \alpha} +   E^{-}_{\kv} \pd_{\kv, \alpha}p_{\kv, \alpha} + \epsilon_{K+\kv} {\bar f} ^\dagger_{\kv,\alpha}{\bar f}_{\kv,\alpha})
\end{align}
where
 \begin{align}\label{eq:quadraticH_1}
  E^{\pm}_{\kv} &= \frac{\epsilon_{\Gamma,\kv} +\epsilon_{\bK+\kv}}{2} \pm \sqrt{\left(\frac{\epsilon_{\Gamma,\kv} -\epsilon_{\bK+\kv}}{2}\right)^2 + 2{\bar M}^2},
  \end{align}
 and ${\bar M}=\sqrt{|\bsdw{r}|^2+|\bsdw{i}|^2}$. The operator ${\bar f}$ is the linear combination of $f$  operators with momenta near ${\bf K}$ and $-{\bf K}$, which does not get coupled to c-operators in the presence of SDW order.
  Because of  the last term in Eq.~\ref{eq:quadraticH}  the system  remains a metal in the SDW phase, even in case of perfect nesting $\epsilon_{\Gamma,\kv} = - \epsilon_{\bK+\kv}$, when excitations described by $E^{\pm}_{\kv}$ are all gapped.

The self-consistent equation for the order parameter ${\bar M}$  reduces to
\begin{align}\label{eq:SelfConsistentSDW}
1=\frac{g_{sdw}}{2N}\sum_{\kv}\frac{1}{\sqrt{\left(\frac{\epsilon_{\Gamma,\kv} -\epsilon_{\bK+\kv}}{2}\right)^2 + 2{\bar M}^2}}
\end{align}

As the dispersion depends on $\bar{M}$, but not  separately on $\bsdw{r}$ and $\bsdw{i}$,  the SDW ground state is degenerate for all configurations in the manifold of $|\bsdw{r}|^2+|\bsdw{i}|^2=\bar{M}^2$.

The Landau Free energy in terms of $\bar M$ is
\begin{align}\label{eq:actiontot_1}
F =& a(\bsdw{r}^2+\bsdw{i}^2)+ b (\bsdw{r}^2+\bsdw{i}^2)^2 + ....
\end{align}

Without loss of generality we can choose $\sdw{r}$ and $\sdw{i}$ to be in the $x-y$ plane and set $\sdw{r}$ to be along $x$ direction.  We then have  $\sdw{r}=M_{r}\hat{e}_x=M\cos \tau \hat{e}_x$, $\sdw{i}=M_{ix}\hat{e}_x+M_{iy}\hat{e}_y=M \sin\tau\cos \theta \hat{e}_x+M\sin\tau\sin\theta \hat{e}_y$. The SDW order parameter  ${\bf M} ({\bf r})$  in real space is related to $\sdw{r},\sdw{i}$ as ( see Appendix~\ref{app:order} for derivation)
\begin{align}\label{eq:SDW3p}
M^x ({\bf r})&=2(M_{r}\cos {\bK \br} +M_{ix}\sin{ \bK \br})\non\\
&=2(M\cos\tau\cos {\bK \br} +M \sin\tau\cos \theta\sin {\bK \br})\non\\
M^{y} ({\bf r})&=2M_{iy}\sin {\bK \br}=2M \sin\tau\sin \theta\sin {\bK \br}
\end{align}

For example, when $\theta=\pi/2$, $\tau=\pi/4$, i.e. $\sdw{i}\bot \sdw{i}$ and $|\sdw{r}|=|\sdw{i}|$, $M^{x}({\br})=\sqrt{2}~M\cos  {\bK \br},~M^{y}({\br})=\sqrt{2}~M\sin  {\bK \br}$, i.e. the SDW order configuration is $120^\circ$ spiral (see Fig.~\ref{fig:SDWconfig1}). When $\theta,\tau=\pi/2$, $M^{x}({\br})=0,\,M^{y}({\br})=2M \sin {\bK \br}$, the SDW  configuration is antiferromagnetic  on two-thirds of sites, while the remaining one third of sites remains non-magnetic (see Fig.~\ref{fig:SDWconfig2}). This kind of order is peculiar to itinerant systems. A similar partial order has been found in the studies of magnetism in in doped graphene~\cite{Chubukov2012,Batista2012} and in doped FeSCs~\cite{Chubukov2010M,Fernandes2016}.

\subsubsection{The selection of the SDW order}\label{sec:SelectionSDW}

\textit{Selection by the anisotropy of the spectrum -- }  One way to lift the degeneracy is to include the anisotropy of the dispersion near the two electron pockets.
 The points ${\bf K}$ and -${\bf K}$  are highly-symmetric points in the Brillouin zone, but still, the lattice symmetry only implies that the dispersion should remain invariant under the rotation by $120^\circ$.  Then the most generic dispersion near $\pm {\bf K}$ is $\epsilon_{\pm \bK+\pv}=\frac{p^2}{2m_e}-\mu_2 \pm \delta \cos3\theta_{\pv}$, where $\theta_{\pv}$ is the angle between  ${\bf p}$ and $ \bK$. A conventional analysis, similar to the one in Ref.~\cite{Kang2015}, shows that a non-zero $\delta$ gives rise to additional quartic term in Landau Free energy in the form $c (\sdw{r}\times\sdw{i})^2$ with $c <0$.  The minimization of the Free energy then yields
  $\sdw{r}\bot \sdw{i}$ and $|\sdw{r}|=|\sdw{i}|$. This corresponds to  the $120^\circ$ SDW order.

\textit{Selection by the other couplings  -- } Another way to lift the ground state degeneracy is to go beyond mean-field and include the corrections to the ground state energy from four-fermion couplings other than $g_1$ and $g_3$. These other couplings do not contribute to SDW order at the mean-field level, but affect the Free energy beyond mean-field. For simplicity of presentation, we analyze the effect of other couplings assuming  that  $\epsilon_{\Gamma,\kv} = - \epsilon_{\pm \bK+\kv}$ (a perfect nesting).

   In our case, there are two contributions from other interactions. First, the terms $g_4,~g_6$, and $g_7$ have non-zero expectation values in the SDW state. This effect is similar to the one  found in Fe-based systems~\cite{Chubukov2010M,Fernandes12}.  The contribution to the Free energy from an average value of these additional interactions is
\begin{align}
\delta F_a&= 2(g_6-g_7-2g_4)\Big(\frac{\sdw{r}\times\sdw{i}}{M^2}\Big)^2(N_F {\bar M})^2\non\\
&=\frac{1}{2}(g_6-g_7-2g_4)(N_F {\bar M})^2\sin^2\theta\sin^2 2\tau,
\end{align}
where $N_F$ is the density of states near the Fermi surface. The selection of SDW order depends on the relative strength of the couplings.  When $g_6-g_7-2g_4<0$, $\delta F_a$ is minimized when $\theta=\pi/2$ (mod $\pi$) and $\tau=\pi/4$ (mod $\pi/2$), i.e. when $\sdw{r}\bot\sdw{i}$ and $|\sdw{r}|=|\sdw{i}|$.  This gives  $120^{\circ}$ spiral SDW  order.
 When $g_6-g_7-2g_4>0$, $\theta=0$ (mod $\pi$) or $\tau=0$ (mod $\pi/2$). In the first case  $\sdw{r}\parallel\sdw{i}$, in the second  either $\sdw{r}$ or $\sdw{i}$ is equal to zero. In both cases, the SDW order is collinear and the ground state manifold remains infinitely degenerate because for $\sdw{r}\parallel\sdw{i}$, $\delta F_a =0$, and the ratio ${\sdw{i}}/{\sdw{r}}$ is arbitrary ($\sdw{i}=0$  or $\sdw{r}=0$ are the two limits of the degenerate set).

The second effect comes from the $g_8$ term, which gives rise to SDW-mediated coupling between fermions near  ${\bf K}$ and near $-{\bf K}$.
Indeed, the $g_8$ term is
\begin{align}
H_{g_8}=&g_8\sum_{p_1,p_2,p_3,\sigma, \sigma'} \big(\fd_{K+p_1,\sigma}\fd_{K+p_2,\sigma'}f_{-K+p_3,\sigma'}c_{p_1+p_2-p_3\sigma}\non\\
&+\fd_{-K +p_1,\sigma}\fd_{-K +p_2,\sigma'}f_{K +p_3,\sigma'}c_{p_1+p_2-p_3\sigma}+h.c.\big).
\end{align}
In the SDW state, this term acquires a piece quadratic in fermions
\begin{align}\label{eq:QuadraticG8}
H_{g_8} \to 2\gamma_8\sum \fd_{K,\sigma}(\bsdw{K}\cdot \vs)_{\sigma,\sigma'} f_{-K,\sigma'}+h.c.,
\end{align}
where $\gamma_8=g_8/g_{sdw}$. In the second order in perturbation, this term gives the correction to the Free energy, which also scales as $\bar M^2$:
\begin{align}\label{eq:E8gr}
\delta F_{b}=&-N_F(\gamma_8 \bar M)^2(3\cos^2\theta\sin^2 2\tau\big(\cos2\tau+1)\non\\
&+\cos^22\tau(3-\cos 2\tau)\big)
\end{align}
The $\tau,~\theta$ that minimize $\delta F_b$ are
\begin{align}
\theta&=-\pi,~0\text{ and }
\tau=\pm\pi/6,~\pm5\pi/6,\non\\
{\text { or }} \tau&=\pm\pi/2 \,\,\text{ and }\theta {\text { arbitrary}}
\end{align}
 One can verify that both choices for $\theta$ and $\tau$ describe a collinear spin configuration with antiferromagnetic spin ordering on two-thirds of sites, while the remaining one third of sites remain non-magnetic (see Fig.~\ref{fig:SDWconfig2}), i.e. $\delta F_b$ selects SDW configuration which corresponds to ${\bf M}_r=0$. For example, when $\theta,\tau=\pi/2$, we obtain from Eq.~\ref{eq:SDW3p} $M^{x}({\br})=0,\,M^{y}({\br})=2M \sin {\bK \br}$. In other words, $\delta F_b$ lifts the degeneracy of collinear SDW states in favor of the state with  antiferromagnetism  on 2/3 of lattice cites.

\begin{figure}
  \centering
 {\includegraphics[width=0.8\linewidth]{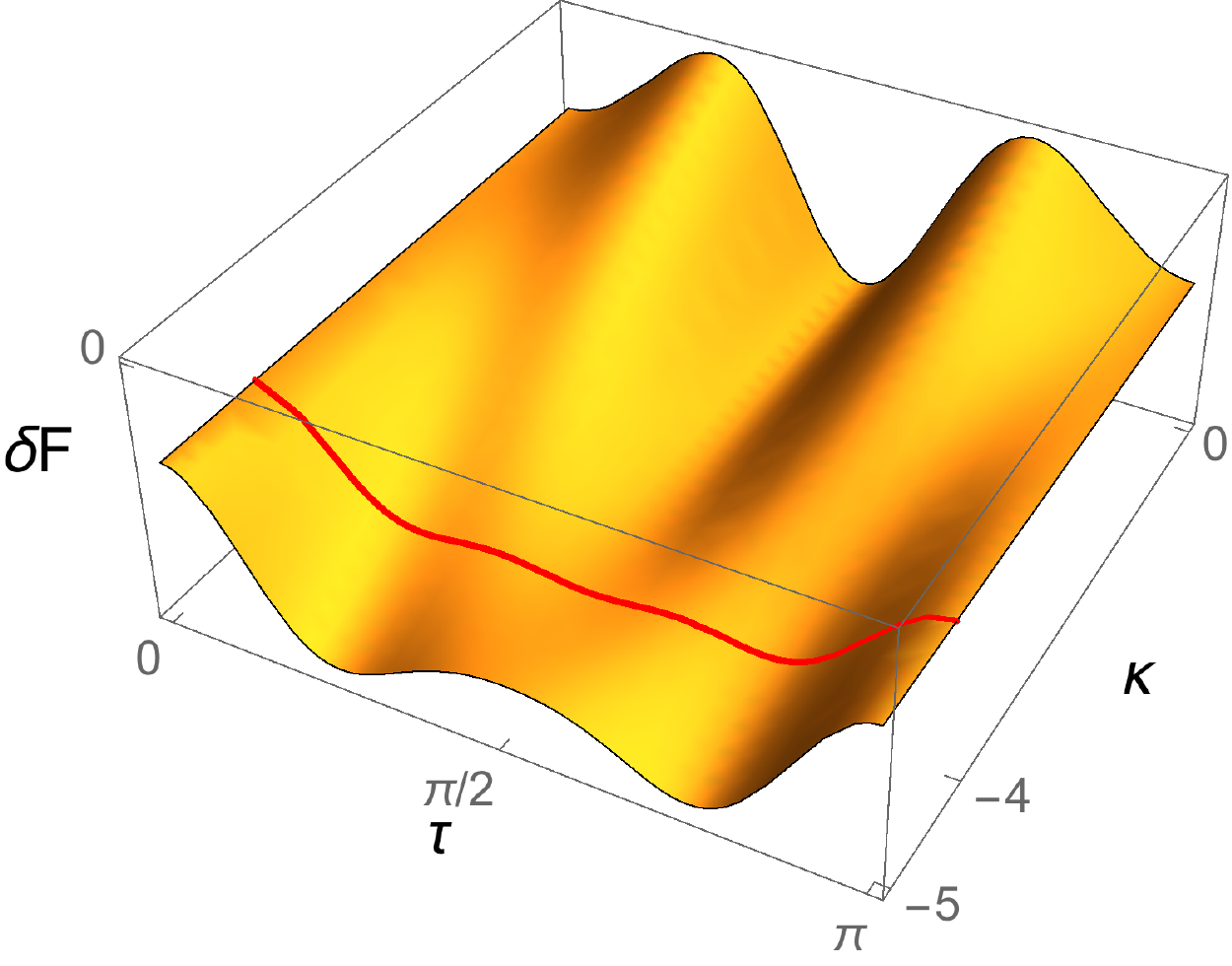}}
  \caption{$\delta F$ (correction to the Free energy from 4-fermion interactions ) at $\theta=\frac{\pi}{2}$. At $\theta=\pi/2$, $\delta F$ can minimized in both SDW order configurations at different $\tau$: $\tau=\frac{\pi}{4},\frac{3\pi}{4}$ for $120^{\circ}$ spiral order (Fig.~\ref{fig:SDWconfig1}) and $\tau=\frac{\pi}{2}$ for collinear order (Fig.~\ref{fig:SDWconfig2}). At $\kappa=-4$ (thick red line), the ground state energy of the two SDW order configurations are the same, indicating a \textit{first order} phase transition. \label{fig:dEmanifold}}
\end{figure}

The SDW ground state configuration is obtained by minimizing the total $\delta F = \delta F_a+\delta F_b$. We define the ratio of the prefactors for $\bar M^2$ terms in
 $\delta F_a$ and $\delta F_b$ as
\begin{align}
\kappa&=\frac{1}{2}N_F \frac{(g_6-g_7-2g_4) }{\gamma_8^2}\non\\
&=\frac{1}{2}N_F g_{sdw}\frac{(g_6-g_7-2g_4) g_{sdw}}{g_8^2}.
\end{align}
We find  that for $\kappa <-4$ the system selects the $120^\circ$ spiral state and for $\kappa>-4$ it selects the  collinear antiferromagnetic state.
  At $\kappa=-4$ (highlighted in red in Fig.~\ref{fig:dEmanifold}), both states correspond to local minima, i.e., the transition between the two is   \textit{first order}.

\subsection{The ISB order}

When $g_3$ is negative, the leading instability in the magnetic channel is towards ISB order ${\bf\Phi}_{\pm K}$.  For this order we have ${\bf \Delta}_{K} = -{\bf \Delta}^*_{-K}$, i.e., ${\bf \Phi}_K = {\bf \Delta}_K$, ${\bf \Phi}_{-K}  = {\bf \Delta}_{- K} = - {\bf \Phi}^*_{ K}$.
  ${\bf \Phi}_{\bf K}$ is also a complex vector ${\bf \Phi}_{\bf K} = {\bf \Phi}_r + i {\bf \Phi}_i$ with  ${\bf \Phi}_{-\bf K} = - {\bf \Phi}_r + i
 {\bf \Phi}_i$. At the mean-field level the Free energy again depends on ${\bf \Phi}^2_r + {\bf \Phi}^2_i$, i.e., the ground state is infinitely degenerate.  The degeneracy is lifted by either the anisotropy of the electron pockets or by other interactions.

 In real space, a non-zero  ${\bf \Phi}_{ K}$ gives rise to a finite value of an imaginary part of an expectation value of a spin operator on a bond between ${\bf r}-{\vect \delta}/2$ and ${\bf r}+{\vect \delta}/2$. The corresponding real order parameter is
  \begin{align}\label{eq:ISB1}
&\Phi ^{\alpha}_{{\bf r},{\bf \delta}}=\frac{i}{\hbar}\hat{\vect\delta}
\langle \fd_{{\bf r}+{\vect \delta}/2}\sigma^{\alpha}c_{{\bf r}-{\vect \delta}/2}+\cd_{{\bf r}+{\vect \delta}/2}\sigma^{\alpha}f_{{\bf r}-{\vect \delta}/2}-h.c.\rangle \nonumber \\
&= \frac{8}{\hbar} \hat{\vect\delta}|\Phi^{\alpha}_{ K}| \sin{ {\bf K}{\vect \delta}} \cos \left({\bf K} {\bf r} - \phi^{\alpha}_K\right)
\end{align}
where ${\Phi^\alpha}_K =  |\Phi^\alpha_K| e^{i \phi^\alpha_K}$ and $\phi^\alpha_{-K} = \pi - \phi^\alpha_{K}$.  This last condition implies that $\Phi^{\alpha}_{{\bf r},{\vect \delta}}$ does not change under ${\bf K} \to - {\bf K}$.

Because $\Phi^{\alpha}_{{\bf r},{\vect \delta}}$ is an odd function of $\vect\delta$, the ISB order is ``directional" in the sense that for a given ${\bf r}$, one can associate $\Phi^{\alpha}_{{\bf r},{\vect \delta}}$ with a vector directed either along or opposite to $\vect\delta$, depending on the sign of $\Phi^{\alpha}_{{\bf r},{\vect \delta}}$.  In Fig.~\ref{fig:ISBconfig} we show ${\Phi}^\alpha_{{\bf r},{\vect \delta}}$ for the two ISB states selected by the lifting of the degeneracy. One is the analog of $120^\circ$  SDW spiral state,  another is the analog of a partially ordered collinear state. In the first case ${\bf \Phi}_r \perp {\bf \Phi}_i$, $|{\bf \Phi}_r| = |{\bf \Phi}_i|$, in the second  ${\bf \Phi}_i=0$. The direction of the arrow on each bond is determined by the sign of ${\Phi}^\alpha_{{\bf r},{\vect \delta}}$  (if it is positive, the arrow goes from ${\bf r} - {\vect \delta}/2$ to ${\bf r} + {\vect \delta}/2$).
 In the  ``$120^\circ$" state (panels (a) and (b)), $\Phi ^{x}_{{\bf r},{\vect \delta}}$ and  $\Phi ^{y}_{{\bf r},{\vect \delta}}$  are both non-zero.  In the ``collinear" state (panels (c) and (d)) only one component of  ${\bf \Phi}_r$ is non-zero.

  We emphasize that ${\Phi}^\alpha_{{\bf r},{\vect \delta}}$ is not a spin current operator (${\Phi}^\alpha_{{\bf r},{\vect \delta}}$ at a given site is not conserved, as it would be required for a current due to local spin conservation). In a generic multi-orbital system a spin current  is expressed in terms of ISB orders and hopping integrals as $J^{\alpha}_{{\bf r},{\vect \delta}} \sim \sum_{(a,b)}t^{(a,b)}_{{\bf r},{\vect \delta}} \Phi ^{\alpha(a,b)}_{{\bf r},{\vect \delta}}$, where $(a,b)$ label the orbital components of $f$- and $c$-fermions (see Ref.~\cite{Klug2017} for a discussion on the orbital currents).
      The hopping parameters $t^{(a,b)}_{{\bf r},{\vect \delta}}$ generally depend on ${\bf r}$ and, for a given ${\bf r}$, may change the sign between different ${\vect\delta}$.
       For a proper choice of $t^{(a,b)}_{{\bf r},{\vect \delta}}$  between orbitals, $J^{\alpha}_{{\bf r},{\vect \delta}} $ may become a spin current. For example, for the ``collinear" $\Phi-$order (panels (c) and (d) in Fig.~\ref{fig:ISBconfig}), a change of the direction and the magnitude on a half of the red bonds  directed towards green sites in Fig.~\ref{fig:ISBconfig}, will give rise to a circulating current, which obeys a local spin conservation. We show this in Fig.~\ref{fig:ISBconfig_1}.

\section{A finite magnetic field: a cone SDW state and a field-induced ISB order}\label{sec:MagneticField}

In this section  we consider the evolution of  the SDW state in a Zeeman magnetic field. In a free electron system a Zeeman field shifts spin-up bands down and spin-down bands up, inducing a net magnetization along the field direction $\hat{z}$. For interacting fermions the effect of a magnetic field  is more complex.  Suppose we start with $120^\circ$ spin ordering in zero field.  For a system of localized spins on a 2D triangular lattice quantum fluctuations select field reorientation in which spins remain in the same plane in a finite field~\cite{Chubukov1991,Ye2017a,*Ye2017b}. We show below that for itinerant fermions  the evolution of the spin configuration with a field ${\bf h} = h\hat{z}$ proceeds differently --  in a finite field the SDW  becomes a non-coplanar cone state in which spins preserve a $120^\circ$ order in the xy plane and simultaneously develop a net magnetization along the field. However, this is not the only effect of the field. We show that a magnetic field triggers the appearance of an ISB order $|{\bf \Phi}_{\pm K}| \propto (h/\mu) |{\bf M}_{\pm K}|$. We remind  that ${\bf \Phi}_{\pm K}$ is even under time-reversal and may give rise to  circulating spin currents.
\subsection{Spin order in a magnetic field}\label{sec:ISBinstability}

\begin{figure}
  \centering
  \includegraphics[width=0.9\linewidth]{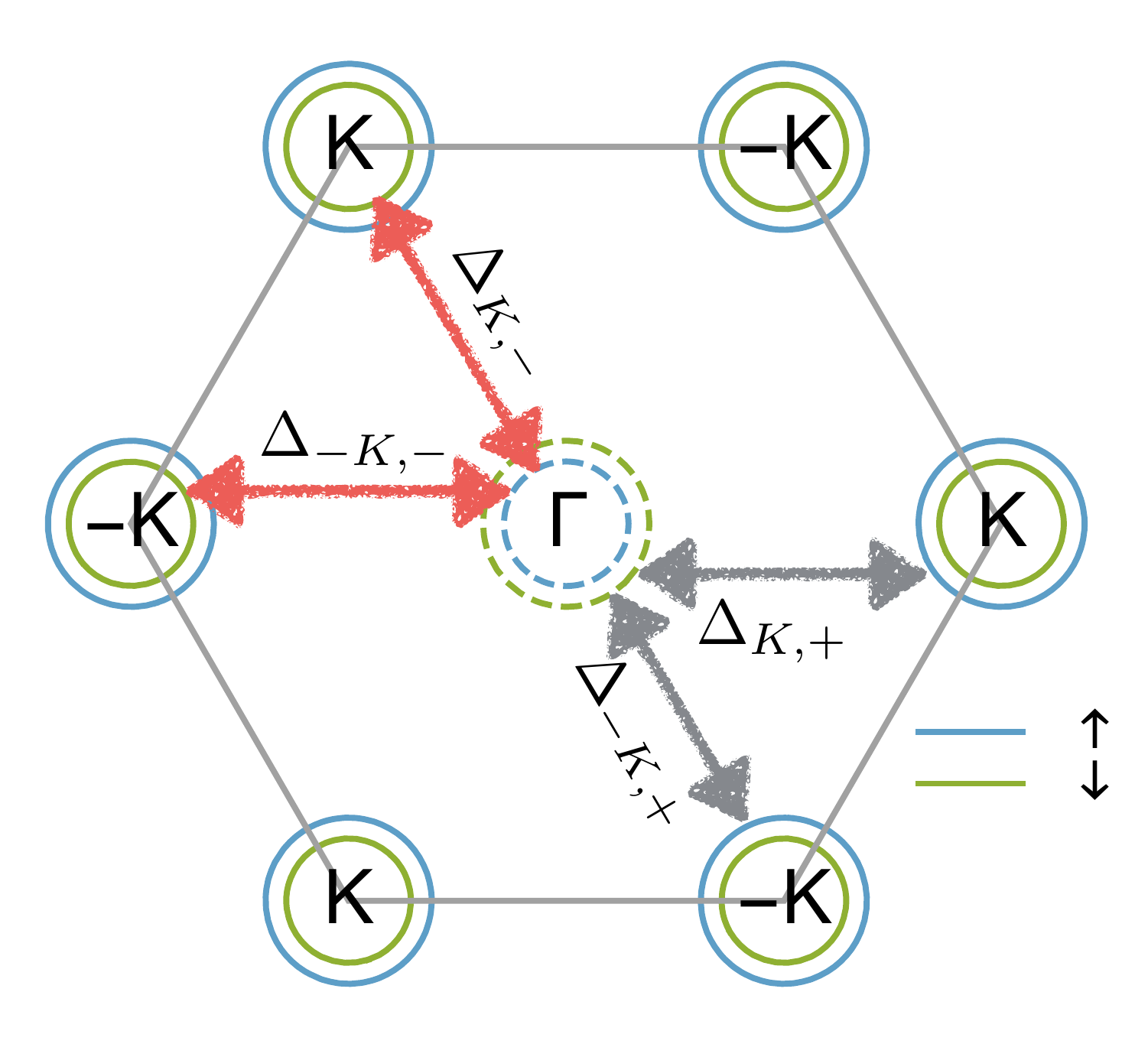}
  \caption{Fermi surface geometry in a magnetic field.  Spin-up (blue) and spin-down (green) bands  split by the Zeeman field. Double arrows connect electronic states that form SDW order in the $\sigma^+$ channel (grey arrow) and $\sigma^-$ channel  (red arrow). The quantity
  $\Delta_{\pm K,\pm}$ is defined in
  Eq.~\ref{eq:SDWtr}. \label{fig:Band3p}}
\end{figure}

When a Zeeman field is applied, say along $\hat{z}$, it splits the spin-up and spin-down bands, as shown in Fig.~\ref{fig:Band3p}. It also breaks  $SU(2)$ spin rotation symmetry down to  $U(1)$, which means that SDW instabilities in $\sigma^{\pm}$ and $\sigma^z$ channels now develop at different temperatures, which we label as $T_{c,tr}$ and $T_{c,z}$ respectively. Only the higher $T_c$ is meaningful.   We show that the SDW order develops in the $\sigma^{\pm}$ channel first, i.e. SDW  is locked in the plane transverse to the field.

To see this we define the order parameters $\Delta_{\pm}$ and $\Delta_z$ as:
\begin{align}\label{eq:SDWtr}
\Delta_{\pm K,\pm}&=\sum_{\kv,\alpha,\beta}\langle \fd_{\kv\pm K,\alpha}\sigma^{\pm}_{\alpha\beta}c_{\kv\beta}\rangle\non\\
\Delta_{\pm K,z}& = \sum_{\kv,\alpha,\beta}\langle \fd_{\kv\pm K,\alpha}\sigma^{z}_{\alpha\beta} c_{\kv\beta}\rangle
\end{align}
where $\alpha,\,\beta=\{\uparrow,\,\downarrow\}$. The linearized equations on $\Delta$ in $\sigma^{\pm}$ channel are  \begin{align}\label{eq:LinearTr}
\Delta_{K,+}=-(g_1\Pi_{+}\Delta_{K,+} + g_3 \Pi_{-}\Delta^*_{-K,-}),\non\\
\Delta^*_{-K,-}=-(g_3\Pi_{+}\Delta_{K,+} + g_1 \Pi_{-}\Delta^*_{-K,-}),
\end{align}
where
 $$\Pi_{\pm}
=T\sum_{\omega_n,\alpha,\beta}\int \frac{\diff^2 k}{\mc{A}_{B.Z.}}\mc{G}^{f,\alpha}(\kv + \bK)\sigma^{\pm}_{\alpha,\beta}\mc{G}^{c,\beta}(\kv).$$ It is essential that $\Pi_{+} \neq \Pi_{-}$ (see below). In the $\sigma^{z}$ channel we have
 \begin{widetext}
\begin{align}\label{eq:LinearL0}
\Delta_{K,z}&=-g_1(\Pi_{z\uparrow}\Delta_{K,\uparrow}-\Pi_{z\downarrow}\Delta_{K,\downarrow})- g_3(\Pi_{z\uparrow}\Delta^*_{-K,\uparrow}-\Pi_{z\downarrow}\Delta^*_{-K,\downarrow})\non\\
\Delta^*_{-K,z}&=-g_1(\Pi_{z\uparrow}\Delta^*_{-K,\uparrow}-\Pi_{z\downarrow}\Delta^*_{-K,\downarrow})- g_3(\Pi_{z\uparrow}\Delta_{K,\uparrow}-\Pi_{z\downarrow}\Delta_{K,\downarrow})
\end{align}
\end{widetext}
where  $\Pi_{z,\alpha}=T\sum_{\omega_n}\int \frac{\diff^2 k}{\mc{A}_{B.Z.}}\mc{G}^{f,\alpha}(\kv + \bK)\sigma^{z}_{\alpha,\alpha}\mc{G}^{c,\alpha}(\kv)$. Both $\Pi_{z,\alpha}$ and  $\Pi_{\pm}$ do not change under ${\bf K} \to - {\bf K}$).

To get qualitative understanding, consider first the case of perfect nesting, i.e. set $m_h=m_e=m$ and $\mu_h=\mu_e=\mu$ such that  $\epsilon_{\Gamma,\kv} = - \epsilon_{\pm K + \kv}$. Then the two larger FSs with Fermi momentum $k_{F}^+=\sqrt{2m(\mu+h)}$  are the electron FS for up spins  and  the hole FS for  down spins. The smaller FSs with $k_{F}^-=\sqrt{2m(\mu-h)}$  are  the electron FS for down spins  and  the hole FS for up spins (see Fig.~\ref{fig:Band3p}).  One can easily verify that in this situation $\Pi_{z\downarrow}=\Pi_{z\uparrow}$.  Eq.~\ref{eq:LinearL0} is then  simplified to
\begin{align}\label{eq:LinearL}
\Delta_{K,z}&=-g_1\Pi_{z}\Delta_{K,z}- g_3\Pi_{z}\Delta^*_{-K,z}\non\\
\Delta^*_{-K,z}&=-g_1\Pi_{z}\Delta^*_{-K,z}- g_3\Pi_{z}\Delta_{K,z}
\end{align}
Solving Eqs.~\ref{eq:LinearTr} and \ref{eq:LinearL} we obtain the SDW instability conditions in (i)  $\sigma^{\pm}$ and (ii) $\sigma^z$ channels as
\begin{widetext}
\begin{align}\label{eq:LinearEq}
\text{(i)}~~&~~ 1+\frac{1}{2}\Big(g_1(\Pi_{+}+\Pi_{-})-\big((\Pi_+-\Pi_-)^2 g_1^2+4\Pi_+\Pi_- g_3^2\big)^{1/2}\Big)=0,\non\\
\text{(ii)}~~&~~ 1+(g_1+g_3)\Pi_{z}=0.
\end{align}
\end{widetext}
Evaluating the expectation value of polarization operators $\Pi_{\pm}$ and $\Pi_z$, we obtain at $h \ll T$ (see Appendix~\ref{app:SpinOrderingInst} for details),
\begin{align}
\Pi_{ph,\pm}&=\Pi_{ph,0} \mp \frac{1}{2}N_F\frac{h}{\mu} ,\nonumber \\
\Pi_{ph,z}&= \Pi_{ph,0} + 0.43 N_F\,\frac{h^2}{T^2},
\end{align}
where $\Pi_{ph,0} \approx -(N_F/2) \log{\mu/T}$ is the polarization at zero field. Substituting into Eq.~\ref{eq:LinearEq} we obtain the critical temperature of SDW order in the transverse and longitudinal channels as
\begin{align}\label{eq:LinearEq_1}
\text{(i)} ~~& T_{c,tr} (h) = T_{c,0} \left[1 - \frac{g_3-g_1}{2 g_3} (g_3+g_1) N_F \big(\frac{h}{\mu}\big)^2\right] ,\non\\
\text{(ii)} ~~&T_{c,z}(h)  =  T_{c,0}  \left[1 - 0.86 \big(\frac{h}{T}\big)^2\right]
\end{align}
where $T_{c,0} = \mu e^{-2/(g_1+g_3)N_F}$.  Because $T \ll \mu$,  $T_{c,tr} > T_{c,z}$ independent on the sign of $g_3-g_1$. For very low $T$, when in a finite field $h \gg T$, the expression for $T_{c,z}$ gets modified (see Appendix D), but still,  $T_{c,tr} > T_{c,z}$. We also computed  $T_{c,tr}$ and $T_{c,z}$ without assuming perfect nesting, by expanding in $\frac{\delta \mu}{\mu}$, and  found that the condition instability temperature in the $\sigma_{\pm}$ channel is larger than  that in the $\sigma_z$ channel.

\subsection{SDW order in a field} \label{sec:SDW_field}

Because  $T_{c,tr} > T_{c,z}$,  the SDW instability  develops in the $\sigma^{\pm}$ channel, i.e,  the spontaneous order remains in xy plane.   A finite field indeed also creates a magnetization component in $z$ direction simply because the total densities of spin-up and spin-down fermions are now different.
 The ratio between $\Delta_{\pm K,\pm}$ and $\Delta^*_{\mp K,\mp}$, however,  changes in the field. We remind that in zero field ${\bf \Delta}_{\pm K} = {\bf \Delta}^*_{\mp K}$, i.e. $\Delta_{\pm K,\pm}=\Delta^*_{\mp K,\mp}$, such that ${\bf M}_{\pm K} = {\bf \Delta}_{\pm K}$ and ${\bf \Phi}_{\pm{ K}} =0$.  At a finite field  the solution of self-consistent equations on $\Delta_{\pm K, + }$ and $\Delta^*_{\mp K,-}$ yields
\begin{align}
\label{eq:ratiols}
\gamma=\frac{\Delta^*_{\mp K,-}}{\Delta_{\pm K,+}}&=1 - \frac{(g_3-g_1)N_F}{2g_3|\Pi_0|}\frac{h}{\mu}
\end{align}
The ground state still remains degenerate  at the mean-field level, i.e., SDW order in xy plane can be either $120^0$ spiral or a collinear state with $2/3$ of lattice sites ordered.  An arbitrary state from a degenerate manifold can be parametrized as
\begin{align}
\Delta_{K,+}&=\cos\phi \,\Delta_+,&~\Delta_{-K,+}&=e^{i\tilde\theta}\sin\phi\, \Delta_+,\non\\
\Delta_{K,-}&=e^{-i\tilde\theta}\sin\phi \,\Delta_-,&~\Delta_{-K,-}&=\cos\phi\, \Delta_-,
\label{wed_2}
\end{align}
where $\phi,\tilde\theta\in (0,2\pi)$, $\Delta_- = \gamma \Delta_+$ and without loss of generality we set $\Delta_+$ to be real. We then have
\begin{widetext}
\begin{align}\label{wed_1}
{\bf{M}}_K&=\frac{1}{2}({\bf\Delta}_{K}+{\bf\Delta}_{-K}^*)=
\frac{(1+\gamma)\Delta_+}{4}\{e^{-i\tilde\theta}(e^{i\tilde\theta}\cos\phi+\sin\phi),i\,e^{-i\tilde\theta}(e^{i\tilde\theta}\cos\phi-\sin\phi),0\},\non\\
\end{align}
\end{widetext}
and ${\bf{M}}_K={\bf{M}}^*_{-K}$. The $120^\circ$ spiral order corresponds to $\phi=0, \pi$ and $\tilde \theta$ arbitrary (and its symmetry equivalents). The collinear state with two-thirds sites ordered corresponds to $\phi=-\frac{\pi}{4}$ and $\tilde \theta=0$ (and symmetry equivalents).  In the real space the SDW order is
\begin{align}\label{eq:SDWr}
\langle\hat{M}^{\alpha}_{\br}\rangle&=4|M^{\alpha}_{K}|\cos (\bK \br -\phi_{\alpha}),
\end{align}
 where $\alpha=x,y,z$, and $\phi_{\alpha}$ is the phase of the $\alpha$ component of ${\bf{M}}_{K}$ in Eq.~\ref{wed_1}. In these notations, the  $120^\circ$ spiral order corresponds to $|\sdw{\bK}^x|=|\sdw{\bK}^y|,\phi_x=0,\phi_y=\pi/2$ and the collinear order corresponds to $|\sdw{\bK}^x|=0,\phi_y=\pi/2$.

To lift the degeneracy, one again has to include into consideration either the $C_3$ anisotropy of electron pockets, or  four-fermion interactions other than $g_1,g_3$. We verified that if in zero field these terms select the  $120^{\circ}$ spiral SDW order, the same order remains at $h \neq 0$, i.e. at least in this case  a  Zeeman field doesn't change the type of the SDW order.

\subsection{Field-induced ISB order}\label{sec:CurrentPattern}

Eq.~\ref{eq:ratiols} has another, more prominent consequence. Because $\Delta_{\pm K, + } \neq \Delta^*_{\mp K, - }$, SDW and ISB channels no longer decouple, i.e., the emergence of a non-zero ${\bf M}_{\pm K} =  \frac{1}{2} ({\bf\Delta}_{\pm K} + {\bf \Delta}^*_{\mp K})$ triggers a non-zero ISB order parameter $\Phi_{\pm K} = \frac{1}{2} ({\bf\Delta}_{\pm K} - {\bf \Delta}^*_{\mp K})$. We remind that ${\bf M}_{\pm K}$  changes sign under time reversal, while ${\bf \Phi}_{\pm K}$ is symmetric under time-reversal.
For a state from a degenerate manifold parametrized by  Eq.~\ref{wed_2}
\begin{widetext}
\begin{align}
{\bf{\Phi}}_K&=\frac{1}{2}({\bf\Delta}_{K}-{\bf\Delta}_{-K}^*)=\frac{(1-\gamma)\Delta_+}{4}\{e^{-i\tilde\theta}(e^{i\tilde\theta}\cos\phi-\sin\phi),i\,e^{-i\tilde\theta}(e^{i\tilde\theta}\cos\phi+\sin\phi),0\},
\end{align}
\end{widetext}
and ${\bf{\Phi}}_K=-{\bf{\Phi}}^*_{-K}$. The ISB order triggered by the $120^\circ$ spiral SDW order is ${\bf{\Phi}}_K=\frac{(1-\gamma)\Delta_+}{4}\{1,i,0\}$. Similarly, the ISB triggered by the partial collinear SDW order is ${\bf{\Phi}}_K=\frac{(1-\gamma)\Delta_+}{4}\{1,0,0\}$.

We emphasize that at small field, when $\gamma = 1 - \mc{O}(h/\mu)$,  the magnitude of ${\bf{\Phi}}_K$ is linearly proportional to that of  ${\bf{M}}_K$:  $|{\bf{\Phi}}_K| \propto (h/\mu) |{\bf{M}}_K|$.  This implies that  a non-zero field mediates a {\it linear} coupling between SDW and ISB order parameters.
This is different (and stronger) effect than a potential generation of ${\bf{\Phi}}_K$  in a field due to non-linear effects, considered in Ref~\cite{Fernandes2016}.

We now derive explicitly the $F_{cross}({\bf{M}}_{\pm K},{\bf{\Phi}}_{\pm K})$ term in the Free energy.

\subsubsection{The Free energy}
\label{subsec:FreeE}

The Free energy in terms of $\bar {\bf{M}}_{\pm K}$ and $\bar {\bf{\Phi}}_{\pm K}$ can be obtained following the standard Hubbard-Stratonovich transformation. We present the details in Appendix B and here quote the result.
\begin{align}
&F [\bar {\bf{M}}_{\pm \bK}, \bar{\bf{\Phi}}_{\pm \bK}]=\non\\
& \frac{2}{g_1+g_3}  (|\bar{\bf{M}}_{K}|^2+|\bar{\bf{M}}_{-K}|^2)+ \frac{2}{g_1-g_3} (|\bar{\bf{\Phi}}_{K}|^2+|\bar{\bf{\Phi}}_{-K}|^2)\non\\
&+\frac{1}{2} \int_k \Tr (\mc{G}_{0,k}\mc{V})^2+\frac{1}{4} \int_k \Tr (\mc{G}_{0,k}\mc{V})^4+\mc{O}(\Delta^6)
\end{align}
where $\int_k$ stands for integration over momentum and frequencies,  $\mc{V}=\mc{V}^M+\mc{V}^{\Phi}$ and
\begin{align}\label{appeq:Vmatrix}
\mc{V}^M=&-
\begin{pmatrix}
0 & \bar{\bf{M}}_K\cdot\vec{\sigma} & \bar{\bf{M}}_{-K}\cdot\vec{\sigma}\\
\bar{\bf{M}}_{-K}\cdot\vec{\sigma} & 0 & 0\\
\bar{\bf{M}}_{K}\cdot\vec{\sigma} & 0 & 0
\end{pmatrix},\non\\
\mc{V}^{\Phi}=&-
\begin{pmatrix}
0 & \bar{\bf{\Phi}}_K\cdot\vec{\sigma} & \bar{\bf{\Phi}}_{-K}\cdot\vec{\sigma}\\
-\bar{\bf{\Phi}}_{-K}\cdot\vec{\sigma} & 0 & 0\\
-\bar{\bf{\Phi}}_{K}\cdot\vec{\sigma} & 0 & 0
\end{pmatrix}.
\end{align}
 The Green's function of free electrons in a field, $\mc{G}_{0,k}$, is:
\begin{align}
\mc{G}_{0,\Gamma}&=\big((i \omega-\epsilon_{\Gamma,\qv})\id+h\sigma_z\big)^{-1}\non\\
\mc{G}_{0,\pm K}&=\big((i \omega-\epsilon_{\pm K,\qv})\id+h\sigma_z\big)^{-1}
\end{align}
The bilinear coupling between ${\bf{M}}_{\pm \bK}$, ${\bf{\Phi}}_{\pm \bK}$ comes from the crossing terms of $\mc{V}^M$ and $\mc{V}^{\Phi}$ in $\frac{1}{2}\Tr (\mc{G}_{0,k}\mc{V})^2$,
\begin{widetext}
\begin{align}\label{wed_4}
F_{cross} =&\frac{1}{2} \int_k \Tr\big(\mc{G}_{0,k}\mc{V}^M\mc{G}_{0,k}\mc{V}^{\Phi}+\mc{G}_{0,k}\mc{V}^{\Phi}\mc{G}_{0,k}\mc{V}^M\big)\non\\
=&\frac{1}{2} \int_k \sum_{i=\pm K}\Tr\big(\mc{G}_{0,\Gamma}\Mi\cdot\vs~\mc{G}_{0,i}\ccur\cdot\vs+\mc{G}_{0,\Gamma}\cMi\cdot\vs~\mc{G}_{0,i}\cur\cdot\vs+(\Mi\leftrightarrow\cur)\big)\non\\
=&4\sum_{i=\pm K}\im(\Mi\times\ccur)\cdot\vec{h}\int_k (\mc{G}_{0,\Gamma}^{(0)2}\mc{G}_{0,i}^{(0)}-\mc{G}_{0,\Gamma}^{(0)}\mc{G}_{0,i}^{(0)2})=-\frac{2N_F}{\mu}\sum_{i=\pm K}\im(\Mi\times\ccur)\cdot\vec{h}
\end{align}
\end{widetext}
To obtain the last line in (\ref{wed_4}) we expanded $\mc{G}$ in powers of $h$ as $\mc{G}_{0,\Gamma}=\mc{G}_{0,\Gamma}^{(0)}-\mc{G}_{0,\Gamma}^{(0)}h\sigma_z\mc{G}_{0,\Gamma}^{(0)}+\mc{O}(h^2)$ and $\mc{G}_{0,i}=\mc{G}_{0,i}^{(0)}-\mc{G}_{0,i}^{(0)}h\sigma_z\mc{G}_{0,i}^{(0)}+\mc{O}(h^2)$. In zero field, $F_{cross}=0$ as the quantities under the trace in the upper line in Eq.~\ref{wed_4} cancel each other. When a magnetic field is applied, $h\sigma_z$ doesn't commute with  $\sigma^{\pm}$ components of SDW and ISB orders, and $F_{cross}$  becomes finite.

\section{Competition between magnetic and other orders}\label{sec:RG}

In this section we return back to the case of zero magnetic field and study the interplay between magnetism, superconductivity and charge density wave order. We remind that  at the mean-field level, SDW magnetism is the leading instability because this channel is attractive and because for positive pair-hopping interaction $g_3$  the attraction is stronger than the one in ISB channel. The strength of the interactions in SC and CDW channels depends on the values of the bare couplings $g_1-g_8$.
If we set all bare couplings to be equal, the interactions in $s^{++}$ SC channel and in CDW channel are repulsive, and the ones in ISB, ``imaginary" charge bond, and $s^{+-}$ SC channel vanish.

 In a system with one type of FSs, a vanishing pairing interaction can be converted into an attraction by going beyond mean-field and adding Kohn-Luttinger-type corrections to the pairing vertex from the particle-hole channel~\cite{*[{For recent review see }][{}] Maiti2013}. However, the corresponding SC $T_c$  is smaller than the one for SDW, except for the case when all couplings are truly small. The situation is different in systems with hole and electron Fermi pockets.  Here,  a particle-hole bubble with the incoming momentum equal to the distance between the pockets ($\pm {\bf K}$ in our case) behaves as $\log W/E$  at energies $E$ smaller than the bandwidth $W$ but larger than, roughly, $E_F$.  As the consequence, Kohn-Luttinger renormalization, as well as the renormalizations of the interaction in CDW channels, become logarithmic. The renormalizations in the particle-particle channel are also logarithmic in 2D  at energies above $E_F$, as long as  fermionic dispersion can be approximated as parabolic.
   The presence of the logarithms  in the particle-hole and  particle-particle channels  implies that at energies between $E_F$ and $W$ the  interactions $g_1-g_8$ flow as one progressively integrates out fermions with higher energies,  and split from each other even if at the bare level  all $g_i$ were set to be equal. This flow can be captured within pRG computational scheme~\cite{Dzyaloshinskii,Metzner1998,*Salmhofer1998,Rice2009,Chubukov2008,Podolsky,Chubukov2009_r,Chubukov2010RG,Vafek2014,Aleiner2012}

  Because $g_i$ flow to different values, the interactions in some SC and CDW channels may flip the sign below a certain $E$ and become attractive. These newly attractive interactions and the attractive interaction in SDW channel compete and mutually affect each other.  SDW order still develops first if there is not enough ``space" in energy domain for the  flow of the couplings.  However, if the system allows the couplings to flow over a sizable range of energies,
   the values of $g_i$ at an energy/temperature,  where the leading instability develops, are in general quite different from the bare ones. Then there is no guarantee that the leading instability will still be in the SDW channel,  and not in one of SC or CDW channels.  To find out which channel wins, one needs to (a) analyze the flow of the couplings, (b) use the running couplings to construct the effective interactions in different channels and compare their strength.   This is what we will do below.  For the full analysis one also has to compute the flow of the vertices in different channels and analyze the corresponding susceptibilities.  This last analysis is important for the selection of subleading instabilities~\cite{Vafek2014,Classen2017,*Xing2017} and for computations in the  channels  where the bare susceptibility is non-logarithmic (e.g., for  a particle-hole channel with zero momentum transfer~\cite{Chubukov_prx,*Khodas2016}).  We will not consider such channels and will only be interested in the leading instability. For such an analysis it will be sufficient to compare  the effective interactions constructed out of the running couplings.

  \subsection{the RG flow}

  As we said, there are 8 different  4-fermion interactions between fermions near hole and electron pockets, allowed by momentum conservation -- the   $g_1-g_8$ terms. These couplings   are shown graphically in Eq.~\ref{eq:diag-interaction}.   The flow of all 8 couplings can be obtained by applying  pRG analysis similar to how this was done for Fe-based materials, which also have hole and electron pockets~\cite{Chubukov2008,Chubukov2009_r,Chubukov2010RG,Chubukov_prx,*Khodas2016,Classen2017,*Xing2017}. We perform one-loop pRG calculation keeping only logarithmically singular terms in the diagrams for the renormalizations of the couplings. In Eq.~\ref{eq:dgs} we  show diagrams for the renormalizations of the representative set of the couplings $g_1, g_2, g_6$ and $g_7$.  The computation of the diagrams  is time-consuming but straightforward, and we just present the result. The flow of the couplings is described by the set of differential equations:
\begin{align}\label{eq:RGflow0}
\dot{g}_1&=g_1^2+g_3^2-g_8^2\non\\
\dot{g}_2&=2g_2(g_1-g_2)-g_8^2\non\\
\dot{g}_3&=g_3(4g_1-2g_2-g_5-g_6-g_7)\non\\
\dot{g}_4&= - g^2_4\non\\
\dot{g}_5&=-g_3^2-g_5^2\non\\
\dot{g}_6&= -g_3^2-g_6^2-g_7^2+2g_8^2\non\\
\dot{g}_7&= -g_3^2-2g_6g_7\non\\
\dot{g}_8&=g_8(3g_1-2g_2+g_3-g_4)
\end{align}
The derivatives are with respect to the RG ``time" $t=\ln(W/E)$, where, we recall, $W$ is the UV cutoff, of order bandwidth, and $E$ is the running pRG scale. The pRG flow terminates at $E\sim \text{max}\{T_{ins}, E_F\}$, below which either an order develops in some channel at $E\sim T_{ins}$, when $T_{ins} > E_F$, or
  the flow equations become different, and the renormalizations of the  interactions in particle-hole and particle-particle channels essentially decouple. The analysis of Eq.~\ref{eq:RGflow0} shows that the equations for the intra-electron pocket coupling $g_4$ and for  $g_{e-} = g_6-g_7$  decouple from the equations for other couplings. As $g_4$ apparently flows to 0 and $g_{e-}$ does not contribute  to the instabilities which we consider here, and we neglect them. The remaining equations are
\begin{align}\label{eq:RGflow}
\dot{g}_1&=g_1^2+g_3^2-g_8^2\non\\
\dot{g}_2&=2g_2(g_1-g_2)-g_8^2\non\\
\dot{g}_3&=g_3(4g_1-2g_2-g_5-2g_e)\non\\
\dot{g}_5&=-g_3^2-g_5^2\non\\
\dot{g}_e&=-g_3^2-2g_e^2+g_8^2\non\\
\dot{g}_8&=g_8(3g_1-2g_2+g_3-g_4),
\end{align}
 where ${g}_e =\frac{ g_6 + g_7}{2}$.
  \begin{widetext}
\begin{align}\label{eq:dgs}
&
\begin{fmffile}{diag-quad1}
\fmfset{arrow_len}{2mm}
\fmfset{dash_len}{2mm}
\begin{gathered}
\begin{fmfgraph*}(60,30)
\fmfleftn{l}{2}\fmfrightn{r}{2}
\fmf{fermion}{l1,v1}
\fmf{scalar}{l2,v2}
\fmf{fermion}{v1,r1}
\fmf{scalar}{v2,r2}
\fmf{dbl_wiggly,label=$\delta {{g}_{1}}$,tension=0}{v1,v2}
\end{fmfgraph*}
\end{gathered}=
\begin{gathered}
\begin{fmfgraph*}(60,30)
\fmfleftn{l}{2}\fmfrightn{r}{2}
\fmf{fermion}{v1,l1}
\fmf{scalar}{l2,v2}
\fmf{fermion}{r1,v3}
\fmf{scalar}{v4,r2}
\fmf{fermion}{v3,v1}
\fmf{scalar}{v2,v4}
\fmf{photon,tension=0}{v1,v2}
\fmf{photon,tension=0}{v3,v4}
\end{fmfgraph*}
\end{gathered}+
\begin{gathered}
\begin{fmfgraph*}(60,30)
\fmfleftn{l}{2}\fmfrightn{r}{2}
\fmf{fermion}{v1,l1}
\fmf{scalar}{l2,v2}
\fmf{fermion}{r1,v3}
\fmf{scalar}{v4,r2}
\fmf{scalar}{v3,v1}
\fmf{heavy}{v2,v4}
\fmf{photon,tension=0}{v1,v2}
\fmf{photon,tension=0}{v3,v4}
\end{fmfgraph*}
\end{gathered}+
\begin{gathered}
\begin{fmfgraph*}(60,30)
\fmfleftn{l}{2}\fmfrightn{r}{2}
\fmf{fermion}{l1,v1}
\fmf{scalar}{l2,v2}
\fmf{fermion}{v3,r1}
\fmf{scalar}{v4,r2}
\fmf{heavy}{v1,v3}
\fmf{heavy}{v2,v4}
\fmf{photon,tension=0}{v1,v2}
\fmf{photon,tension=0}{v3,v4}
\end{fmfgraph*}
\end{gathered}
\end{fmffile}\non\\
&
\begin{fmffile}{diag-quad2}
\fmfset{arrow_len}{2mm}
\fmfset{dash_len}{2mm}
\begin{gathered}
\begin{fmfgraph*}(60,30)
\fmfleftn{l}{2}\fmfrightn{r}{2}
\fmf{fermion}{l1,v1}
\fmf{scalar}{l2,v2}
\fmf{scalar}{v1,r1}
\fmf{fermion}{v2,r2}
\fmf{dbl_wiggly,label=$\delta {{g}_{2}}$,tension=0}{v1,v2}
\end{fmfgraph*}
\end{gathered}=
\begin{gathered}
\begin{fmfgraph*}(60,30)
\fmfleftn{l}{2}\fmfrightn{r}{2}
\fmf{fermion,tension=2}{l1,v1}
\fmf{scalar,tension=2}{l2,v4}
\fmf{scalar,tension=2}{v1,r1}
\fmf{fermion,tension=2}{v4,r2}
\fmf{phantom,tension=0.05}{v1,v2,v3,v4}
\fmf{photon,tension=0.05}{v1,v2}
\fmf{photon,tension=0.05}{v3,v4}
\fmf{scalar,left=1,tension=0}{v2,v3}
\fmf{heavy,left=1,tension=0}{v3,v2}
\end{fmfgraph*}
\end{gathered}+
\begin{gathered}
\begin{fmfgraph*}(60,30)
\fmfleftn{l}{2}\fmfrightn{r}{2}
\fmf{fermion,tension=2}{l1,v1}
\fmf{scalar,tension=2}{l2,v4}
\fmf{scalar,tension=2}{v1,r1}
\fmf{fermion,tension=2}{v4,r2}
\fmf{phantom,tension=0.05}{v1,v2,v3,v4}
\fmf{photon,tension=0.05}{v1,v2}
\fmf{photon,tension=0.05}{v3,v4}
\fmf{fermion,left=1,tension=0}{v2,v3}
\fmf{scalar,left=1,tension=0}{v3,v2}
\end{fmfgraph*}
\end{gathered}+2
\begin{gathered}
\begin{fmfgraph*}(60,30)
\fmfleftn{l}{2}\fmfrightn{r}{2}
\fmf{fermion,tension=4}{l1,v1}
\fmf{scalar,tension=4}{v1,r1}
\fmf{scalar,tension=2}{l2,v2}
\fmf{heavy,tension=0.06}{v2,v3}
\fmf{scalar,tension=0.06}{v3,v4}
\fmf{fermion,tension=2}{v4,r2}
\fmf{photon,tension=2}{v2,v4}
\fmf{photon,tension=0.1}{v1,v3}
\end{fmfgraph*}
\end{gathered}+2
\begin{gathered}
\begin{fmfgraph*}(60,30)
\fmfleftn{l}{2}\fmfrightn{r}{2}
\fmf{fermion,tension=4}{l1,v1}
\fmf{scalar,tension=4}{v1,r1}
\fmf{scalar,tension=2}{l2,v2}
\fmf{scalar,tension=0.06}{v2,v3}
\fmf{fermion,tension=0.06}{v3,v4}
\fmf{fermion,tension=2}{v4,r2}
\fmf{photon,tension=2}{v2,v4}
\fmf{photon,tension=0.1}{v1,v3}
\end{fmfgraph*}
\end{gathered}+
\begin{gathered}
\begin{fmfgraph*}(60,30)
\fmfleftn{l}{2}\fmfrightn{r}{2}
\fmf{fermion}{l1,v1}
\fmf{scalar}{l2,v2}
\fmf{scalar}{v3,r1}
\fmf{fermion}{v4,r2}
\fmf{heavy}{v1,v3}
\fmf{heavy}{v2,v4}
\fmf{photon,tension=0}{v1,v2}
\fmf{photon,tension=0}{v3,v4}
\end{fmfgraph*}
\end{gathered}
\end{fmffile}\non\\
&
\begin{fmffile}{diag-quad3}
\fmfset{arrow_len}{2mm}
\fmfset{dash_len}{2mm}
\begin{gathered}
\begin{fmfgraph*}(60,30)
\fmfleftn{l}{2}\fmfrightn{r}{2}
\fmf{heavy}{l1,v1}
\fmf{fermion}{l2,v2}
\fmf{heavy}{v1,r1}
\fmf{fermion}{v2,r2}
\fmf{dbl_wiggly,label=$\delta {{g}_{6}}$,tension=0}{v1,v2}
\end{fmfgraph*}
\end{gathered}=
\begin{gathered}
\begin{fmfgraph*}(60,30)
\fmfleftn{l}{2}\fmfrightn{r}{2}
\fmf{heavy}{l1,v1}
\fmf{fermion}{l2,v2}
\fmf{heavy}{v3,r1}
\fmf{fermion}{v4,r2}
\fmf{heavy}{v1,v3}
\fmf{fermion}{v2,v4}
\fmf{photon,tension=0}{v1,v2}
\fmf{photon,tension=0}{v3,v4}
\end{fmfgraph*}
\end{gathered}+
\begin{gathered}
\begin{fmfgraph*}(60,30)
\fmfleftn{l}{2}\fmfrightn{r}{2}
\fmf{heavy}{l1,v1}
\fmf{fermion}{l2,v2}
\fmf{heavy}{v3,r1}
\fmf{fermion}{v4,r2}
\fmf{scalar}{v1,v3}
\fmf{scalar}{v2,v4}
\fmf{photon,tension=0}{v1,v2}
\fmf{photon,tension=0}{v3,v4}
\end{fmfgraph*}
\end{gathered}+
\begin{gathered}
\begin{fmfgraph*}(60,30)
\fmfleftn{l}{2}\fmfrightn{r}{2}
\fmf{heavy}{l1,v1}
\fmf{fermion}{l2,v2}
\fmf{heavy}{v3,r1}
\fmf{fermion}{v4,r2}
\fmf{fermion}{v1,v3}
\fmf{heavy}{v2,v4}
\fmf{photon,tension=0}{v1,v2}
\fmf{photon,tension=0}{v3,v4}
\end{fmfgraph*}
\end{gathered}+
\begin{gathered}
\begin{fmfgraph*}(60,30)
\fmfleftn{l}{2}\fmfrightn{r}{2}
\fmf{heavy}{v1,l1}
\fmf{fermion}{l2,v2}
\fmf{heavy}{r1,v3}
\fmf{fermion}{v4,r2}
\fmf{fermion}{v3,v1}
\fmf{scalar}{v2,v4}
\fmf{photon,tension=0}{v1,v2}
\fmf{photon,tension=0}{v3,v4}
\end{fmfgraph*}
\end{gathered}+
\begin{gathered}
\begin{fmfgraph*}(60,30)
\fmfleftn{l}{2}\fmfrightn{r}{2}
\fmf{heavy}{v1,l1}
\fmf{fermion}{l2,v2}
\fmf{heavy}{r1,v3}
\fmf{fermion}{v4,r2}
\fmf{scalar}{v3,v1}
\fmf{heavy}{v2,v4}
\fmf{photon,tension=0}{v1,v2}
\fmf{photon,tension=0}{v3,v4}
\end{fmfgraph*}
\end{gathered}
\end{fmffile}\non\\
&
\begin{fmffile}{diag-quad41}
\fmfset{arrow_len}{2mm}
\fmfset{dash_len}{2mm}
\begin{gathered}
\begin{fmfgraph*}(60,30)
\fmfleftn{l}{2}\fmfrightn{r}{2}
\fmf{heavy}{l1,v1}
\fmf{fermion}{l2,v2}
\fmf{fermion}{v1,r1}
\fmf{heavy}{v2,r2}
\fmf{dbl_wiggly,label=$\delta {{g}_{7}}$,tension=0}{v1,v2}
\end{fmfgraph*}
\end{gathered}=
\begin{gathered}
\begin{fmfgraph*}(60,30)
\fmfleftn{l}{2}\fmfrightn{r}{2}
\fmf{heavy}{l1,v1}
\fmf{fermion}{l2,v2}
\fmf{fermion}{v3,r1}
\fmf{heavy}{v4,r2}
\fmf{heavy}{v1,v3}
\fmf{fermion}{v2,v4}
\fmf{photon,tension=0}{v1,v2}
\fmf{photon,tension=0}{v3,v4}
\end{fmfgraph*}
\end{gathered}+
\begin{gathered}
\begin{fmfgraph*}(60,30)
\fmfleftn{l}{2}\fmfrightn{r}{2}
\fmf{heavy}{l1,v1}
\fmf{fermion}{l2,v2}
\fmf{fermion}{v3,r1}
\fmf{heavy}{v4,r2}
\fmf{scalar}{v1,v3}
\fmf{scalar}{v2,v4}
\fmf{photon,tension=0}{v1,v2}
\fmf{photon,tension=0}{v3,v4}
\end{fmfgraph*}
\end{gathered}+
\begin{gathered}
\begin{fmfgraph*}(60,30)
\fmfleftn{l}{2}\fmfrightn{r}{2}
\fmf{heavy}{l1,v1}
\fmf{fermion}{l2,v2}
\fmf{fermion}{v3,r1}
\fmf{heavy}{v4,r2}
\fmf{fermion}{v1,v3}
\fmf{heavy}{v2,v4}
\fmf{photon,tension=0}{v1,v2}
\fmf{photon,tension=0}{v3,v4}
\end{fmfgraph*}
\end{gathered}+
\begin{gathered}
\begin{fmfgraph*}(60,30)
\fmfleftn{l}{2}\fmfrightn{r}{2}
\fmf{fermion,tension=2}{l1,v1}
\fmf{scalar,tension=2}{l2,v4}
\fmf{heavy,tension=2}{v1,r1}
\fmf{heavy,tension=2}{v4,r2}
\fmf{phantom,tension=0.05}{v1,v2,v3,v4}
\fmf{photon,tension=0.05}{v1,v2}
\fmf{photon,tension=0.05}{v3,v4}
\fmf{scalar,left=1,tension=0}{v2,v3}
\fmf{fermion,left=1,tension=0}{v3,v2}
\end{fmfgraph*}
\end{gathered}+
\begin{gathered}
\begin{fmfgraph*}(60,30)
\fmfleftn{l}{2}\fmfrightn{r}{2}
\fmf{fermion,tension=2}{l1,v1}
\fmf{scalar,tension=2}{l2,v4}
\fmf{heavy,tension=2}{v1,r1}
\fmf{heavy,tension=2}{v4,r2}
\fmf{phantom,tension=0.05}{v1,v2,v3,v4}
\fmf{photon,tension=0.05}{v1,v2}
\fmf{photon,tension=0.05}{v3,v4}
\fmf{heavy,left=1,tension=0}{v2,v3}
\fmf{scalar,left=1,tension=0}{v3,v2}
\end{fmfgraph*}
\end{gathered}
\end{fmffile}\non\\
&\qquad\qquad\qquad\qquad
\begin{fmffile}{diag-quad42}
\fmfset{arrow_len}{2mm}
\fmfset{dash_len}{2mm}
+2
\begin{gathered}
\begin{fmfgraph*}(60,30)
\fmfleftn{l}{2}\fmfrightn{r}{2}
\fmf{heavy,tension=4}{l1,v1}
\fmf{fermion,tension=4}{v1,r1}
\fmf{fermion,tension=2}{l2,v2}
\fmf{scalar,tension=0.06}{v2,v3}
\fmf{fermion,tension=0.06}{v3,v4}
\fmf{heavy,tension=2}{v4,r2}
\fmf{photon,tension=2}{v2,v4}
\fmf{photon,tension=0.1}{v1,v3}
\end{fmfgraph*}
\end{gathered}+2
\begin{gathered}
\begin{fmfgraph*}(60,30)
\fmfleftn{l}{2}\fmfrightn{r}{2}
\fmf{heavy,tension=4}{l1,v1}
\fmf{fermion,tension=4}{v1,r1}
\fmf{fermion,tension=2}{l2,v2}
\fmf{heavy,tension=0.06}{v2,v3}
\fmf{scalar,tension=0.06}{v3,v4}
\fmf{heavy,tension=2}{v4,r2}
\fmf{photon,tension=2}{v2,v4}
\fmf{photon,tension=0.1}{v1,v3}
\end{fmfgraph*}
\end{gathered}
\end{fmffile}
\end{align}
\end{widetext}
Comparing this set with the corresponding pRG equations for the $3p$ model on a square lattice (one hole pocket at $\Gamma$ and two electron pockets at $(0,\pi)$ and $(\pi,0)$),  we note that in our case the r.h.s of the flow equations contain additional terms due to the presence of the  Umklapp $g_8$ term, which couples particle-particle and particle-hole channels~\cite{Ganesh2014}.

  To analyze the fixed trajectories of the pRG flow we rewrite interactions as $g_i=\gamma_i g$, where we choose  $g$ as one of the couplings, which increases under pRG and eventually diverges along the fixed trajectory (as we verify a'posteriori), and  assume that $\gamma_i$  tend to some constant values $\gamma^*_i$ at the fixed trajectory~\cite{Chubukov2009_r}. We then search for
  the solutions
\begin{equation}
\beta_i=\dot{\gamma_i}=\frac{1}{g}(\dot{g_i}-\gamma_i \dot{g}) =0
\end{equation}
The fixed trajectory is stable if small perturbations around it do not grow, i.e. the real parts of the eigenvalues of the matrix $T_{ij}=\partial{\beta_i}/\partial{\gamma_j}|_{\gamma^*}$ are negative.

\begin{widetext}
  \begin{figure*}[t]
  \centering
  \subfigure[]{\includegraphics[width=0.45\linewidth]{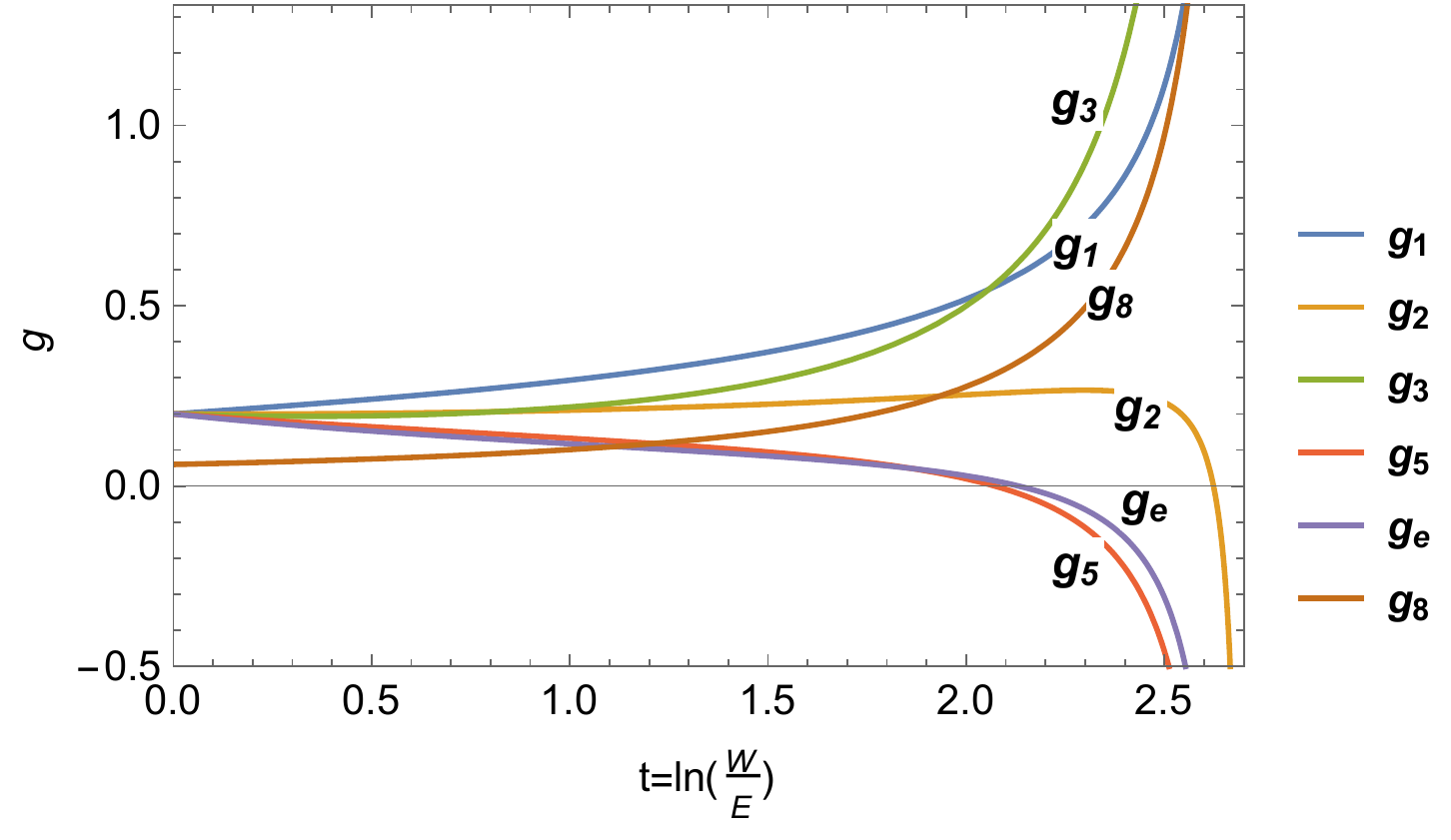}\label{fig:RG1}}
  \quad~~\subfigure[]{\includegraphics[width=0.45\linewidth]{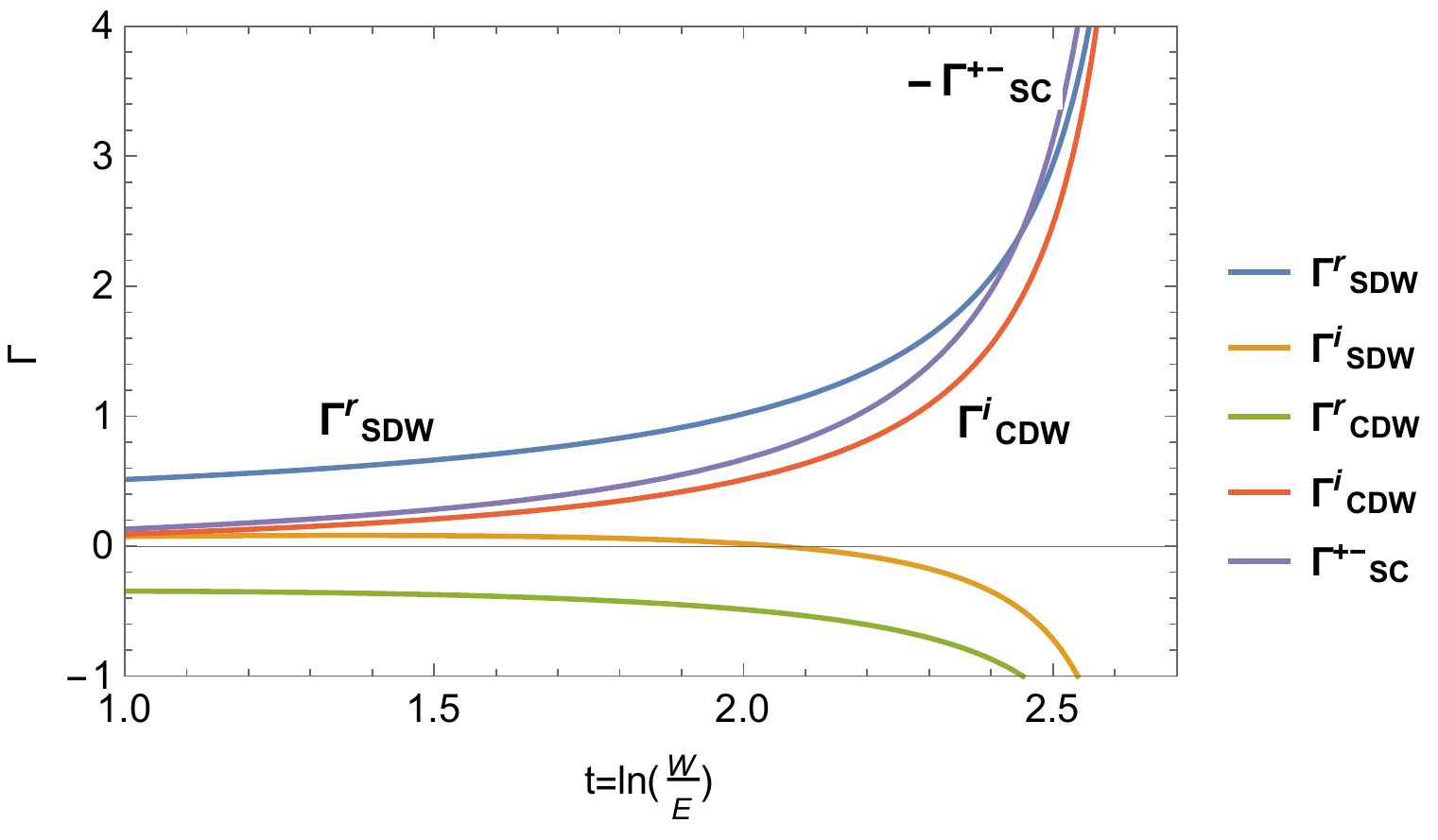}\label{fig:RG1v}}\\
  \quad\subfigure[]{\includegraphics[width=0.45\linewidth]{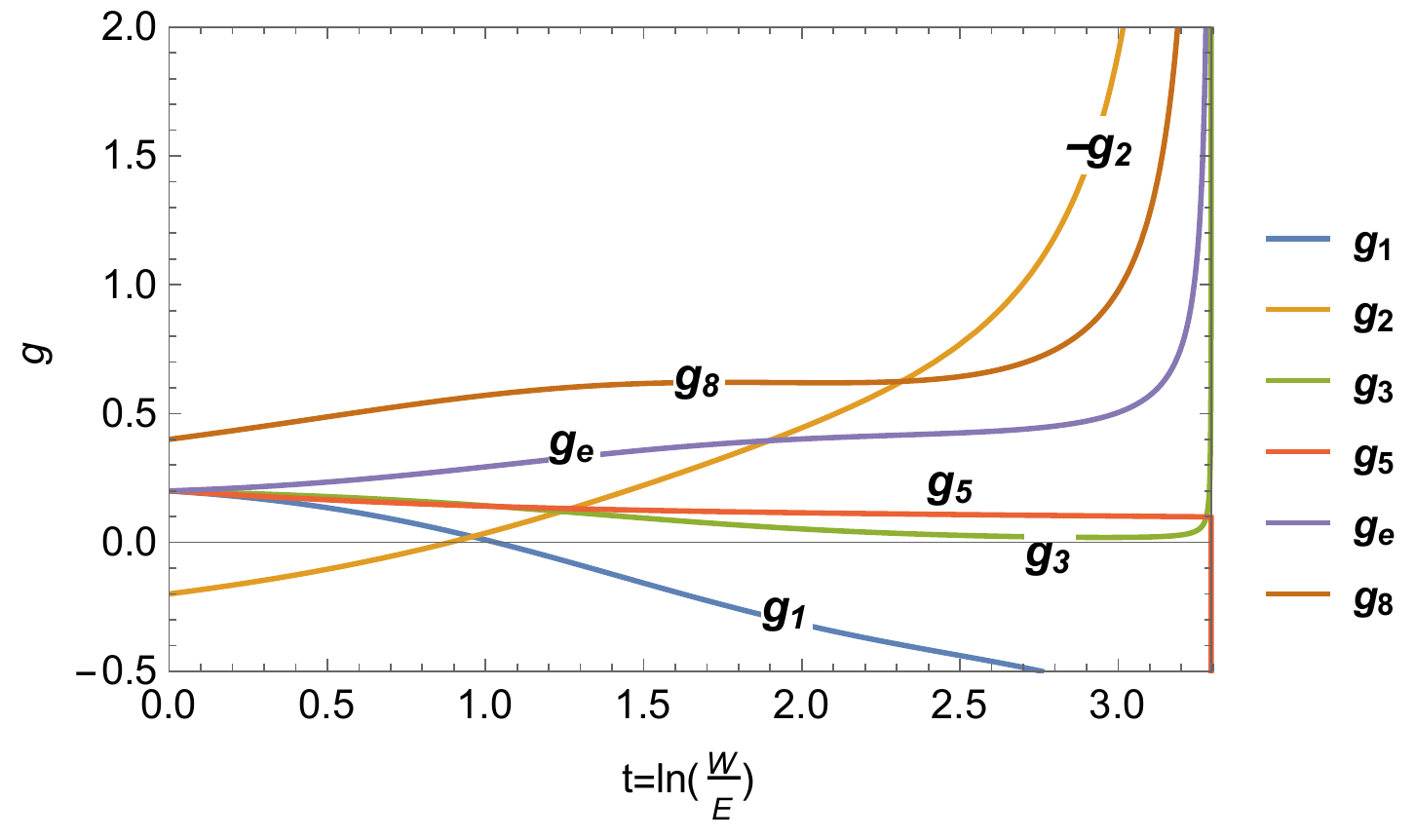}\label{fig:RG2}}
  \quad~~\subfigure[]{\includegraphics[width=0.45\linewidth]{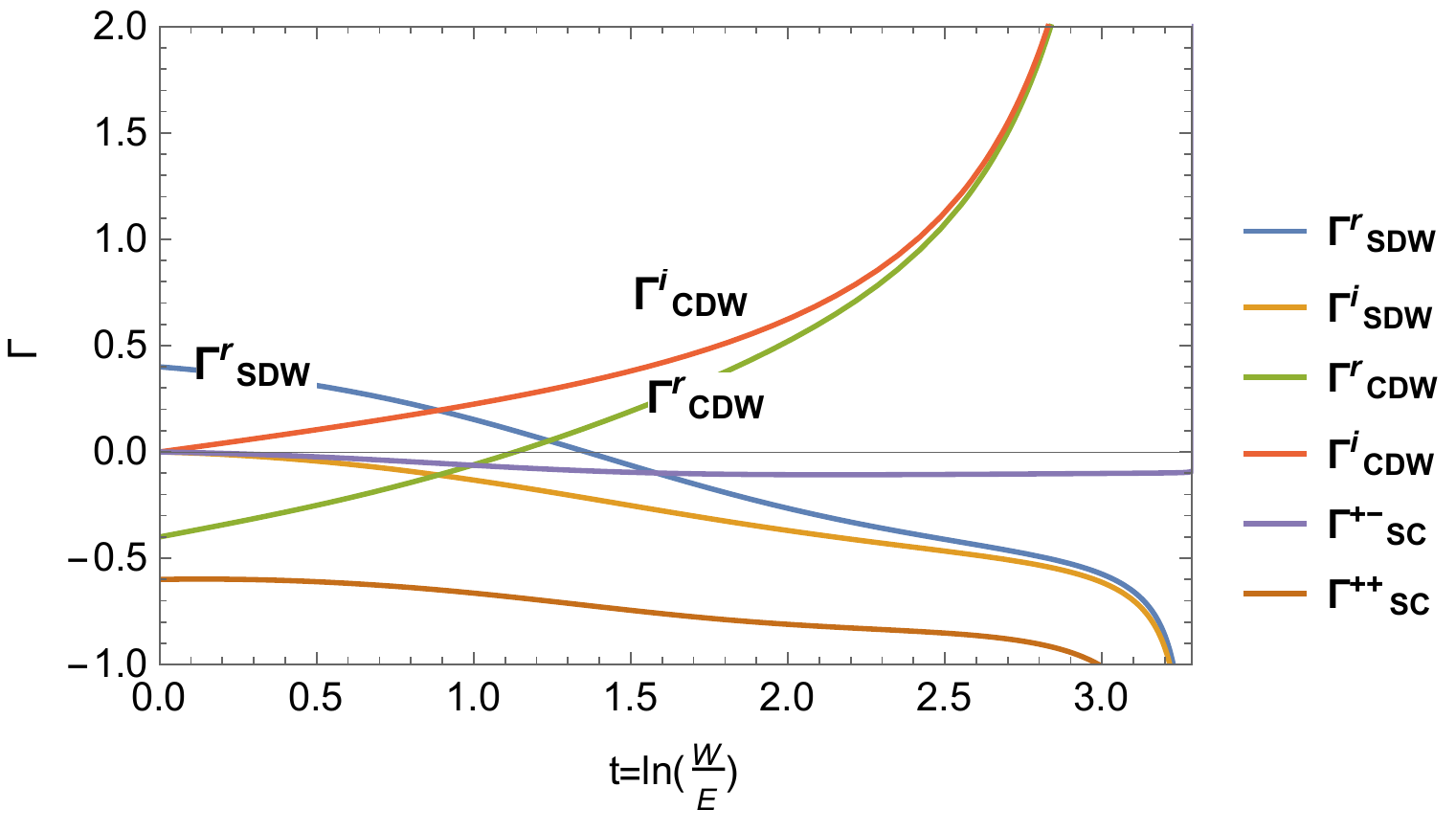}\label{fig:RG2v}}
  \caption{ The renormalization group (RG) flow of the interactions and the effective vertices. We assume that system parameters are such that parquet RG flow runs over a wide range of energies.
   Panels  $({\bf a})$ and  $({\bf b})$ -- the flow when the initial values of the couplings are
   $g_1^{(0)}=g_2^{(0)}=g_3^{(0)}=g_5^{(0)}=g_e^{(0)}=g^{(0)}=0.2,~g_8^{(0)}=0.3g^{(0)}$. At the beginning of the flow SDW vertex $\Gamma^r_{SDW}$ is the largest, but near the fixed trajectory the vertex  $\Gamma^{+-}_{sc}$ in superconducting $s^{+-}$ channel diverges stronger than other vertices.  Panels  {\bf{(c) and (d)}}: the flow when the initial values of the couplings are $g_1^{(0)}=g_2^{(0)}=g_3^{(0)}=g_5^{(0)}=g_e^{(0)}=g^{(0)}=0.2, g_8^{(0)}=2g^{(0)}$. The SDW vertex $\Gamma^r_{SDW}$  is again the largest one at the beginning of the flow, but near the fixed trajectory
      the vertex $\Gamma^i_{CDW}$ in "imaginary" charge density wave channel becomes the largest.  The divergence of $\Gamma^i_{CDW}$ signals an instability into a state with non-zero magnitude of the imaginary part of the expectation value of a charge operator on a bond.
  \label{fig:RGflow}}
\end{figure*}
\end{widetext}

  We focus on the effects of $g_8$ in the RG flow and study the fixed trajectories obtained by varying $g_8$ from weak to strong relative to other interactions $g_1-g_7$. For  definiteness we set the bare values of all other interactions to be equal and positive, i.e. set $g_i^{(0)}=g^{(0)}$, $i=1,2,3, 5, e$. We find two stable fixed trajectories by varying $g_8^{(0)}$.  The pRG flow is towards one fixed trajectory when $g_{8}^{(0)} < g_{8,c}^{(0)} = \frac{1}{2}g^{(0)}$ and towards the other when $g_{8}^{(0)} > g_{8,c}^{(0)}$. We show the pRG flow of the couplings for  $g_{8}^{(0)} < g_{8,c}^{(0)}$ and $g_{8}^{(0)} > g_{8,c}^{(0)}$  in Fig. \ref{fig:RGflow}. We checked that these two fixed trajectories are stable.  We didn't search for other possible fixed trajectories in the 6-dimensional space of the bare couplings.

The couplings along these two trajectories are:
\begin{itemize}
\item[(1)] $g_8^{(0)}<g_{8,c}^{(0)}$.
  We choose $g_1 =g$, $g_i=\gamma_i g_1$.  We find $g=(3/23) \frac{1}{(t_0-t)}$, $\gamma_2=\gamma_8=0,~\gamma_3=2\sqrt{5/3},~\gamma_5=-1,~\gamma_e=-4/3$. On a more careful look we find that $g_8$ still diverges, but  with a smaller exponent,  as $g_8\sim \frac{1}{(t_0-t)^{0.56}}$.
    \item[(2)] $g_8^{(0)}>g_{8,c}^{(0)}$. Now the system flows to another fixed point where $g_2$ remains the only leading divergent interaction and it changes sign in the process of pRG flow and becomes negative along the fixed trajectory. We choose  $g_2=g$, $g_i=\gamma_i g_2$, and obtain
        $g= (-1/2) \frac{1}{(t_0-t)}$, $\gamma_1=\gamma_3=\gamma_5=\gamma_e=\gamma_8=0$. Again, on a more careful look we find that
          $g_8$ and $g_1$ actually also diverge and are only logarithmically smaller than $g$:
         $g_8= \frac{1}{\sqrt{6}}\frac{1}{t_0-t}(\log\frac{1}{t_0-t})^{-0.5}$, $g_1=\frac{-1}{6}\frac{1}{t_0-t}(\log\frac{1}{t_0-t})^{-1}$.
    \end{itemize}
\subsection{Interactions in different channels}

We now need to relate pRG results to the competition between different ordering tendencies.  To do this, we introduce infinitesimal vertices for various bilinear combinations of fermions and find which combination of $g_i$ contributes to the  renormalization of each of these vertices.  To be more specific, we introduce SDW and CDW vertices with incoming momentum $\pm {\bf K}$ and SC vertices for fermions near hole or electron pockets, with zero total momentum.  These vertices are
\begin{align}
&{\text {SDW}} ~~ {\bf{\Delta}}_{\pm K}^{s}\cdot\sum_{\kv} \cd_{\kv}\vs f_{\kv\pm K},\non\\
&{\text {CDW}} ~~ \Delta_{\pm K}^{c}\sum_{\kv} \cd_{\kv}\sigma^{0} f_{\kv\pm K},\non\\
&{\text {SC}} ~~~~~~\Delta_{h}^{sc}\sum_{\kv} \cd_{\kv} i\sigma^{y}\cd_{-\kv},~\Delta_{e}^{sc}\sum_{\kv} \fd_{\kv+ K} i\sigma^{y}\fd_{-\kv-K}
\end{align}
where $\sigma^{0},~\vs$ are the identity and the Pauli matrices in spin space, respectively. The equations for different vertices  are presented diagrammatically in Eqs.~\ref{eq:LinearSDW}-\ref{eq:LinearSC}.
\begin{align}
&\begin{fmffile}{diag-tri1}
\fmfset{arrow_len}{2mm}
\fmfset{dash_len}{2mm}
\begin{gathered}
\begin{fmfgraph*}(60,40)
\fmfleftn{l}{1}\fmfrightn{r}{2}
\fmfpolyn{shaded}{G}{3}
\fmf{photon,label=${{\bf{\Delta}}_{K}^{s*}}$,tension=2}{l1,G1}
\fmf{phantom,label=${\vec{\sigma}}$,tension=0}{G2,G3}
\fmf{scalar,tension=0.65}{r1,G2}\fmf{fermion,tension=0.65}{G3,r2}
\end{fmfgraph*}
\end{gathered}=
\begin{gathered}
\begin{fmfgraph*}(60,40)
\fmfleftn{l}{1}\fmfrightn{r}{2}
\fmfpolyn{shaded}{G}{3}
\fmf{photon,label=${{\bf{\Delta}}_{K}^{s*}}$,tension=2}{l1,G1}
\fmf{scalar,tension=0.65}{v1,G2}\fmf{fermion,tension=0.65}{G3,v2}
\fmf{photon,label=$g_1$,tension=0}{v2,v1}
\fmf{scalar,tension=0.65}{r1,v1}\fmf{fermion,tension=0.65}{v2,r2}
\end{fmfgraph*}
\end{gathered}+
\begin{gathered}
\begin{fmfgraph*}(60,40)
\fmfleftn{l}{1}\fmfrightn{r}{2}
\fmfpolyn{shaded}{G}{3}
\fmf{photon,label=${{\bf{\Delta}}_{-K}^{s}}$,tension=2}{l1,G1}
\fmf{heavy,tension=0.65}{v1,G2}\fmf{scalar,tension=0.65}{G3,v2}
\fmf{photon,label=$g_3$,tension=0}{v2,v1}
\fmf{scalar,tension=0.65}{r1,v1}\fmf{fermion,tension=0.65}{v2,r2}
\end{fmfgraph*}
\end{gathered}
\end{fmffile},\non\\
&\begin{fmffile}{diag-tri2}
\fmfset{arrow_len}{2mm}
\fmfset{dash_len}{2mm}
\begin{gathered}
\begin{fmfgraph*}(60,40)
\fmfleftn{l}{1}\fmfrightn{r}{2}
\fmfpolyn{shaded}{G}{3}
\fmf{photon,label=${{\bf{\Delta}}_{-K}^{s}}$,tension=2}{l1,G1}
\fmf{heavy,tension=0.65}{r1,G2}\fmf{scalar,tension=0.65}{G3,r2}
\fmf{phantom,label=${\vec{\sigma}}$,tension=0}{G2,G3}
\end{fmfgraph*}
\end{gathered}=
\begin{gathered}
\begin{fmfgraph*}(60,40)
\fmfleftn{l}{1}\fmfrightn{r}{2}
\fmfpolyn{shaded}{G}{3}
\fmf{photon,label=${{\bf{\Delta}}_{-K}^{s}}$,tension=2}{l1,G1}
\fmf{heavy,tension=0.65}{v1,G2}\fmf{scalar,tension=0.65}{G3,v2}
\fmf{photon,label=$g_1$,tension=0}{v2,v1}
\fmf{heavy,tension=0.65}{r1,v1}\fmf{scalar,tension=0.65}{v2,r2}
\end{fmfgraph*}
\end{gathered}+
\begin{gathered}
\begin{fmfgraph*}(60,40)
\fmfleftn{l}{1}\fmfrightn{r}{2}
\fmfpolyn{shaded}{G}{3}
\fmf{photon,label=${{\bf{\Delta}}_{K}^{s*}}$,tension=2}{l1,G1}
\fmf{scalar,tension=0.65}{v1,G2}\fmf{fermion,tension=0.65}{G3,v2}
\fmf{photon,label=$g_3$,tension=0}{v2,v1}
\fmf{heavy,tension=0.65}{r1,v1}\fmf{scalar,tension=0.65}{v2,r2}
\end{fmfgraph*}
\end{gathered}
\end{fmffile}\label{eq:LinearSDW}\\
&\begin{fmffile}{diag-tri-cdw1a}
\fmfset{arrow_len}{2mm}
\fmfset{dash_len}{2mm}
\begin{gathered}
\begin{fmfgraph*}(60,40)
\fmfleftn{l}{1}\fmfrightn{r}{2}
\fmfpolyn{shaded}{G}{3}
\fmf{photon,label=${{\bf{\Delta}}_{K}^{c*}}$,tension=2}{l1,G1}
\fmf{phantom,tension=0}{G2,G3}
\fmf{scalar,tension=0.65}{r1,G2}\fmf{fermion,tension=0.65}{G3,r2}
\end{fmfgraph*}
\end{gathered}=
\begin{gathered}
\begin{fmfgraph*}(60,40)
\fmfleftn{l}{1}\fmfrightn{r}{2}
\fmfpolyn{shaded}{G}{3}
\fmf{photon,label=${{\bf{\Delta}}_{K}^{c*}}$,tension=2}{l1,G1}
\fmf{scalar,tension=0.65}{v1,G2}\fmf{fermion,tension=0.65}{G3,v2}
\fmf{photon,label=$g_1$,tension=0}{v2,v1}
\fmf{scalar,tension=0.65}{r1,v1}\fmf{fermion,tension=0.65}{v2,r2}
\end{fmfgraph*}
\end{gathered}+
\begin{gathered}
\begin{fmfgraph*}(60,40)
\fmfleftn{l}{1}\fmfrightn{r}{2}
\fmfpolyn{shaded}{G}{3}
\fmf{photon,label=${{\bf{\Delta}}_{-K}^{c}}$,tension=2}{l1,G1}
\fmf{heavy,tension=0.65}{v1,G2}\fmf{scalar,tension=0.65}{G3,v2}
\fmf{photon,label=$g_3$,tension=0}{v2,v1}
\fmf{scalar,tension=0.65}{r1,v1}\fmf{fermion,tension=0.65}{v2,r2}
\end{fmfgraph*}
\end{gathered}
\end{fmffile}\non\\
&\qquad\qquad\qquad\quad+
\begin{fmffile}{diag-tri-cdw1b}
\fmfset{arrow_len}{2mm}
\fmfset{dash_len}{2mm}
\begin{gathered}
\begin{fmfgraph*}(60,40)
\fmfleftn{l}{1}
\fmfrightn{r}{2}
\fmfpolyn{shaded}{G}{3}
\fmf{photon,label=${{\bf{\Delta}}_{-K}^{c}}$,tension=2}{l1,G1}
\fmf{scalar,tension=0.5}{r1,v2}\fmf{fermion,tension=0.5}{v2,r2}
\fmf{scalar,right=0.5,tension=0.3}{G2,v1}\fmf{heavy,right=0.5,tension=0.3}{v1,G3}
\fmf{photon,label=$g_3$,tension=2}{v1,v2}
\end{fmfgraph*}
\end{gathered}+
\begin{gathered}
\begin{fmfgraph*}(60,40)
\fmfleftn{l}{1}
\fmfrightn{r}{2}
\fmfpolyn{shaded}{G}{3}
\fmf{photon,label=${{\bf{\Delta}}_{K}^{c*}}$,tension=2}{l1,G1}
\fmf{scalar,tension=0.5}{r1,v2}\fmf{fermion,tension=0.5}{v2,r2}
\fmf{fermion,left=0.5,tension=0.3}{G3,v1}\fmf{scalar,left=0.5,tension=0.3}{v1,G2}
\fmf{photon,label=$g_2$,tension=2}{v1,v2}
\end{fmfgraph*}
\end{gathered}
\end{fmffile}\non\\
&\begin{fmffile}{diag-tri-cdw2a}
\fmfset{arrow_len}{2mm}
\fmfset{dash_len}{2mm}
\begin{gathered}
\begin{fmfgraph*}(60,40)
\fmfleftn{l}{1}\fmfrightn{r}{2}
\fmfpolyn{shaded}{G}{3}
\fmf{photon,label=${{\bf{\Delta}}_{-K}^{c}}$,tension=2}{l1,G1}
\fmf{phantom,tension=0}{G2,G3}
\fmf{heavy,tension=0.65}{r1,G2}\fmf{scalar,tension=0.65}{G3,r2}
\end{fmfgraph*}
\end{gathered}=
\begin{gathered}
\begin{fmfgraph*}(60,40)
\fmfleftn{l}{1}\fmfrightn{r}{2}
\fmfpolyn{shaded}{G}{3}
\fmf{photon,label=${{\bf{\Delta}}_{-K}^{c}}$,tension=2}{l1,G1}
\fmf{heavy,tension=0.65}{v1,G2}\fmf{scalar,tension=0.65}{G3,v2}
\fmf{photon,label=$g_1$,tension=0}{v2,v1}
\fmf{heavy,tension=0.65}{r1,v1}\fmf{scalar,tension=0.65}{v2,r2}
\end{fmfgraph*}
\end{gathered}+
\begin{gathered}
\begin{fmfgraph*}(60,40)
\fmfleftn{l}{1}\fmfrightn{r}{2}
\fmfpolyn{shaded}{G}{3}
\fmf{photon,label=${{\bf{\Delta}}_{K}^{c*}}$,tension=2}{l1,G1}
\fmf{scalar,tension=0.65}{v1,G2}\fmf{fermion,tension=0.65}{G3,v2}
\fmf{photon,label=$g_3$,tension=0}{v2,v1}
\fmf{heavy,tension=0.65}{r1,v1}\fmf{scalar,tension=0.65}{v2,r2}
\end{fmfgraph*}
\end{gathered}
\end{fmffile}\non\\
&\qquad\qquad\qquad\quad+
\begin{fmffile}{diag-tri-cdw2b}
\fmfset{arrow_len}{2mm}
\fmfset{dash_len}{2mm}
\begin{gathered}
\begin{fmfgraph*}(60,40)
\fmfleftn{l}{1}
\fmfrightn{r}{2}
\fmfpolyn{shaded}{G}{3}
\fmf{photon,label=${{\bf{\Delta}}_{-K}^{c}}$,tension=2}{l1,G1}
\fmf{heavy,tension=0.5}{r1,v2}\fmf{scalar,tension=0.5}{v2,r2}
\fmf{fermion,left=0.5,tension=0.3}{G3,v1}\fmf{scalar,left=0.5,tension=0.3}{v1,G2}
\fmf{photon,label=$g_3$,tension=2}{v1,v2}
\end{fmfgraph*}
\end{gathered}+
\begin{gathered}
\begin{fmfgraph*}(60,40)
\fmfleftn{l}{1}
\fmfrightn{r}{2}
\fmfpolyn{shaded}{G}{3}
\fmf{photon,label=${{\bf{\Delta}}_{-K}^{c}}$,tension=2}{l1,G1}
\fmf{heavy,tension=0.5}{r1,v2}\fmf{scalar,tension=0.5}{v2,r2}
\fmf{scalar,left=0.5,tension=0.3}{G3,v1}\fmf{heavy,left=0.5,tension=0.3}{v1,G2}
\fmf{photon,label=$g_2$,tension=2}{v1,v2}
\end{fmfgraph*}
\end{gathered}
\end{fmffile}\label{eq:LinearCDW}\\
&\begin{fmffile}{diag-SC1}
\fmfset{arrow_len}{2mm}
\fmfset{dash_len}{2mm}
\begin{gathered}
\begin{fmfgraph*}(60,40)
\fmfleftn{l}{1}\fmfrightn{r}{2}
\fmfpolyn{shaded}{G}{3}
\fmf{photon,label=${{\Delta}_{h}^{sc}}$,tension=2}{l1,G1}
\fmf{scalar,tension=0.65}{G2,r1}\fmf{scalar,tension=0.65}{G3,r2}
\end{fmfgraph*}
\end{gathered}=
\begin{gathered}
\begin{fmfgraph*}(60,40)
\fmfleftn{l}{1}\fmfrightn{r}{2}
\fmfpolyn{shaded}{G}{3}
\fmf{photon,label=${{\Delta}_{h}^{sc}}$,tension=2}{l1,G1}
\fmf{scalar,tension=0.65}{G2,v1}\fmf{scalar,tension=0.65}{G3,v2}
\fmf{photon,label=$g_5$,tension=0}{v2,v1}\fmf{scalar,tension=0.65}{v1,r1}\fmf{scalar,tension=0.65}{v2,r2}
\end{fmfgraph*}
\end{gathered}+
\begin{gathered}
\begin{fmfgraph*}(60,40)
\fmfleftn{l}{1}\fmfrightn{r}{2}
\fmfpolyn{shaded}{G}{3}
\fmf{photon,label=${{\Delta}_{e}^{sc}}$,tension=2}{l1,G1}
\fmf{fermion,tension=0.65}{G2,v1}\fmf{heavy,tension=0.65}{G3,v2}
\fmf{photon,label=$g_3$,tension=0}{v2,v1}\fmf{scalar,tension=0.65}{v1,r1}\fmf{scalar,tension=0.65}{v2,r2}
\end{fmfgraph*}
\end{gathered}
\end{fmffile}\non\\
&\qquad\qquad\qquad\quad+
\begin{fmffile}{diag-SC1b}
\fmfset{arrow_len}{2mm}
\fmfset{dash_len}{2mm}
\begin{gathered}
\begin{fmfgraph*}(60,40)
\fmfleftn{l}{1}\fmfrightn{r}{2}
\fmfpolyn{shaded}{G}{3}
\fmf{photon,label=${{\Delta}_{e}^{sc}}$,tension=2}{l1,G1}
\fmf{heavy,tension=0.65}{G2,v1}\fmf{fermion,tension=0.65}{G3,v2}
\fmf{photon,label=$g_3$,tension=0}{v2,v1}\fmf{scalar,tension=0.65}{v1,r1}\fmf{scalar,tension=0.65}{v2,r2}\end{fmfgraph*}
\end{gathered}
\end{fmffile}\non\\
&\begin{fmffile}{diag-SC2a}
\fmfset{arrow_len}{2mm}
\fmfset{dash_len}{2mm}
\begin{gathered}
\begin{fmfgraph*}(60,40)
\fmfleftn{l}{1}\fmfrightn{r}{2}
\fmfpolyn{shaded}{G}{3}
\fmf{photon,label=${{\Delta}_{e}^{sc}}$,tension=2}{l1,G1}
\fmf{heavy,tension=0.65}{G2,r1}\fmf{fermion,tension=0.65}{G3,r2}
\end{fmfgraph*}
\end{gathered}=
\begin{gathered}
\begin{fmfgraph*}(60,40)
\fmfleftn{l}{1}\fmfrightn{r}{2}
\fmfpolyn{shaded}{G}{3}
\fmf{photon,label=${{\Delta}_{e}^{sc}}$,tension=2}{l1,G1}
\fmf{heavy,tension=0.65}{G2,v1}\fmf{fermion,tension=0.65}{G3,v2}
\fmf{photon,label=$g_6$,tension=0}{v2,v1}
\fmf{heavy,tension=0.65}{v1,r1}\fmf{fermion,tension=0.65}{v2,r2}
\end{fmfgraph*}
\end{gathered}+
\begin{gathered}
\begin{fmfgraph*}(60,40)
\fmfleftn{l}{1}\fmfrightn{r}{2}
\fmfpolyn{shaded}{G}{3}
\fmf{photon,label=${{\Delta}_{e}^{sc}}$,tension=2}{l1,G1}
\fmf{fermion,tension=0.65}{G2,v1}\fmf{heavy,tension=0.65}{G3,v2}
\fmf{photon,label=$g_7$,tension=0}{v2,v1}
\fmf{heavy,tension=0.65}{v1,r1}\fmf{fermion,tension=0.65}{v2,r2}
\end{fmfgraph*}
\end{gathered}
\end{fmffile}\non\\
&\qquad\qquad\qquad\quad+
\begin{fmffile}{diag-SC2b}
\fmfset{arrow_len}{2mm}
\fmfset{dash_len}{2mm}
\begin{gathered}
\begin{fmfgraph*}(60,40)
\fmfleftn{l}{1}\fmfrightn{r}{2}
\fmfpolyn{shaded}{G}{3}
\fmf{photon,label=${{\Delta}_{h}^{sc}}$,tension=2}{l1,G1}
\fmf{scalar,tension=0.65}{G2,v1}\fmf{scalar,tension=0.65}{G3,v2}
\fmf{photon,label=$g_3$,tension=0}{v2,v1}
\fmf{heavy,tension=0.65}{v1,r1}\fmf{fermion,tension=0.65}{v2,r2}
\end{fmfgraph*}
\end{gathered}
\end{fmffile}\label{eq:LinearSC}
\end{align}

We defined the couplings in the magnetic, charge, and SC channels as $\Gamma_{s}, \Gamma_{c}$, and $\Gamma_{sc}$. The sign convention is such that the corresponding interaction is attractive if  $\Gamma_{c}, \Gamma_{s} >0$  and $\Gamma_{sc} <0$.

In the magnetic channel, the result is the same as in our  earlier consideration -- the two order parameters are SDW and ISB, and the corresponding couplings are
\begin{align}
\Gamma^{r,i}_{s}=& g_1\pm g_3,
\label{th_1}
\end{align}
where the superscript $r$ stands for SDW and $i$ stands for ISB ( symmetric and antisymmetric combinations of $\Delta^s_{\pm K}$ and $(\Delta^s_{\pm K})^*$, respectively).

In the charge channel we have
\begin{align}
\Gamma^{r,i}_{c}=& g_1 \mp g_3 -2 g_2,
\label{th_2}
\end{align}
where $r$ and $i$ again stand for \textit{symmetric}  and  \textit{antisymmetric}  combinations of $\Delta^c_{\pm K}$ and $(\Delta^c_{\pm K})^*$. The symmetric solution describes a conventional CDW order and the antisymmetric solution describes imaginary charge bond (ICB)  order~\cite{Chubukov2008,Schmalian2015}.  The latter may give rise to circulating charge currents, if the hopping integrals have proper symmetry properties.

In the SC channel we have
\begin{align}
\Gamma^{+}_{sc}=& \frac{(g_5+2g_e)+\sqrt{8 g_3^2+(g_5-2g_e)^2}}{2}, \nonumber \\
\Gamma^{-}_{sc}=&\frac{(g_5+2g_e)-\sqrt{8 g_3^2+(g_5-2g_e)^2}}{2}
\label{th_3}
\end{align}
The solution with $\Gamma^{+}_{sc}$ is a  conventional $s^{++}$ pairing with ${\Delta}_{h}^{sc},{\Delta}_{e}^{sc}$ having the same sign. The  solution with $\Gamma^{-}_{sc}$ is a $s^{+-}$ pairing for which  ${\Delta}_{h}^{sc}$ and ${\Delta}_{e}^{sc}$ having opposite signs.

 The transition temperatures of potential density-wave and pairing instabilities  are
\begin{align}
1&=-T^{r,i}_{s}\Gamma^{r,i}_{s} \Pi_{ph} (\pm K),~~1=-T^{r,i}_{c}\Gamma^{r,i}_{c} \Pi_{ph} (\pm K),\non\\
1&=-T^{+,-}_{sc}\Gamma^{+,-}_{sc} \Pi_{pp} (0)
\end{align}
where
\begin{align}
\Pi_{ph} (\pm \bf K) &=\sum_{\omega_m}\int \diff\epsilon_{\kv}\mc{G}^c(\kv,\omega_m)\mc{G}^f(\kv\pm K,\omega_m), \non\\
\Pi_{pp} (0) &= \sum_{\omega_m}\int \diff\epsilon_{\kv}\mc{G}^c(\kv,\omega_m)\mc{G}^c(-\kv,-\omega_m).
\end{align}
At a perfect nesting,  $\Pi_{ph}  (\pm {\bf K})= - \Pi_{pp} (0)$. Then the leading instability will be in the channel for which $\Gamma$ is of proper sign and the largest by magnitude.   Away from perfect nesting, $\Pi_{ph} (\pm {\bf K})$ and $-\Pi_{pp} (0)$ differ by the ratio of the masses $m_h/m_e$, but still are logarithmic.
 For simplicity, below we assume $m_h=m_e$.

If we set the bare values of the couplings to be the same, the interactions in $s^{++}$ SC channel and in CDW channel are repulsive, the ones in ISB, ICB, and $s^{+-}$ SC channel vanish, and the interaction in SDW channel is attractive. At this level, the SDW is the leading instability.

 If, however, we allow RG to run and compare $\Gamma$'s  for the couplings along the fixed trajectory, we obtain different results. For the first fixed trajectory (smaller $g^{(0)}_8$) we have
 \begin{align}
 &\Gamma^{r}_{s} = \Gamma^{i}_{c} = 3.58 g_1,~~ \Gamma^{i}_{s} =    \Gamma^{r}_{c} = -1.58 g_1, \nonumber \\
& \Gamma^{+}_{sc} = 1.91g_1, ~~ \Gamma^{-}_{sc} = -5.58 g_1,~~g_1=\frac{3}{23}\frac{1}{t_0-t}
 \end{align}
 We see that the largest coupling is in $s^{+-}$ superconducting channel.
 For the  second fixed trajectory (larger $g^{(0)}_8$) we have
 \begin{align}
& \Gamma^{r}_{s} = \Gamma^{i}_{s} =0,~~ \Gamma^{r}_{c} = \Gamma^{r}_{c} = 2|g_2| = \frac{1}{t_0-t}, \nonumber \\
& \Gamma^{+}_{sc} = \Gamma^{-}_{sc} = 0
 \end{align}
Now the largest vertex is in CDW and ICB channels. To lift the degeneracy between the two we  notice that the condition $\gamma_1 = \gamma_3 =0$ along this fixed trajectory does not imply that $g_1$ and $g_3$ vanish but rather that they are parametrically smaller than $|g_2|$. For our purpose, it is sufficient to note that   $\Gamma^{r,i}_{c}=g_1\pm g_3-2g_2$, and  $g_3>0$  remains positive in the pRG flow. As the consequence,  $\Gamma^i_{c} > \Gamma^r_{c}$, i.e., the leading instability is towards an unconventional  ICB order. A similar instability has been previously found in  $4p$ model on a hexagonal lattice~\cite{Ganesh2014}.

\section{Summary}\label{sec:Summary}
In this work we studied the three-pocket itinerant fermion system on a 2D triangular lattice.
 We assumed that there is a small hole pocket centered at $\Gamma = (0,0)$ and two electron pockets centered at $\pm {\bf K} = \pm (4\pi/3,0)$.  Our goals were to study in detail the magnetic order in such a system in zero and a finite magnetic field, and the interplay between magnetism and another potential orders like superconductivity and charge order.
  We first analyzed  Stoner type magnetism in zero field. We found that for purely repulsive interaction the leading instability is towards a conventional SDW order with
   momentum ${\pm \bK}$. The SDW order parameter ${\bf M}_{ K}$ satisfies ${\bf M}_{-{ K}} = {\bf M}^*_{ K}$, but ${\bf M}_{ K}$ is a complex order parameter
   ${\bf M}_{ K} = {\bf M}_r + i {\bf M}_i$. In mean-field approximation the Free energy depends on ${\bf M}^2_r + {\bf M}^2_i$, i.e., the ground state is infinitely divergent. Different choices of
   ${\bf M}_r$ and ${\bf M}_i$, subject to ${\bf M}^2_r + {\bf M}^2_i = {\text const}$, yield different spin configurations from a degenerate manifold. Beyond mean-field, we found that the ground state degeneracy is lifted.  Depending on parameters, the ground state configuration is either $120^{\circ}$ ``triangular" structure (same as for localized spins), or a collinear state with antiferromagnetic spin order on 2/3 of sites and no magnetic  order on the remaining 1/3 sites. Such partial  order with non-equal magnitude of magnetization on different sites cannot be realized in a localized spin system.

  When some interactions are repulsive and some attractive, the system develops another type of order, which we labeled as ISB order. The corresponding order parameter is the imaginary part of the (complex) expectation value of a spin operator on a bond.  This order parameter is even under time reversal. We argued that an ISB state can possess circulating spin currents if the hopping integrals have a certain symmetry.

  We then returned to a system with purely repulsive interactions and considered a magnetic order in a non-zero field.  We  found that  $120^{\circ}$ ``triangular" spin configuration becomes a non-coplanar cone state with   $120^{\circ}$  spin order in the plane perpendicular to the field and ferromagnetic order along the field.  We also found that a field generates a bilinear coupling between SDW and ISB order parameters, i.e.,  a SDW order in a field immediately triggers an ISB order.  This is one of the central results of our work.

We next considered the interplay between magnetism  and superconductivity and charge order. For this, we analyzed the flow of the couplings within pRG and used the running couplings to analyze the flow of the effective interactions in magnetic, SC, and charge channels.
 We argued that magnetic order develops if there is little space for pRG, however if the system parameters are such that pRG runs over a wide window of energies,
    the couplings flow towards one of the two fixed trajectories (depending on the values of the bare couplings), and for both fixed trajectories  magnetism is not the leading instability.  For one fixed trajectory we found that the leading instability is towards
    $s^{\pm}$ superconductivity, for the other the leading instability is towards ICB order, which may support circulating charge currents. This  highly unconventional charge order is induced by the Umklapp scattering process ($g_8$ term), which couples particle-hole and particle-particle channels.

  We call for the extension of our work to multi-orbital models of fermions on a triangular lattice. Among other things, these studies should settle the issue whether the ISB/ICB  orders, which we found, support circulating spin/charge currents.

\begin{acknowledgements}
We acknowledge with thanks useful conversations with Cristian Batista, Rafael Fernandes, and Jian Kang. The work was supported by the Office of Basic Energy Sciences U. S. Department of Energy under the award de-sc0014402. M.Y. acknowledges support from the KITP graduate fellowship program.
\end{acknowledgements}

\bibliography{FM3p}

\clearpage
\appendix
\onecolumngrid
\section{Effective action for the spin order}
\label{app:HubbardStratonovich}
In this section, we follow the standard Hubbard-Stratonovich transformation and derive the effective action for the spin order. We show that the symmetric and antisymmetric component of $\{\bsdw{K}, \bsdw{-K}^*\}$ naturally decouple in zero field, and are coupled by the magnetic field.

Consider interactions restricted to the spin channel, Eq.~\ref{eq:H4Spin} in the main text,
\begin{align}\label{eq:interactionSpin}
\mc{H}_4=\sum_{\qv}-\frac{g_3}{2}\big(\hat{\bf \Delta}_{K-q}\hat{\bf \Delta}_{-K+q}+h.c.)\big)-\frac{g_1}{2}\big(\hat{\bf \Delta}^{\dg}_{K-q}\hat{\bf \Delta}_{K+q}+(K\rightarrow -K)\big)+...,
\end{align}
We apply the identity $e^{w^{\dg}A w}=\int \mc{D}v~e^{-v^{\dg}A^{-1}v+w^{\dg}v+v^{\dg}w}$ ($A$ should be positive definite for convergence), and obtain the partition function in terms of 6-component fermionic field $\Psi$ and bosonic field $v$:
\begin{align}
Z=\int \mc{D}\bar{\Psi}\mc{D}\Psi\mc{D}v~e^{-S[\Psi,v]}.
\end{align}
From Eq.~\ref{eq:interactionSpin}, $w=\{\hat{{\bf \Delta}}_{K}, \hat{{\bf \Delta}}_{-K}, \hat{{\bf \Delta}}^{\dg}_{K},\hat{{\bf \Delta}}^{\dg}_{-K}\}^T$ and
\begin{align}
A=\frac{1}{4}
\begin{pmatrix}
g_1 & 0 & 0 & g_3\\
0 & g_1 & g_3 & 0\\
0 & g_3 & g_1 & 0\\
g_3 & 0 & 0 & g_1
\end{pmatrix}
\end{align}
The action written in compact form as :
\begin{align}\label{appeq:action}
S[\Psi,v]=\int_k-\Psi^{\dg}_k\mc{G}_{0,k}^{-1}\Psi_k+v^{\dg}A^{-1}v-w^{\dg}v-v^{\dg}w
\end{align}
We express the bosonic field $v$ as $v=\frac{1}{2}\{\bar {\bf \Delta}_{K}, \bar {\bf \Delta}_{-K}, \bar {\bf \Delta}_{K}^*,\bar {\bf \Delta}_{-K}^*\}^T$ to relate it with the order parameter field at mean field level. Eq.~\ref{appeq:action} becomes:
\begin{align}
S[\Psi,v]=\int_k-\Psi^{\dg}_k\mc{G}_k^{-1}\Psi_k+v^{\dg}A^{-1}v
\end{align}
where $\mc{G}_k^{-1}=\mc{G}_{0,k}^{-1}-\mc{V}$, with
\begin{align} \label{appeq:noninteracting}
\mc{V}=
-\begin{pmatrix}
0 & \bar {\bf \Delta}_{K}\cdot\vec{\sigma}\ & \bar {\bf \Delta}_{-K}\cdot\vec{\sigma}\\
\bar {\bf \Delta}_{K}^*\cdot\vec{\sigma} & 0 & 0\\
\bar {\bf \Delta}_{-K}^*\cdot\vec{\sigma} & 0 & 0
\end{pmatrix},~~
A^{-1}=\frac{4}{g_1^2-g_3^2}
\begin{pmatrix}
g_1 & 0 & 0 & -g_3\\
0 & g_1 & -g_3 & 0\\
0 & -g_3 & g_1 & 0\\
-g_3 & 0 & 0 & g_1
\end{pmatrix}
\end{align}
The canonical bosonic fields can be obtained by diagonalizing $A^{-1}$, and are
\begin{align}
\bar{\bf{M}}_{\pm K}&=\frac{1}{2}(\bar {\bf \Delta}_{\pm K}+\bar {\bf \Delta}_{\mp K}^*),\non\\
\bar{\bf{\Phi}}_{\pm K}&=\frac{1}{2}(\bar {\bf \Delta}_{\pm K}-\bar {\bf \Delta}_{\mp K}^*).
\end{align}
We note that under time-reversal, $\bar {\bf \Delta}_{\pm K}\rightarrow -\bar {\bf \Delta}^*_{\mp K}$. As a result, $\bar{\bf{M}}_{\pm K}$ is odd under time-reversal and $\bar{\bf{\Phi}}_{\pm K}$ is time-reversal symmetric. From App.~\ref{app:order}, ${\bf{M}}_{\pm K}$ and ${\bf{\Phi}}_{\pm K}$, defined in momentum space, contributes to SDW and ISB order in real space, respectively. $v^\dg A^{-1} v$ becomes:
\begin{align}
v^\dg A^{-1} v=\frac{2}{g_1+g_3}(|\bar{\bf{M}}_K|^2+|\bar{\bf{M}}_{-K}|^2)+\frac{2}{g_1-g_3}(|\bar{\bf{\Phi}}_K|^2+|\bar{\bf{\Phi}}_{-K}|^2)
\end{align}
The quadratic coupling of fermions $\mc{V}$ can be written as $\mc{V}=\mc{V}^{M}+\mc{V}^{\Phi}$, with
\begin{align}\label{appeq:Vmatrix}
\mc{V}^M=-
\begin{pmatrix}
0 & \bar{\bf{M}}_K\cdot\vec{\sigma} & \bar{\bf{M}}_{-K}\cdot\vec{\sigma}\\
\bar{\bf{M}}_{-K}\cdot\vec{\sigma} & 0 & 0\\
\bar{\bf{M}}_{K}\cdot\vec{\sigma} & 0 & 0
\end{pmatrix},~~
\mc{V}^{\Phi}=-
\begin{pmatrix}
0 & \bar{\bf{\Phi}}_K\cdot\vec{\sigma} & \bar{\bf{\Phi}}_{-K}\cdot\vec{\sigma}\\
-\bar{\bf{\Phi}}_{-K}\cdot\vec{\sigma} & 0 & 0\\
-\bar{\bf{\Phi}}_{-K}\cdot\vec{\sigma} & 0 & 0
\end{pmatrix}.
\end{align}

Since the action is quadratic in fermion operators, it is straight forward to integrate out the fermion fields and obtain the effective action in terms of bosonic fields as
\begin{align}
\label{appeq:action2}
S_{eff}[\bsdw{K},\bsdw{-K}]=-\Tr\ln\big(1-\mc{G}_{0,k}\mc{V}\big)+\int_{\qv}\frac{2}{g_1+g_3}(|\bar{\bf{M}}_K|^2+|\bar{\bf{M}}_{-K}|^2)+\frac{2}{g_1-g_3}(|\bar{\bf{\Phi}}_K|^2+|\bar{\bf{\Phi}}_{-K}|^2)
\end{align}
Right below the transition temperature that the ordering instability starts developing, $\Tr\ln\big(1-\mc{G}_{0,k}\mc{V}\big)$ can be expanded in powers of $\mc{V}$ as
\begin{align}
S_{eff}[\bsdw{K},\bsdw{-K}]=&\sum_{n}\frac{1}{n}\Tr (\mc{G}_{0,k}\mc{V})^n+\int_{\qv}\frac{2}{g_1+g_3}(|\bar{\bf{M}}_K|^2+|\bar{\bf{M}}_{-K}|^2)+\frac{2}{g_1-g_3}(|\bar{\bf{\Phi}}_K|^2+|\bar{\bf{\Phi}}_{-K}|^2),
\end{align}
where $\Tr(...)$ sums over momentum, frequency and spin indices. By solving self-consistency equations, we verified that $\bar{\bf \Delta}_{\pm \bK}=\frac{g_{sdw}}{2}{\bf \Delta}_{\pm \bK}$, $\bar{\bf \Phi}_{\pm \bK}=\frac{g_{sdw}}{2}{\bf \Phi}_{\pm \bK}$, $\bar{\bf M}_{\pm \bK}=\frac{g_{sdw}}{2}{\bf M}_{\pm \bK}$, where ${\bf \Delta}_{\pm \bK}, {\bf M}_{\pm \bK}, {\bf \Phi}_{\pm \bK}$ are defined in the main text, e.g. Eq.~\ref{me_1}.

\subsection{Effective action in zero field}
In zero field, evaluation of the trace $\frac{1}{2}\Tr (\mc{G}_{0,k}\mc{V})^2$ yields identical quadratic coefficients for both $|\bar{\bf{M}}|^2$ and $|\bar{\bf{\Phi}}|^2$, i.e.\begin{align}\label{eqapp:effectiveQuadratic}
S_{eff,2}=\int_{\qv}(\frac{2}{g_1+g_3}+\xi_0)(|\bar{\bf{M}}_K|^2+|\bar{\bf{M}}_{-K}|^2)+(\frac{2}{g_1-g_3}+\xi_0)(|\bar{\bf{\Phi}}_K|^2+|\bar{\bf{\Phi}}_{-K}|^2)
\end{align}
where $\xi_0=T\sum_{\omega_m}\int \diff\epsilon_{\kv}\mc{G}^c(\kv,\omega_m)\mc{G}^f(\kv\pm K,\omega_m)<0$.

At mean field level, due to the repulsive Coulomb interaction, $g_1+g_3>g_1-g_3$. As a result, the quadratic coefficient for ${\bf{M}}_{\pm K}$ becomes negative first, i.e. the leading instability should be SDW order. Beyond mean field, the four-fermion interactions are strongly renormalized by the logarithmically singular fluctuations in particle-particle and particle-hole channel. From the pRG analysis shown in Sec.~\ref{sec:RG}, in the interaction range that stabilize spin ordering, $g_1+g_3>g_1-g_3$, again SDW order wins over ISB order.

To be precise, if $g_1-g_3<0$, i.e. the effective interaction in the antisymmetric spin ordering channel is repulsive, ${\bf{\Phi}}_{\pm K}$ condensates are impossible to develop in any case. In this case, the formulation should be modified as the Hubbard-Stratonovich for the channel with repulsion should be $e^{-w^{\dg}A w}=\int \mc{D}v~e^{-v^{\dg}A^{-1}v+i w^{\dg}v-i v^{\dg}w}$, $A$ positive definite. As there is no essential change of physics, we don't consider this possibility further.

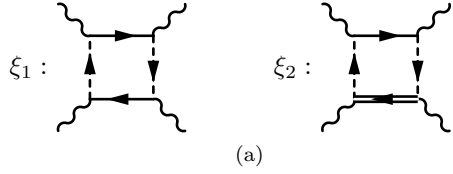
\begin{figure}
  \centering
  \subfigure[]{
  \begin{fmffile}{diag-box}
  \fmfset{arrow_len}{3mm}
  \fmfset{dash_len}{2mm}
  \begin{fmfgraph*}(60,48)
  \fmfleft{i1,i2}
  \fmfright{o1,o2}
  \fmf{photon,tension=2}{i1,v1}
  \fmf{photon,tension=2}{i2,v2}
  \fmf{photon,tension=2}{o1,v3}
  \fmf{photon,tension=2}{o2,v4}
  \fmf{fermion}{v2,v4}
  \fmf{fermion}{v3,v1}
  \fmf{scalar}{v1,v2}
  \fmf{phantom,label=$\xi_1:\quad$,tension=0}{v2,v1}
  \fmf{scalar}{v4,v3}
  \end{fmfgraph*}\qquad\qquad
  \begin{fmfgraph*}(60,48)
  \fmfleft{i1,i2}
  \fmfright{o1,o2}
  \fmf{photon,tension=2}{i1,v1}
  \fmf{photon,tension=2}{i2,v2}
  \fmf{photon,tension=2}{o1,v3}
  \fmf{photon,tension=2}{o2,v4}
  \fmf{fermion}{v2,v4}
  \fmf{double}{v3,v1}
  \fmf{phantom_arrow,tension=0}{v3,v1}
  \fmf{scalar}{v1,v2}
  \fmf{scalar}{v4,v3}
  \fmf{phantom,label=$\xi_2:\quad$,tension=0}{v2,v1}
  \end{fmfgraph*}
  \end{fmffile}
  }
\caption{Feynman diagrams for the quartic terms in the Landau Free energy in Eq.~\ref{appeq:actionquartic}.\label{fig:QuarticCoefficient}
 }
\end{figure}

As terms linear in ${\bf{\Phi}}_{\pm K}$ should vanish in the expansion due to time-reversal symmetry, the ISB instability cannot be triggered by the SDW order in zero field.
We restrict to the SDW channel, and calculate the quartic term by evaluating $\frac{1}{4}\Tr (\mc{G}_k^{(0)}\mc{V})^4$. It is convenient to express the $\bar{\bf{M}}_{\pm K}$ in terms of real and imaginary component of SDW order, and $\bar{\bf{M}}_K=\frac{1}{\sqrt{2}}(\bsdw{r}+i\bsdw{i})$, $\bar{\bf{M}}_{-K}=\frac{1}{\sqrt{2}}(\bsdw{r}-i\bsdw{i})$. $\frac{1}{4}\Tr (\mc{G}_k^{(0)}\mc{V})^4$ in terms of $\{\bsdw{r},\bsdw{i}\}$ is:
\begin{align}\label{appeq:actionquartic}
S_{eff,4}=2(\xi_1+\xi_2)(\bsdw{r}^2+\bsdw{i}^2)^2+8(\xi_1-\xi_2)(\bsdw{r}\times\bsdw{i})^2
\end{align}
where $\xi_1=\int_k (\mc{G}^c_k)^2(\mc{G}^f_{K+k})^2,~\xi_2=\int_k (\mc{G}^c_k)^2\mc{G}^f_{K+k}\mc{G}^f_{-K+k}$ and are shown diagrammatically in Fig.~\ref{fig:QuarticCoefficient}.

For circular Fermi surface, $\mc{G}^f_{K+k}=\mc{G}^f_{-K+k}$, the second term in Eq.~\ref{appeq:actionquartic} vanishes and the degeneracy of SDW order cannot be lifted by the quartic term, consistent with the analysis of Eq.~\ref{eq:quadraticH_1}.

Anisotropy in the Fermi surface breaks the degeneracy similar to the analysis of the iron-based materials on a square lattice~\cite{}. Consider the quadratic spectrum $\epsilon_{\Gamma,\kv}=\frac{k^2}{2m}-\mu$, $\epsilon_{\pm K,\kv}=\frac{k^2}{2m}-\mu+\delta _{\mu} \pm \delta_{ m} \cos3\theta_{\kv}$, we find $\xi_2-\xi_1>0$. Thus to lower the free energy in Eq.~\ref{appeq:actionquartic}, $\sdw{r}\bot\sdw{i}$ and $|\sdw{r}|=|\sdw{i}|$, i.e. the SDW in real space is the $120^{\circ}$ spiral order due to anisotropy of the electron Fermi surface.
\subsection{Effective action in a Zeeman field}
We now derive the effective action in a Zeeman field, and show that the Zeeman field introduces bilinear coupling between $\bar{\bf{M}}$ and $\bar{\bf{\Phi}}$ as $F_{cross}=-\frac{2N_F}{\mu}\sum_{i=\pm K}\im(\bar\Mi\times\bar{\bf \Phi}^*_i)\cdot\vec{h}$.

The Green's function of free electrons in the normal state is:
\begin{align}
\mc{G}_{0,\Gamma}&=\big((i \omega-\epsilon_{\Gamma,\qv})\id+h\sigma_z\big)^{-1}\non\\
\mc{G}_{0,\pm K}&=\big((i \omega-\epsilon_{\pm K,\qv})\id+h\sigma_z\big)^{-1}
\end{align}
The bilinear coupling comes from the crossing terms of $\mc{V}^M$ and $\mc{V}^{\Phi}$ in $\frac{1}{2}\Tr (\mc{G}_{0,k}\mc{V})^2$.
\begin{align}
F_{cross}=&\frac{1}{2\beta}\Tr\big(\mc{G}_{0,k}\mc{V}^M\mc{G}_{0,k}\mc{V}^{\Phi}+\mc{G}_{0,k}\mc{V}^{\Phi}\mc{G}_{0,k}\mc{V}^M\big)\non\\
=&\frac{1}{2\beta}\sum_{i=\pm K}\Tr\big(\mc{G}_{0,\Gamma}\bar\Mi\cdot\vs~\mc{G}_{0,i}\bar{\bf \Phi}^*_i\cdot\vs+\mc{G}_{0,\Gamma}\bar{\bf M}^*_i\cdot\vs~\mc{G}_{0,i}\bar\cur\cdot\vs+(\bar\Mi\leftrightarrow\bar\cur)\big)\non\\
=&4\sum_{i=\pm K}\im(\bar\Mi\times\bar{\bf \Phi}^*_i)\cdot\vec{h}\int(\mc{G}_{0,\Gamma}^{(0)2}\mc{G}_{0,i}^{(0)}-\mc{G}_{0,\Gamma}^{(0)}\mc{G}_{0,i}^{(0)2})
\end{align}
From the second to the third line, we expand $\mc{G}$ in powers of $h$ as $\mc{G}_{0,\Gamma}=\mc{G}_{0,\Gamma}^{(0)}-\mc{G}_{0,\Gamma}^{(0)}h\sigma_z\mc{G}_{0,\Gamma}^{(0)}+\mc{O}(h^2)$ and $\mc{G}_{0,i}=\mc{G}_{0,i}^{(0)}-\mc{G}_{0,i}^{(0)}h\sigma_z\mc{G}_{0,i}^{(0)}+\mc{O}(h^2)$, and use the identities for tracing spin index
\begin{align}
\Tr[(\sigma_z\vec{a}\cdot\vs)(\vec{b}\cdot\vs)]=2i(\vec{a}\times\vec{b})\cdot\hat{z},~~
\Tr[(\vec{a}\cdot\vs)(\sigma_z\vec{b}\cdot\vs)]=-2i(\vec{a}\times\vec{b})\cdot\hat{z}.
\end{align}

The integral $\mc{I}^{(3)}=\int_k(\mc{G}_{\Gamma}^{(0)2}\mc{G}_i^{(0)}-\mc{G}_{\Gamma}^{(0)}\mc{G}_i^{(0)2})$ is:
\begin{align}
\mc{I}^{(3)}=\int\frac{\diff\omega}{2\pi}\frac{\Diff{2}k}{\mc{B}}(\mc{G}_{\Gamma}^{(0)2}\mc{G}_i^{(0)}-\mc{G}_{\Gamma}^{(0)}\mc{G}_i^{(0)2})
=N_F\int\frac{\diff\omega}{2\pi}\diff\epsilon\frac{1}{i\omega+\epsilon}\frac{1}{i\omega-\epsilon}(\frac{1}{i\omega+\epsilon}-\frac{1}{i\omega-\epsilon})
=-\frac{N_F}{2\mu}
\end{align}
\section{Selection of SDW order by electronic correlations}\label{app:Diagonalization}
\subsection{Diagonalize the quadratic Hamiltonian in an SDW state}
The quadratic Hamiltonian of the SDW state Eq.~\ref{eq:HSDW3p} can be diagonalized in two steps. Without loss of generality, we choose $\bsdw{r}$ and $\bsdw{i}$ to be on $x-y$ plane, and set $\bsdw{r}=M_{r}\hat{e}_x$, $\bsdw{i}=\bar M_{ix}\hat{e}_x+\bar M_{iy}\hat{e}_y$. For simplicity, we consider first nesting of the two electron pockets, i.e. $\epsilon_{\bK+\kv}=\epsilon_{-\bK+\kv}=\epsilon_{e,\kv}$. First, the quadratic Hamiltonian is block diagonalized under a rotation of basis from $\{f_{\bK+\kv,\sigma}, f_{-\bK+\kv,\sigma}\}^T$ to $\{f_{a\kv,\sigma}, \bar f_{\kv,\sigma}\}^T$, which mixes fermions around $\bK$ and $-\bK$.
\begin{align}\label{appeq:rot}
f_{\bK+\kv,\uparrow}=b^* f_{a\kv,\uparrow}-a \bar f_{\kv,\uparrow},\non\\
f_{-\bK+\kv,\uparrow}=a^* f_{a\kv,\uparrow}+b \bar f_{\kv,\uparrow},\non\\
f_{\bK+\kv,\downarrow}=a f_{a\kv,\downarrow}-b^* \bar f_{\kv,\downarrow},\non\\
f_{-\bK+\kv,\downarrow}=b f_{a\kv,\downarrow}+a^* \bar f_{\kv,\downarrow},
\end{align}
where $a=\frac{\bar M_r+\bar M_{iy}-i \bar M_{ix}}{\sqrt{2}  \bar M }$, $b=\frac{\bar M_r-\bar M_{iy}+i \bar M_{ix}}{\sqrt{2} \bar M} $. The block diagonalized Hamiltonian is:
\begin{align}  \label{appeq:noninteracting}
\mc{H}_{M}&=\tilde \Psi_{1\kv}^{\dg}H_1\tilde\Psi_{1\kv}+\tilde \Psi_{2\kv}^{\dg}H_2\tilde\Psi_{2\kv}\non\\
H_1&=H_2=
\begin{pmatrix}
\epsilon_{\Gamma,\kv} & -\sqrt{2}\bar M & 0\\
-\sqrt{2}\bar M & \epsilon_{e,\kv} & 0\\
0 & 0 & \epsilon_{e,\kv}
\end{pmatrix},
\end{align}
where $\bar M=\sqrt{|\bsdw{r}|^2+|\bsdw{i}|^2}$, $\tilde \Psi_{1\kv}=\{c_{\kv,\downarrow},~f_{a\kv,\uparrow},~\bar f_{\kv,\uparrow}\}$, and $\tilde\Psi_{2\kv}=\{c_{\kv,\uparrow},~f_{a\kv,\downarrow},~\bar f_{\kv,\downarrow}\}$. From Eq.~\ref{appeq:noninteracting}, one can see that the SDW order couples electron and hole pockets with the opposite spin defined perpendicular to the plane of the SDW order. Moreover, the SDW state is a half-metal with two degenerate bands labeled by fermions $\bar f_{\sigma}$, $\sigma=\uparrow,\downarrow$. The fully diagonalized Hamiltonian can be obtained straight forwardly by the standard Bogolyubov transfromation,
\begin{align}
c_{\kv,\sigma}&=\cos \psi_{\kv}~ p_{\kv,\alpha}+\sin \psi_{\kv} ~e_{\kv,\alpha}\non\\
f_{a\kv,\tilde\sigma}&=-\sin \psi_{\kv} ~p_{\kv,\alpha}+\cos\psi_{\kv} ~e_{\kv,\alpha}
\end{align}
where
\begin{align}
\cos \psi_{\kv}=\sqrt{\frac{E_{\kv}-\epsilon_{\Gamma,\kv}}{2\sqrt{\left(\frac{\epsilon_{\Gamma,\kv} -\epsilon_{\bK+\kv}}{2}\right)^2 + 2{\bar M}^2}}},\,~~
\sin \psi_{\kv}=\sqrt{\frac{E_{\kv}-\epsilon_{\bK+\kv}}{2\sqrt{\left(\frac{\epsilon_{\Gamma,\kv} -\epsilon_{\bK+\kv}}{2}\right)^2 + 2{\bar M}^2}}},
\end{align}
and $\sigma,\tilde\sigma=\uparrow,\downarrow \text{ or } \downarrow,\uparrow$, $\alpha$ labels the pseudo-spin up and down in the new basis. The fully diagonalized quadratic Hamiltonian is expressed as Eq.~\ref{eq:quadraticH}.

We also note that the composition of $\bar f_{\sigma}$ in terms of $f_{\pm \bK+\kv,\sigma}$ depends on SDW order configurations. Interestingly, for the $120^\circ$ SDW order, $a=1,\,b=0$ in Eq.~\ref{appeq:rot}. The metallic bands in the SDW state are $\bar f_{\kv,\uparrow}=-f_{\bK+\kv,\uparrow}, \bar f_{\kv,\downarrow}=f_{-\bK+\kv,\downarrow}$. Such $\pm \bK$ dependent splitting of spin up- and down- bands of electron pockets can also be realized by Ising type spin-orbit coupling, and the interesting superconductivity state of the remaining spin up and down pockets at $\bK$ and $-\bK$, respectively, has been discussed in Ref.~\cite{Hsu2017}.
\subsection{Correction to the ground state energy}\label{sec:}
\begin{figure}[htb!]
  \centering
{\includegraphics[width=0.48\linewidth]{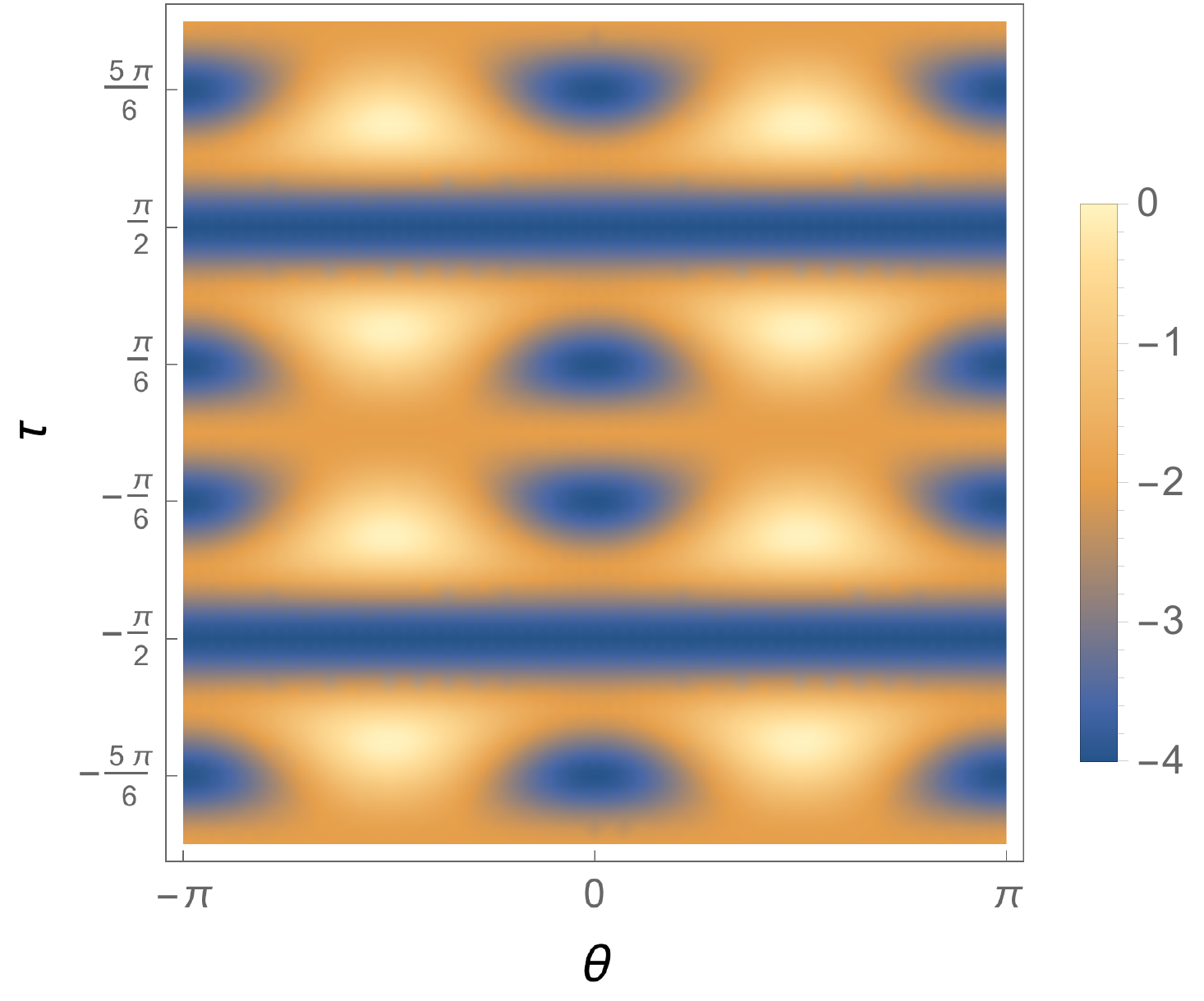}}
  \caption{Correction to the Free energy from $g_8$ interaction - the $\delta F_b (\theta,\tau)$ term, Eq.~\ref{appeq:E8gr}. \label{fig:dEmanifold8}}
\end{figure}

\textit{From $g_8$ -- }
For simplicity, we perform the calculation assuming perfect nesting between electron and hole pockets, i.e. $\epsilon_{\Gamma,\kv}=-\epsilon_{\pm \bK+\kv}$. From Eq.~\ref{eq:QuadraticG8}, the corrections to the free energy $\delta F_b$ obtained from second order perturbation is:
\begin{align}\label{appeq:E8gr}
\delta F_b&=-(\gamma_{8}  \bar M)^2\sum_{\kv}\frac{\Delta^2}{E_{\kv}}\big(\frac{|\kappa_1|^2}{4E_{\kv}^2}+\frac{|\kappa_2|^2}{(E_{\kv}+|\epsilon_{\kv}|)^2}\big)\non\\
&=-N_F(\gamma_{8}  \bar M)^2(\frac{|\kappa_1|^2}{2}+|\kappa_2|^2)
\end{align}
where $\kappa_1,~\kappa_2$ come from the vertex corrections from $H_{g_8}$ in the canonical basis of $\{e,~p,~f_b\}$ that couple $\{e,~p\}$ and $\{e \text{ or } p, ~ f_b\}$, respectively. $\kappa_1,~\kappa_2$ depends on the SDW order configuration, which is characterized by the magnitude $M$ and two angles $\tau,~\theta$ defined above Eq.~\ref{eq:SDW3p}. We found
\begin{align}
\kappa_1&=2   \cos \tau\left(\cos ^2\tau -\left(2+e^{-2 i \theta }\right) \sin ^2\tau \right),\non\\
\kappa_2&=2  \sin \tau  (\sin \theta  \cos 2 \tau -i \cos \theta  (2 \cos 2 \tau +1)).
\end{align}
Plug $\kappa_1,~\kappa_2$ into Eq.~\ref{appeq:E8gr}, we obtain Eq.~\ref{eq:E8gr} in the main text. $\delta F_b(\theta,\tau)$ is plotted in Fig.~\ref{fig:dEmanifold8}, the minimum are located at values of $\theta,\tau$ that $\theta=-\pi,~0$ and $\tau=\pm\pi/6,~\pm5\pi/6$ or $\tau=\pm\pi/2$ and all $\theta$. The correction to ground state energy at these $\tau,\theta$ is $\delta F_b=-4N_F(\gamma_{8}  \bar M)^2$.

\section{Real space  SDW and ISB orders }
\label{app:order}
The real space SDW order  ${\bf M}_{\bf r}$  and ISB order ${\bf \Phi} _{{\bf r}, {\vect \delta}}$
 are related to the SDW and ISB order parameters in momentum space by
\begin{align}\label{appeq:SDWOP}
M^{\alpha}_{\bf r}&=\langle\fd_{ r}\sigma^{\alpha}c_{ r}+\cd_{ r}\sigma^{\alpha}f_{ r}\rangle\non\\
&=\sum_{k,k'}\langle e^{i (k'-k) r}\fd_{k}\sigma^{\alpha}c_{k'}+e^{-i (k'-k) r}\cd_{k'}\sigma^{\alpha}f_{k}\rangle\non\\
&=\sum_{q',q,Q_i}\langle e^{i (q'-q-Q_i) r}\fd_{Q_i+q}\sigma^{\alpha}c_{q'}+e^{-i (q'-q-Q_i) r}\cd_{q'}\sigma^{\alpha}f_{q+Q_i}\rangle\non\\
&=\sum_{q,Q_i}\langle e^{-i Q_i r}\fd_{Q_i+q}\sigma^{\alpha}c_{q}+e^{i Q_i r}\cd_{q}\sigma^{\alpha}f_{q+Q_i}\rangle\non\\
&=\sum_{Q_i}(e^{-i Q_i r}\Delta^{\alpha}_{Q_i}+h.c.)\non\\
&=\frac{1}{2}\sum_{Q_i}(e^{-i Q_i r}(\Delta^{\alpha}_{Q_i}+\Delta^{\alpha*}_{-Q_i})+h.c.)
\end{align}
where $\br$ is the coordinate of site, $\Delta^{\alpha}_{Q_i}=\sum_{q}\langle\fd_{Q_i+q}\sigma^{\alpha}c_{q}\rangle$, $Q_i=\pm K$ for the 3p model on a hexagonal lattice. The last line is obtained by averaging over condensates with $Q_i$ and $-Q_i$. From the last line, it is clear that only the \textit{symmetric} component of $\{\Delta^{\alpha}_{Q_i},\Delta^{\alpha*}_{-Q_i}\}$ contributes to the density-wave order. Similarly, only the {\it{symmetric }}component contributes to bond-SDW order defined as $M^{\alpha}_{\br,\vect{\delta}}\sim\langle\fd_{ r+\vect\delta/2}\sigma^{\alpha}c_{ r-\vect\delta/2}+\cd_{ r+\vect\delta/2}\sigma^{\alpha}f_{ r-\vect\delta/2}\rangle+h.c.$ On the other hand, the {\it{antisymmetric}} component contributes to ISB order parameter
\begin{align}\label{appeq:BSCOP}
\Phi^{\alpha}_{{\bf r},{\vect \delta}}&=\frac{i}{\hbar}\hat{\delta}\langle (\fd_{r+\delta/2}\sigma^{\alpha}c_{r-\delta/2}-\cd_{ r-\delta/2}\sigma^{\alpha}f_{ r+\delta/2})+(\cd_{r+\delta/2}\sigma^{\alpha}f_{r-\delta/2}-\fd_{ r-\delta/2}\sigma^{\alpha}c_{ r+\delta/2})\rangle\non\\
&=\frac{i}{\hbar}\hat{\delta}\sum_{k,k'}\langle ( e^{i (k'-k) r}e^{-i(k'+k)\delta/2}\fd_{k}\sigma^{\alpha}c_{k'}+ e^{i (k-k') r}e^{-i(k'+k)\delta/2}\cd_{k'}\sigma^{\alpha}f_{k})-h.c.\rangle\non\\
&=\frac{i}{\hbar}\hat{\delta}\sum_{q',q,Q_i}\langle ( e^{i (q'-q-Q_i) r}e^{-iQ_i\delta/2}\fd_{Q_i+q}\sigma^{\alpha}c_{q'}+ e^{i (q+Q_i-q') r}e^{-iQ_i\delta/2}\cd_{q'}\sigma^{\alpha}f_{q+Q_i})-h.c.\rangle\non\\
&=\frac{i}{\hbar}\hat{\delta}\sum_{q,Q_i}\langle ( e^{-iQ_i r}e^{-iQ_i\delta/2}\fd_{Q_i+q}\sigma^{\alpha}c_{q}+ e^{iQ_i r}e^{-iQ_i\delta/2}\cd_{q}\sigma^{\alpha}f_{q+Q_i})-h.c.\rangle\non\\
&=\frac{i}{\hbar}\hat{\delta}\sum_{Q_i}( e^{-iQ_i r}e^{-iQ_i\delta/2}\Delta^{\alpha}_{Q_i}+ e^{iQ_ir}e^{-iQ_i\delta/2}\Delta^{\alpha *}_{Q_i})-h.c.\non\\
&=\frac{i}{2\hbar}\hat{\delta}\sum_{Q_i} (e^{-iQ_i r}e^{-iQ_i\delta/2}\Delta^{\alpha}_{Q_i}+ e^{iQ_ir}e^{-iQ_i\delta/2}\Delta^{\alpha *}_{Q_i}+e^{iQ_i r}e^{iQ_i\delta/2}\Delta^{\alpha}_{-Q_i}+ e^{-iQ_ir}e^{iQ_i\delta/2}\Delta^{\alpha *}_{-Q_i})-h.c.\non\\
&=\frac{i}{2\hbar}\hat{\delta}\sum_{Q_i}( e^{-iQ_i r}e^{-iQ_i\delta/2}- e^{-iQ_i r}e^{iQ_i\delta/2})(\Delta^{\alpha}_{Q_i}-\Delta^{\alpha*}_{-Q_i})-h.c.,
\end{align}
where the bond is defined as from site $\br-\vect\delta/2$ to site $\br+\vect\delta/2$. In transforming from the second to the third line we used the fact that $e^{i q \delta}\approx 1$ because Fermi pockets are small.

We define the \textit{symmetric} component of $\Delta^{\alpha}_{Q_i}$ as $M^{\alpha}_{Q_i}=\frac{\Delta^{\alpha}_{Q_i}+\Delta^{\alpha*}_{-Q_i}}{2}$ for the density-wave part, define the \textit{antisymmetric} component of $\Delta^{\alpha}_{Q_i}$ as $\Phi^{\alpha}_{Q_i}=\frac{\Delta^{\alpha}_{Q_i}-\Delta^{\alpha*}_{-Q_i}}{2}$ for the imaginary bond-order part. In particular, when $Q_i=-Q_i$, $M^{\alpha}_{Q_i}=\frac{\Delta^{\alpha}_{Q_i}+\Delta^{\alpha*}_{Q_i}}{2}$ must be real, and $\Phi^{\alpha}_{Q_i}=\frac{\Delta^{\alpha}_{Q_i}-\Delta^{\alpha*}_{Q_i}}{2}$ must be imaginary.

The SDW and ISB order parameter in real space can be re-expressed in terms of $M$ and $\Phi$ as
\begin{align}\label{appeq:RealSO}
M^{\alpha}(\br)&=2\sum_{Q_i}|M^{\alpha}_{Q_i}|\cos (Q_i r -\phi_{i,\alpha})\non\\
\Phi^{\alpha}_{\br,\vect\delta}&=\frac{4}{\hbar}\hat{\delta}\sum_{Q_i}|\Phi^{\alpha}_{Q_i}|\sin{(Q_i\delta/2)}\cos{(Q_i r -\varphi_{i,\alpha})}
\end{align}
where $\phi_{i,\alpha}$ and $\varphi_{i,\alpha}$ are the phase of $M^{\alpha}_{Q_i}$ and $\Phi^{\alpha}_{Q_i}$ respectively.

\section{Spin ordering instability in a magnetic field}\label{app:SpinOrderingInst}
In this section, we show details of solving the linearized spin ordering equations in (i) $\sigma^{\pm}$ and (ii) $\sigma^z$ channel  assuming perfect nesting between electron and hole pockets (Eq.~\ref{eq:LinearEq} in the main text),
\begin{align}\label{appeq:LinearEq}
\text{(i)} ~~&~~1+\frac{1}{2}\Big(g_1(\Pi_{+}+\Pi_{-})-\big((\Pi_+-\Pi_-)^2 g_1^2+4\Pi_+\Pi_- g_3^2\big)^{1/2}\Big)=0,\non\\
\text{(ii)} ~~& ~~1+(g_1+g_3)\Pi_{z}=0.
\end{align}
To solve for Eq.~\ref{appeq:LinearEq} requires calculating the particle-hole polarization $\Pi_{ph}$ for different spin channels, where
\begin{align}
\Pi_{ph}&=T\sum_{\omega_n}\int \frac{\diff^2 k}{\mc{A}_{B.Z.}}\mc{G}^f(\kv\pm K)\mc{G}^c(\kv)=\int \frac{\diff^2 k}{\mc{A}_{B.Z.}}\frac{n_F(\epsilon_{\kv})-n_F(\epsilon_{\kv\pm K})}{\epsilon_{\kv}-\epsilon_{\kv\pm K}}=N_F\int \diff \epsilon_{\kv}\frac{n_F(\epsilon_{\kv})-n_F(\epsilon_{\kv\pm K})}{\epsilon_{\kv}-\epsilon_{\kv\pm K}}.
\end{align}
In the following, we discuss the result of the integral for different band structure configurations.

In zero field, $\Pi_{ph}$ is
\begin{align}
\Pi_{ph,0}=N_F\int_{-\mu}^{\Lambda} \diff \epsilon\frac{n_F(\epsilon)-n_F(-\epsilon)}{2\epsilon}=-\frac{1}{2}N_F\int_{-\mu}^{\Lambda}\diff \epsilon\frac{\tanh \frac{\beta \epsilon}{2}}{\epsilon}\sim-\frac{1}{2}N_F\ln \frac{\mu}{T}+const.
\end{align}

In a Zeeman field, with $\mc{H}_Z=-\vect{h}\cdot\sum_{\kv} (\cd_{\kv}\vect{\sigma} c_{\kv}+\fd_{\kv}\vect{\sigma} f_{\kv})$, the particle and hole pockets involved in the spin ordering in the $\sigma^{\pm}$ channel remain perfectly nested, and $\epsilon_{\kv\pm K,\uparrow}=-\epsilon_{\kv,\downarrow}=\frac{k^2}{2m}-\mu-h$, $\epsilon_{\kv\pm K,\downarrow}=-\epsilon_{\kv,\uparrow}=\frac{k^2}{2m}-\mu+h$. As a result, the band splitting only modifies the energy at the bottom of the band, i.e. the high energy cutoff in the integral from $\mu\rightarrow \mu\pm h$.
\begin{align}
\Pi_{ph,\pm}=-\frac{1}{2}N_F\int_{-(\mu\pm h)}^{\Lambda\mp h}\diff \epsilon\frac{\tanh \frac{\beta \epsilon}{2}}{\epsilon}\sim-\frac{1}{2}N_F\big(\ln \frac{\mu\pm h}{T}+const.\big)=-(|\Pi_{ph,0}|\pm\frac{1}{2}N_F\frac{h}{\mu})
\end{align}
Plug it into Eq.~\ref{appeq:LinearEq}, to the leading order in $h/\mu$, the solution to the linearized ordering equation in $\sigma^{\pm}$ channel becomes
\begin{align}\label{appeq:OrderingTr}
1+(g_1+g_3)\Pi_0(T)\Big(1-\frac{g_3-g_1}{4g_3}\big(\frac{N_F}{\Pi_0}\big)^2\big(\frac{h}{\mu}\big)^2\Big)=0.
\end{align}

In the $\sigma_z$ channel, the particle-hole symmetry between the involved bands is broken. For example, in the evaluation of $\Pi_{ph,\uparrow}$, as $\epsilon_{\kv\pm K,\uparrow}=\frac{k^2}{2m}-\mu-h$, $\epsilon_{\kv,\uparrow}=-(\frac{k^2}{2m}-\mu+h)=-\epsilon_{\kv\pm K,\uparrow}-2h$,
\begin{align}\label{appeq:PolarH}
\Pi_{ph,\uparrow}&=\int \frac{\diff^2 k}{\mc{A}_{B.Z.}}\frac{n_F(\epsilon_{\kv,\uparrow})-n_F(\epsilon_{\kv\pm K,\uparrow})}{\epsilon_{\kv,\uparrow}-\epsilon_{\kv\pm K,\uparrow}}=N_F\int_{-(\mu+h)}^{\Lambda} \diff \epsilon\frac{n_F(-\epsilon-2h)-n_F(\epsilon)}{-2\epsilon-2h}\non\\
&=-\frac{N_F}{2}\int_{-(\mu+h)}^{\Lambda} \diff \epsilon\frac{1}{\epsilon+h}\big(\frac{1}{e^{\beta(-\epsilon-2h)}+1}-\frac{1}{e^{\beta\epsilon}+1}\big)\non\\
&=-\frac{N_F}{2}\int_{-\mu}^{\Lambda} \diff \epsilon\frac{1}{\epsilon}\big(\frac{1}{e^{-\beta\epsilon-\beta h}+1}-\frac{1}{e^{\beta\epsilon-\beta h}+1}\big)=-\frac{N_F}{2}\int_{-\mu\beta}^{\Lambda\beta} \diff x\frac{1}{x}\big(\frac{1}{e^{-x-\beta h}+1}-\frac{1}{e^{x-\beta h}+1}\big)
\end{align}
Similarly, $\Pi_{ph,\downarrow}=-\frac{N_F}{2}\int_{-\mu}^{\Lambda} \diff \epsilon\frac{1}{\epsilon}\big(\frac{1}{e^{-\beta\epsilon+\beta h}+1}-\frac{1}{e^{\beta\epsilon+\beta h}+1}\big)=\Pi_{ph,\uparrow}$. To evaluate the integral in Eq.~\ref{appeq:PolarH}, we note that the integrand is a function  of $\beta h$, and it behaves differently in the cases of $\beta h\ll 1$ and $\beta h\gg 1$.

\textit{The limit $h\ll T$ --} The integral is suppressed only near $\epsilon=0$, to the leading order in $\beta h$, we have
\begin{align}
\Pi_{ph,\uparrow}&=-\frac{N_F}{2}\int_{-\mu}^{\Lambda} \diff \epsilon\frac{1}{\epsilon}\big(\frac{1}{e^{-\beta\epsilon-\beta h}+1}-\frac{1}{e^{\beta\epsilon-\beta h}+1}\big)=-\frac{N_F}{2}\int_{-\mu}^{\Lambda} \diff \epsilon\frac{1}{\epsilon}\tanh\frac{\beta\epsilon}{2}\Big(1-\frac{(\beta h)^2}{4}\frac{1}{\cosh^2\frac{\beta\epsilon}{2}}\Big)+\mc{O}(\beta h)^3\non\\
&=-\frac{N_F}{2}\Big(\int_{-\mu}^{\Lambda} \diff \epsilon\frac{1}{\epsilon}\tanh\frac{\beta\epsilon}{2}-\beta^2 h^2\int_{-\mu}^{\Lambda} \diff \epsilon\frac{1}{4\epsilon}\tanh\frac{\beta\epsilon}{2}\frac{1}{\cosh^2\frac{\beta\epsilon}{2}}\Big)=-(|\Pi_{ph,0}|-\frac{0.85}{2}N_F\,\frac{h^2}{T^2})
\end{align}
where $\int_{-\mu}^{\Lambda} \diff \epsilon\frac{1}{4\epsilon}\tanh\frac{\beta\epsilon}{2}\frac{1}{\cosh^2\frac{\beta\epsilon}{2}}=0.85$ is evaluated numerically in the limit $\beta \mu,\beta \Lambda\gg 1$. Plug $\Pi_{ph,\uparrow},\Pi_{ph,\downarrow}$ into Eq.~\ref{appeq:LinearEq}, to the leading order in $h/T$, the solution to the linearized ordering equation in $\sigma^{z}$ channel becomes
\begin{align}\label{appeq:OrderingZ}
1+(g_1+g_3)\Pi_0(T)\big(1-0.43\frac{N_F}{|\Pi_0|}\,\frac{h^2}{T^2}\big)=0
\end{align}
Comparing Eq.~\ref{appeq:OrderingTr} and Eq.~\ref{appeq:OrderingZ}, as $\frac{N_F}{|\Pi_0|}\ll 1$, $\frac{h}{\mu}\ll \frac{h}{T}$, we conclude that the ordering instability in the $\sigma^{\pm}$ channel develops first.

\textit{The limit $h\gg T$ --} $\beta h$ modifies the integrand non perturbatively and sets the cutoff in the integral as $\beta h$. As a result, $\Pi_{ph,\uparrow}$ simply changes to $\Pi_{ph,\uparrow}=-\frac{1}{2}N_F(\ln \frac{\mu}{h}+const.)$. Because $h\gg T$, $\Pi_{ph,\uparrow}\ll \Pi_{ph,0}$ in this limit. The correction to $\Pi_{ph,\pm}$ remains the same dependance on $h/\mu$ as in the limit $h\ll T$. Due to the further non-perturbative suppression of $\Pi_{ph}$ in $\sigma^z$ channel, the ordering instability must also first develop in the $\sigma^{\pm}$ channel in the $h\gg T$ limit.

We also did a similar analysis when the particle-hole symmetry of the band structure in zero field is slightly broken, i.e. $\epsilon_{\kv}=-\epsilon_{\kv\pm K}+\delta \mu$, and $\delta \mu \ll \mu$. The conclusion remains unchanged.

\end{document}